\documentclass[aps,pra,reprint,floatfix,superscriptaddress]{revtex4-2}
\usepackage{amsmath,amssymb}
\usepackage{graphicx}
\usepackage[bookmarks=false,colorlinks=true,linkcolor=blue,filecolor=blue,citecolor=blue,urlcolor=blue]{hyperref}
\usepackage{bm}
\usepackage{physics}
\usepackage{soul}
\usepackage{dcolumn}
\usepackage{appendix}
\usepackage{here}
\usepackage{color}
\usepackage{ulem}

\begin{document}

\title{Charge fluctuation and charge-resolved entanglement in a monitored quantum circuit with $U(1)$ symmetry}

\author{Hisanori Oshima}
\affiliation{Department of Applied Physics, University of Tokyo, Tokyo 113-8656, Japan}

\author{Yohei Fuji}
\affiliation{Department of Applied Physics, University of Tokyo, Tokyo 113-8656, Japan}

\date{\today}

\begin{abstract}
We study a (1+1)-dimensional quantum circuit consisting of Haar-random unitary gates and projective measurements, both of which conserve a total $U(1)$ charge and thus have $U(1)$ symmetry. 
In addition to a measurement-induced entanglement transition between a volume-law and an area-law entangled phase, we find a phase transition between two phases characterized by bipartite charge fluctuation growing with the subsystem size or staying constant. 
At this charge-fluctuation transition, steady-state quantities obtained by evolving an initial state with a definitive total charge exhibit critical scaling behaviors akin to Tomonaga-Luttinger-liquid theory for equilibrium critical quantum systems with $U(1)$ symmetry, such as logarithmic scaling of bipartite charge fluctuation, power-law decay of charge correlation functions, and logarithmic scaling of charge-resolved entanglement whose coefficient becomes a universal quadratic function in a flux parameter. 
These critical features, however, do not persist below the transition in contrast to a recent prediction based on replica field theory and mapping to a classical statistical mechanical model.
\end{abstract}

\maketitle
\tableofcontents

%%%%%%%%%%%%%%%%%%%%%%%%%%%%%%%%%%%%%%%%%%%%%%%%%%%%%%%%%%%%%%%%%%%%%%%%%%%%%%%%%%%%%
%~~~~~~~~~~~~~~~~~~~~~~~~~~~~~~~~~~~~~ SECTION ~~~~~~~~~~~~~~~~~~~~~~~~~~~~~~~~~~~~~~
% Introduction
\section{Introduction}\label{sec:Introduction}
%%%%%%%%%%%%%%%%%%%%%%%%%%%%%%%%%%%%%%%%%%%%%%%%%%%%%%%%%%%%%%%%%%%%%%%%%%%%%%%%%%%%%

Quantum many-body systems evolved under repeated measurements have recently been shown to harbor a rich variety of phases and phase transitions that have no counterparts in equilibrium \cite{Potter22, Fisher22}. 
The unitary dynamics for typical thermalizing systems causes linear growth of bipartite entanglement entropy, whose saturation value after a long time scales extensively with the volume of the subsystem. 
Projective measurements of local operators, on the other hand, disentangle local degrees of freedom from the rest of the system and suppress the growth of the entanglement. 
The competition between unitary time evolution and repeated measurements leads to an \textit{entanglement transition} between a volume-law and an area-law entangled phase \cite{Li18, Skinner19, Chan19}. 
Surprisingly, the entanglement transition in (1+1) dimensions exhibits emergent conformal invariance, akin to equilibrium critical phenomena, which is not captured by physical quantities linear in the density matrix but revealed in characteristic behaviors of nonlinear quantities, such as logarithmic scaling of the entanglement entropy and algebraic decays of squared correlation functions \cite{Li19}. 
While such measurement-induced phase transitions (MIPTs) have been intensively studied in unitary-measurement-hybrid quantum circuits \cite{Szyniszewski19, Choi20, Iaconis20, Tang20, Turkeshi20, Zabalo20, Fan21, Lavasani21a, Lavasani21b, Li21a, Lu21, Lunt21, Sang21, Cote22, Sierant22b, Zabalo22}, they are expected to occur in a diverse range of monitored quantum systems, such as measurement-only quantum circuits \cite{Lang20, Ippoliti21, Klocke22, Lavasani22, Sriam22}, free fermion systems \cite{Alberton21, Buchhold21, Kells21, Turkeshi21, Piccitto22, Turkeshi22}, interacting systems subject to continuous monitoring \cite{Fuji20, Szyniszewski20, SKJian21, VanRegemortel21, Boorman22, Doggen22}, and long-range interacting systems \cite{Sahu21, Block22, Hashizume22, Minato22, Muller22, Sierant22a, Sharma22}. 

However, universal properties of generic MIPTs remain to be well understood, except for certain models that can be mapped to classical percolation problems \cite{Bao20, Jian20a}. 
As indicated by the emergent conformal invariance, the MIPTs appear to share common features with equilibrium phase transitions, for which symmetry plays an indispensable role in understanding their universality classes. 
In fact, hybrid quantum circuits with two competing measurements that preserve a global $\mathbb{Z}_2$ symmetry have been shown to possess two distinct area-law phases separated by an entanglement transition \cite{Sang21, Lang20} and thus bear a strong resemblance with the Ising model in equilibrium (see also Refs.~\cite{Lavasani21a, Nahum20, Ippoliti21, Li21b, Han22, Klocke22} for related studies). 
It has also been argued that, although measurement outcomes are intrinsically random, translation symmetry of the corresponding statistical ensemble in combination with a global symmetry, such as an $SU(2)$ spin rotation symmetry, gives rise to super-area-law entanglement \cite{Nahum20}, in analogy with the Lieb-Schultz-Mattis theorem for ground states of quantum many-body systems \cite{Tasaki22}. 
Furthermore, it has been shown that interplay between global symmetry and dynamically generated symmetry acting on a replica space leads to a variety of exotic measurement-induced phases \cite{Bao21}. 

Recently, it has been predicted that hybrid quantum circuits with $U(1)$ symmetry, or equivalently particle-number conservation, undergo a novel type of MIPT distinguished from the entanglement transition as the measurement rate is increased \cite{Agrawal22, Barratt22a, Barratt22b}. 
For a hybrid quantum circuit consisting of charged qubits and neutral qudits with $d$ levels, mapping to a classical statistical mechanical model, which becomes analytically tractable in the limit of large $d$, has been employed to show the presence of \textit{charge-sharpening transition} within the volume-law phase of entanglement \cite{Agrawal22, Barratt22a}. 
For the $d=1$ case, which reduces to the monitored Haar-random circuit with $U(1)$ symmetry, Ref.~\cite{Agrawal22} has numerically shown that the charge-sharpening transition can be dynamically characterized when the initial state mixes different charge sectors; an ancilla probe or charge variance can be used to quantify a time duration required for the initial state to collapse into a single charge sector, which grows linearly with the system size in a charge fuzzy phase below the transition whereas sublinearly in a charge sharp phase above the transition. 
In Ref.~\cite{Barratt22a}, the statistical mechanical model in the $d \to \infty$ limit has been studied by both numerical and field-theoretical approaches to show that the charge-sharpening transition is of Berezinskii-Kosterlitz-Thouless (BKT) type and the charge fuzzy phase below the transition exhibits critical steady-state properties described by Tomonaga-Luttinger-liquid (TLL) theory.

In this paper, we numerically investigate TLL-like critical phenomena emerging from the monitored Haar-random circuit with $U(1)$ symmetry. 
While this model has already been studied in Ref.~\cite{Agrawal22}, we exclusively focus on static, steady-state properties obtained by evolving an initial state within a single charge sector at a given filling fraction. 
Besides the entanglement transition between the volume-law and area-law phase, we identify another phase transition, dubbed \textit{charge-fluctuation transition}, which separates two phases where bipartite charge fluctuation grows with the subsystem size below the transition whereas stays constant above the transition. 
In the vicinity of the charge-fluctuation transition, we find that bipartite charge fluctuation, (unsquared) charge correlation functions, and charge-resolved entanglement all exhibit scaling behaviors peculiar to critical systems described by TLL theory. 
While one may think that the charge-fluctuation transition coincides with the charge-sharpening transition dynamically located in Ref.~\cite{Agrawal22}, as the former also exists slightly below the entanglement transition, we cannot find clear signatures of the BKT-type universality at the charge-fluctuation transition or an extended critical phase described by TLL theory below the transition as predicted from mapping to the classical statistical mechanical model in Ref.~\cite{Barratt22a}. 
Our results thus call for more careful studies on universal properties of measurement-induced criticality in the presence of $U(1)$ symmetry.

The rest of this paper is organized as follows. 
In Sec.~\ref{sec:Model}, we describe our monitored quantum circuit with $U(1)$ symmetry and simulation protocol.
In Sec.~\ref{sec:Numerical results}, we present our numerical results with particular focus on the half-filling case and discuss the presence of an entanglement transition and a charge-fluctuation transition. 
We then analyze scaling properties of various steady-state quantities at and below the transitions. 
Numerical results for other filling fractions are provided in Appendix~\ref{app:Numerical results for N=L/4 system}.
We conclude in Sec.~\ref{sec:Disucussion} with discussions and future directions.

%%%%%%%%%%%%%%%%%%%%%%%%%%%%%%%%%%%%%%%%%%%%%%%%%%%%%%%%%%%%%%%%%%%%%%%%%%%%%%%%%%%%%
%~~~~~~~~~~~~~~~~~~~~~~~~~~~~~~~~~~~~~ SECTION ~~~~~~~~~~~~~~~~~~~~~~~~~~~~~~~~~~~~~~
% Model
\section{Model}\label{sec:Model}
%%%%%%%%%%%%%%%%%%%%%%%%%%%%%%%%%%%%%%%%%%%%%%%%%%%%%%%%%%%%%%%%%%%%%%%%%%%%%%%%%%%%%

\begin{figure}[tbp]
\includegraphics[width=0.4\textwidth]{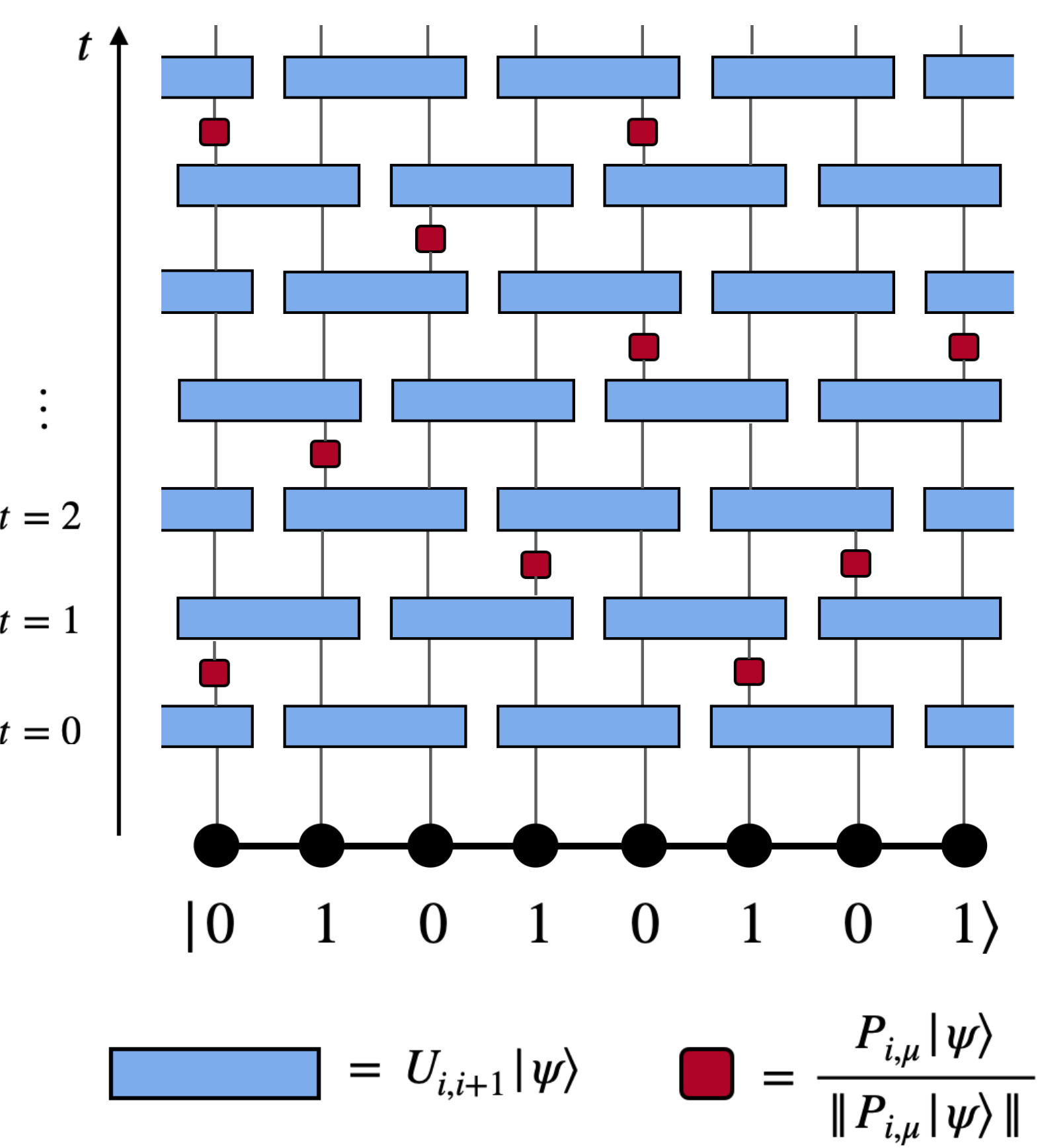}
\caption{\label{fig:circuit}
Schematic picture of the hybrid quantum circuit. 
The horizontal and vertical axis correspond to a spatial and temporal direction, respectively. 
An initial N\'eel state $\ket{1010 \cdots 10}$ is evolved by two-site Haar-random unitary gates preserving the $U(1)$ symmetry (blue rectangles) and by projective measurements of qubits in the Pauli $Z$ basis (red squares).
}
\end{figure}

We consider a (1+1)-dimensional [(1+1)D] hybrid quantum circuit consisting of local unitary gates and interspersed local projective measurements, both of which preserve a global $U(1)$ symmetry \cite{Agrawal22}, as schematically shown in Fig.~\ref{fig:circuit}. 
%====================================================================================
%~~~~~~~~~~~~~~~~~~~~~~~~~~~~~~~~~~~~~~ FIGURE ~~~~~~~~~~~~~~~~~~~~~~~~~~~~~~~~~~~~~~
%====================================================================================
The system is defined on a one-dimensional chain of qubits $q_i \in \{0,1\}$ with the length $L$ where $i \in [1,L]$ denotes the site index. 
We impose the periodic boundary condition such that $q_{L+1} \equiv q_1$. 
We introduce a charge operator $n_i$ acting on a local Hilbert space labeled by site $i$ as $n_i \ket{q_i} = q_i \ket{q_i}$, which can be written as
\begin{align}
    n_i = \frac{\mathbb{I}_i-Z_i}{2},
\end{align}
in terms of the Pauli operator $Z_i$ and the identity operator $\mathbb{I}_i$.
We then define the total charge operator by
\begin{align}
    n_\textrm{tot} = \sum_{i=1}^L n_i \label{def:charge},
\end{align}
which is conserved during time evolution by the hybrid quantum circuits. 
Since $q_i$ can also be interpreted as the local particle number for a boson, we use the terms of charge and particle number interchangeably throughout this paper. 

The local unitary gate is a $4 \times 4$ unitary matrix acting on two neighboring sites and is chosen to take the block-diagonal form,
\begin{align}
    U_{i,i+1} = \mqty( \dmat{U_{1\times1}, U_{2\times2}, U_{1\times1}} ),
    \label{unitarygates}
\end{align}
with $U_{n\times n}$ being an $n \times n$ unitary matrix, such that it commutes with the total charge operator $n_\textrm{tot}$ in Eq.~\eqref{def:charge}. 
These unitary gates are arranged in a brick-wall fashion in spacetime (see Fig.~\ref{fig:circuit}). 
Thus, in the absence of measurements, a state $\ket{\psi(t)}$ at time $t$ is evolved by
\begin{align}
   \ket{\tilde{\psi}(t)} = \left( \bigotimes_{i \in \mathcal{S}_t} U_{i,i+1} \right) \ket{\psi(t)},
\end{align}
within a single time step. 
Here, $\mathcal{S}_t=\{1,3,...,L-1\}$ for odd $t>0$,  and $\mathcal{S}_t=\{2,4,...,L\}$ for even $t>0$. 
At each time and for each link $(i,i+1)$, the unitary matrices $U_{n\times n}$ in Eq.~\eqref{unitarygates} are independently drawn from a Haar-random distribution, which can be generated by following Ref.~\cite{Mezzadri06}: 
First, we create a random $n\times n$ matrix $M$ such that each element follows a complex normal distribution (i.e., $M$ belongs to the Ginibre ensemble). 
We then apply the QR decomposition $M=QR$ to obtain a unitary matrix $Q$ and an upper triangular matrix $R$. 
By multiplying the diagonal matrix $\Lambda = \mathrm{diag}(R_{11} /|R_{11}|, \cdots, R_{nn}/|R_{nn}|)$ to $Q$, we finally obtain $Q'=Q\Lambda$ whose distribution is given by the Haar measure on $U(n)$.

At every time step after applications of the local unitary gates, each qubit is measured with probability $p$ in the Pauli $Z$ basis. 
The measurement outcome $\mu = \{ +1, -1 \}$ for a state $\ket{\tilde{\psi}(t)}$ is obtained with the Born probability 
\begin{align}
p_\mu = \bra{\tilde{\psi}(t)} P_{i,\mu} \ket{\tilde{\psi}(t)},
\end{align}
where we have defined projectors onto the eigenstates of $Z_i$ by
\begin{align}
    \label{projectors}
    P_{i,\mu} = \frac{\mathbb{I}_i + \mu Z_i}{2}.
\end{align}
According to the measurement outcome $\mu$, the state is updated after the measurement to be
\begin{align}
    \ket{\tilde{\psi}(t)} \to \frac{P_{i,\mu}\ket{\tilde{\psi}(t)}}{\norm{P_{i,\mu}\ket{\tilde{\psi}(t)}}}.
\end{align}
When this process of local projective measurements runs over all sites, the time evolution within a single time step is completed and yields a state $\ket{\psi(t+1)}$.

Since both unitary gates in Eq.~\eqref{unitarygates} and projectors in Eq.~\eqref{projectors} commute with the total charge operator in Eq.~\eqref{def:charge}, if the initial state $\ket{\psi(0)}$ is an eigenstate of $n_\textrm{tot}$ with eigenvalue $N$, the evolved state $\ket{\psi(t)}$ is kept an eiganstate of $n_\textrm{tot}$ with the same eigenvalue. 
Indeed, we only consider such initial states with fixed $N$ in the following analysis. 
Specifically, for a given filling fraction $\bar{n} = N/L$ with $L$ divisible by $N$, we choose the initial state to be a ``N\'eel state'', which is a product state formed by alternating single $\ket{1}$'s and $1/\bar{n}-1$ consecutive $\ket{0}$'s:
\begin{align}
    \ket{\psi(0)} = \prod_{n=0}^{N-1} X_{n/\bar{n} +1} \ket{00 \cdots 0},
\end{align}
where $X_i$ is the Pauli $X$ operator acting on a single qubit as $X_i \ket{q_i} = \ket{1-q_i}$.
For instance, we have $\ket{\psi(0)} = \ket{1010 \cdots 10}$ for $\bar{n}=1/2$. 

Starting from an initial pure state $\ket{\psi(0)}$, we repeat the above procedures of unitary evolution and projective measurement to obtain a pure state $\ket{\psi(t)}$ at time $t$. 
Such a pure state is called the \textit{quantum trajectory} and specified by a given choice of unitary gates and measurement positions and also by measurement outcomes. 
Given the pure-state density matrix corresponding to a trajectory $\rho(t) = \ket{\psi(t)}\bra{\psi(t)}$ at time $t$, any physical quantity $O_A[\rho(t)]$, such as entanglement entropy or correlation functions, supported on a spatial region $A$, is computed after application of unitary gates and subsequent projective measurements. 
We then take an average $\overline{O_A[\rho(t)]}$ over different trajectories, which are generated for randomly drawn unitary gates and measurement positions and intrinsically random measurement outcomes. 
We note that such a quantity averaged over different trajectories conditioned on measurement outcomes is generally different from the unconditional average $O_A[\overline{\rho(t)}]$ calculated from a usually mixed, averaged density matrix $\overline{\rho(t)}$; they coincide with each other only when $O_A(\rho)$ is linear in $\rho$. 
In our circuit model, the averaged density matrix $\overline{\rho(t)}$ is expected to reach a unique, infinite-temperature mixed state $\rho \propto \mathbb{I}$ within a single charge sector with total charge $N$ in the long-time limit $t \to \infty$, irrespective of the measurement probability $p$. 
Thus, MIPTs are revealed only in dynamics of the conditional average of physical quantities $O_A[\rho(t)]$ nonlinear in $\rho(t)$ or, in other words, correlation among different trajectories. 

In addition to the average over trajectories as explained above, we also take a spatial average for physical quantities $O_A[\rho(t)]$.
Since our unitary gates are arranged in the brick-wall fashion, physical quantities averaged over trajectories still exhibit even-odd effects depending on the choice of a region $A$. 
In order to suppress this effect, we further take an average of $O_A[\rho(t)]$ over all translations of $A$ for each trajectory. 
Therefore, any physical quantity $\overline{O_A[\rho(t)]}$ shown in the following discussions is the average over (i) translations of $A$ and (ii) different trajectories and is simply denoted by $O_A[\rho(t)]$ hereafter.

%%%%%%%%%%%%%%%%%%%%%%%%%%%%%%%%%%%%%%%%%%%%%%%%%%%%%%%%%%%%%%%%%%%%%%%%%%%%%%%%%%%%%
%~~~~~~~~~~~~~~~~~~~~~~~~~~~~~~~~~~~~~ SECTION ~~~~~~~~~~~~~~~~~~~~~~~~~~~~~~~~~~~~~~
% Numerical results for half filling system
\section{Numerical results}\label{sec:Numerical results}
%%%%%%%%%%%%%%%%%%%%%%%%%%%%%%%%%%%%%%%%%%%%%%%%%%%%%%%%%%%%%%%%%%%%%%%%%%%%%%%%%%%%%

In this section, we show our numerical results for hybrid quantum circuits with a fixed filling $\bar{n} = 1/2$. 
We first examine entanglement quantities to confirm an entanglement transition between an area-law and a volume-law phase (Sec.~\ref{subsec:Entanglement transition}). 
We then focus on charge fluctuation (Sec.~\ref{subsec:Charge fluctuation}) and charge-resolved entanglement entropy (Sec.~\ref{subsec:Charge-resolved entanglement}) to diagnose a charge-fluctuation transition with TLL-like criticality peculiar to (1+1)D systems with $U(1)$ symmetry. 
We use the system sizes ranging from $L=8$ to $24$ for entanglement quantities and those from $L=8$ to $26$ for charge correlations. 
All physical quantities shown in this section are averaged over 1000 trajectories. 
We have also performed similar numerical analyses for filling fractions $\bar{n}=1/4$ and $\bar{n}=1/6$, whose details are provided in Appendix~\ref{app:Numerical results for N=L/4 system}.

%%%%%%%%%%%%%%%%%%%%%%%%%%%%%%%%%%%%%%%%%%%%%%%%%%%%%%%%%%%%%%%%%%%%%%%%%%%%%%%%%%%%%
%~~~~~~~~~~~~~~~~~~~~~~~~~~~~~~~~~~~~~ SECTION ~~~~~~~~~~~~~~~~~~~~~~~~~~~~~~~~~~~~~~
% Numerical results for half filling system
    % Entanglement transition
    \subsection{Entanglement transition}\label{subsec:Entanglement transition}
        % Entanglement entropy
        \subsubsection{Entanglement entropy}\label{subsubsec:Entanglement entropy}
%%%%%%%%%%%%%%%%%%%%%%%%%%%%%%%%%%%%%%%%%%%%%%%%%%%%%%%%%%%%%%%%%%%%%%%%%%%%%%%%%%%%%
        
\begin{figure}[tbp]
        \includegraphics[width=0.48\textwidth]{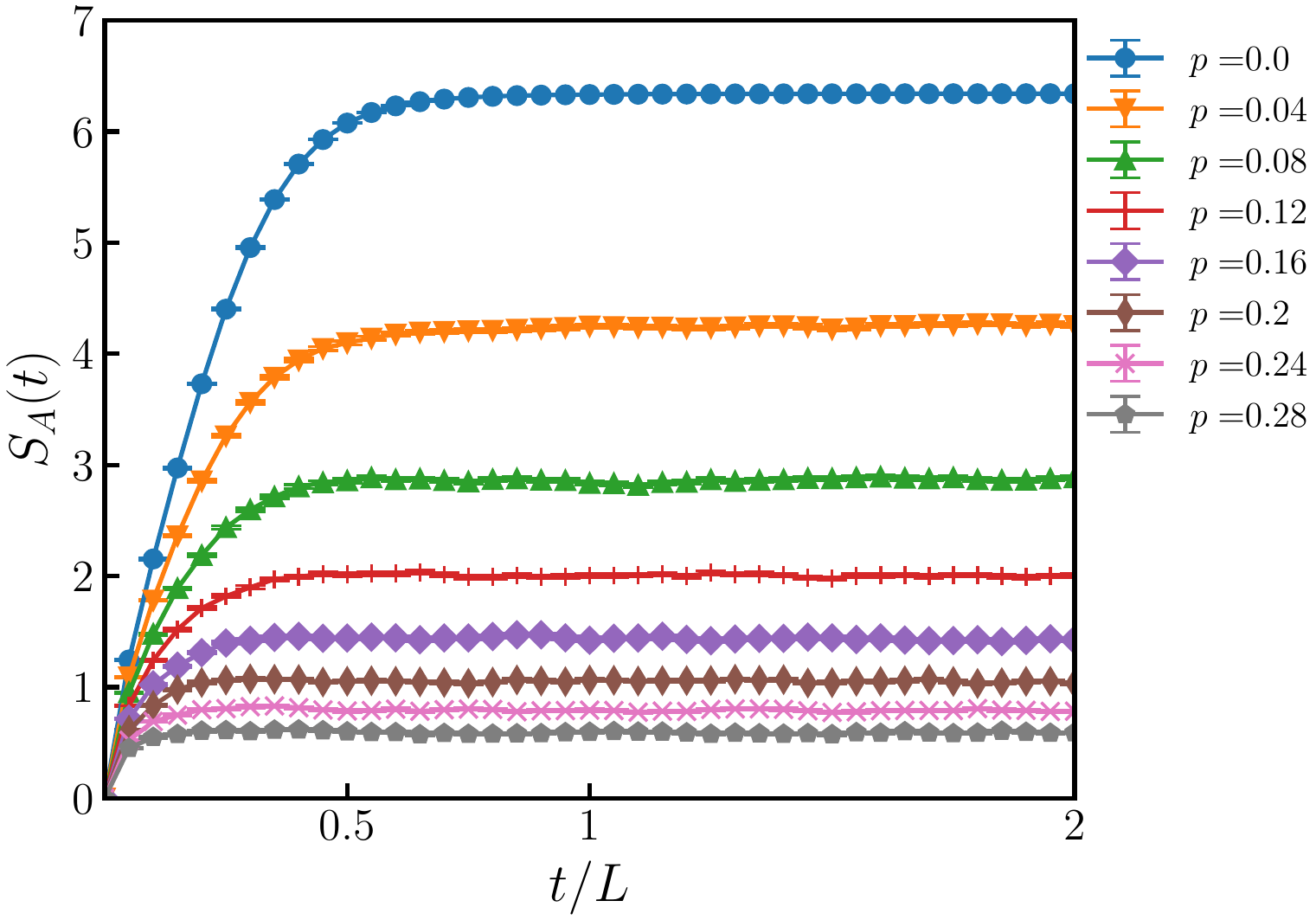}
        \caption{\label{fig:ent_timeevol_linear} Time evolution of the von Neumann entanglement entropy $S_A(t)$ for $L=20$, $|A|=L/2$, and $\bar{n}=1/2$.
        Error bars indicate the standard errors on trajectory average.}
        \end{figure}

        We first look at time evolution of the entanglement entropy under bipartition of the system into a contiguous region $A$ and its complement $\bar{A}$. 
        Given a pure-state trajectory $\rho(t) = \ket{\psi(t)}\bra{\psi(t)}$, the von Neumann entanglement entropy is defined by
        \begin{align}
            S_A(t) = -\mathrm{Tr}_A [\rho_A(t) \ln \rho_A(t)],
        \end{align}
        where $\rho_A(t)$ is the reduced density matrix given by $\rho_A(t)=\mathrm{Tr}_{\bar{A}} [\rho(t)]$. 
        In Fig.~\ref{fig:ent_timeevol_linear}, we show time evolution of the (trajectory averaged) von Neumann entropy under a half cut $(|A|=L/2)$ for $L=20$ and for various values of the measurement rate $p$. 
%====================================================================================
%~~~~~~~~~~~~~~~~~~~~~~~~~~~~~~~~~~~~~~ FIGURE ~~~~~~~~~~~~~~~~~~~~~~~~~~~~~~~~~~~~~~
%===================================================================================
        As the initial state is a product state, the von Neumann entropy is initially zero, but it grows in time by unitary dynamics and saturates to a steady-state value for sufficiently long time $t \sim L$. 
        It is also clear that the von Neumann entropy in the steady-state regime decreases as the measurement rate $p$ is increased. 
        
\begin{figure}[tbp]
        \includegraphics[width=0.48\textwidth]{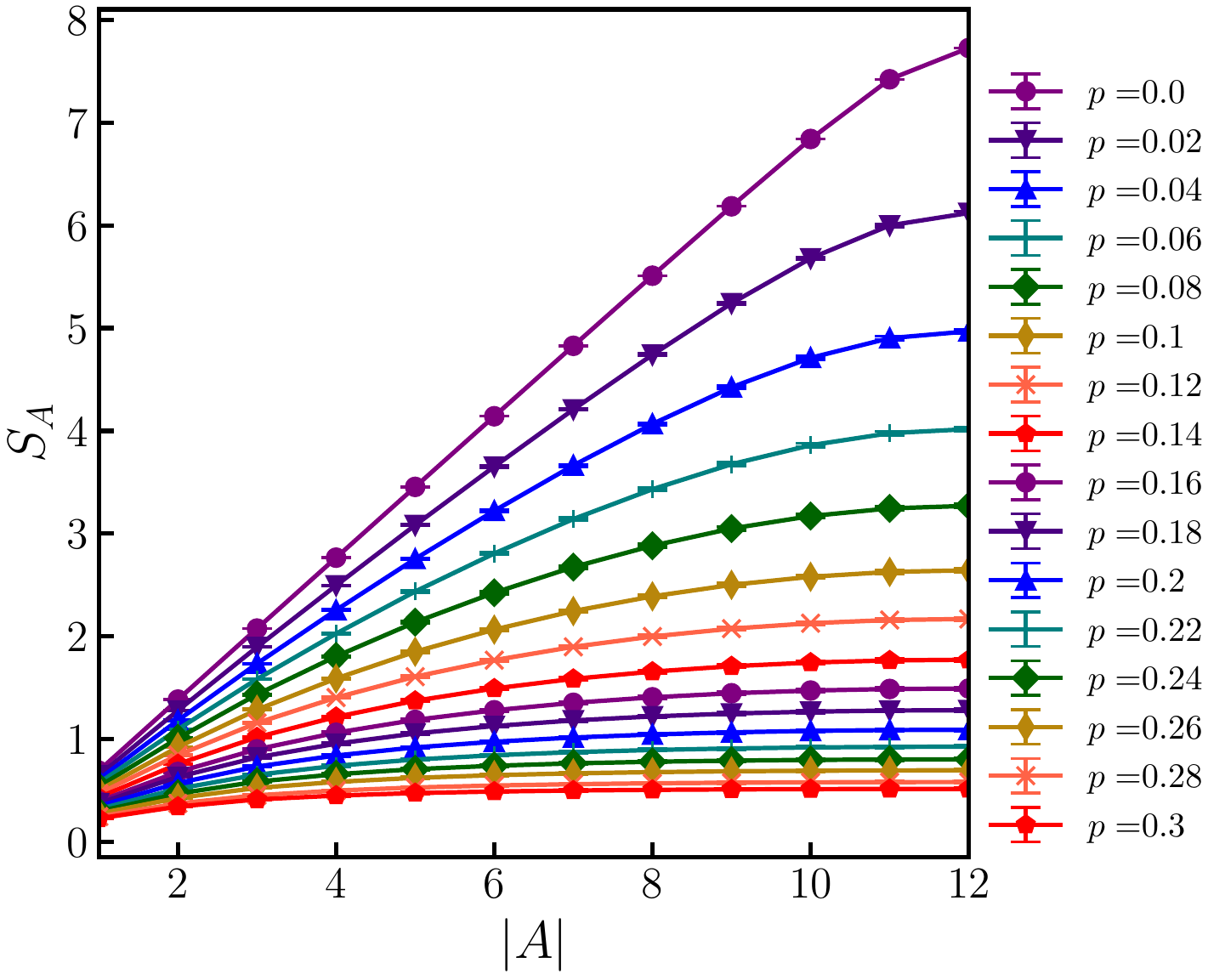}
        \caption{\label{fig:meanEEvsA} Steady-state value of the von Neumann entanglement entropy $S_A$ as a function of the subsystem size $|A|$ for $L=24$ and $\bar{n}=1/2$. 
        }
        \end{figure}

        In order to study the subsystem-size dependence of the steady-state values of $S_A(t)$, we pick up $t=2L$, which is deep inside the steady-state regime for various values of $p$ and the system length $L$.
        Figure~\ref{fig:meanEEvsA} shows the von Neumann entropy at $t=2L$ for $L=24$ as functions of the subsystem size $|A|$. 
%====================================================================================
%~~~~~~~~~~~~~~~~~~~~~~~~~~~~~~~~~~~~~~ FIGURE ~~~~~~~~~~~~~~~~~~~~~~~~~~~~~~~~~~~~~~
%====================================================================================
        When the measurement rate $p$ is sufficiently small, $S_A$ increases linearly in $|A|$ and thus exhibits a volume-law scaling. 
        For $p \sim 0.3$, $S_A$ takes a constant value for large $|A|$ and shows an area-law scaling. 
        We thus expect that a measurement-induced entanglement transition between a volume-law and an area-law phase takes place at some $p=p_c$, as observed in $(1+1)$D monitored circuits with \cite{Agrawal22} or without $U(1)$ symmetry \cite{Li18, Skinner19, Li19}. 
        While the entanglement entropy is expected to scale logarithmically with the subsystem size due to emergent conformal invariance, directly resorting to the scaling behavior of entanglement entropy does not seem to be an accurate way for locating the transition point $p=p_c$ for small size systems. 
        Instead, we consider the two-site mutual information and the squares of two-site correlations functions, whose peak positions can be used as a rough indicator of the transition (see Sec.~\ref{subsubsec:Mutual information and correlation function}). 
        We also perform a scaling analysis for tripartite mutual information to more accurately estimate $p_c$ (see Sec.~\ref{subsubsec:Bipartite and tripartite mutual information}). 
        
        In the rest of this section, we always focus on the trajectory averages of steady-state quantities $O_A(t)$ at $t=2L$. 
        We thus suppress the time dependence of $O_A(t)$ and simply write it as $O_A$ hereafter. 

%%%%%%%%%%%%%%%%%%%%%%%%%%%%%%%%%%%%%%%%%%%%%%%%%%%%%%%%%%%%%%%%%%%%%%%%%%%%%%%%%%%%%
%~~~~~~~~~~~~~~~~~~~~~~~~~~~~~~~~~~~~~ SECTION ~~~~~~~~~~~~~~~~~~~~~~~~~~~~~~~~~~~~~~
% Numerical results for half filling system
    % Entanglement transition
        % Mutual information and correlation function
        \subsubsection{Two-site mutual information and squared correlation functions}\label{subsubsec:Mutual information and correlation function}
%%%%%%%%%%%%%%%%%%%%%%%%%%%%%%%%%%%%%%%%%%%%%%%%%%%%%%%%%%%%%%%%%%%%%%%%%%%%%%%%%%%%%
        
        We here focus on the von Neumann mutual information between two subsystems $A$ and $B$, which is defined by
        \begin{align}
            I(A:B)=S_A+S_B-S_{A\cup B},
            \label{eq:mutual_inf}
        \end{align}
        where $S_A$, $S_B$, and $S_{A\cup B}$ are the von Neumann entanglement entropies of the subsystem $A$ and $B$ and their disjoint union $A\cup B$, respectively.
        The mutual information gives an upper bound for correlation functions through the inequality \cite{Wolf08}, 
        \begin{align}
            I(A:B) \geq \frac{|\langle O_A O_B\rangle_c|^2}{2\norm{O_A}^2\norm{O_B}^2}. 
            \label{eq:MIbound}
        \end{align}
        Here, $O_A$ and $O_B$ are arbitrary operators supported on the subsystem $A$ and $B$, respectively, $\langle O_A O_B \rangle_c$ is the connected correlation function, 
         \begin{align}
            \langle O_A O_B\rangle_c = \textrm{Tr}(\rho O_A O_B)-\textrm{Tr}(\rho O_A) \textrm{Tr}(\rho O_B),
        \end{align}
        and $\norm{O}$ denotes the operator norm of an operator $O$, which is equivalent to the largest singular value of $O$. 
        Here, we focus on the mutual information and squared correlation functions between two antipodal sites $A=\{ i \}$ and $B=\{ j \}$ on a ring of the length $L$ (i.e., $|i-j|=L/2$). 
        
\begin{figure*}[tbp]
        \includegraphics[width=0.9\textwidth]{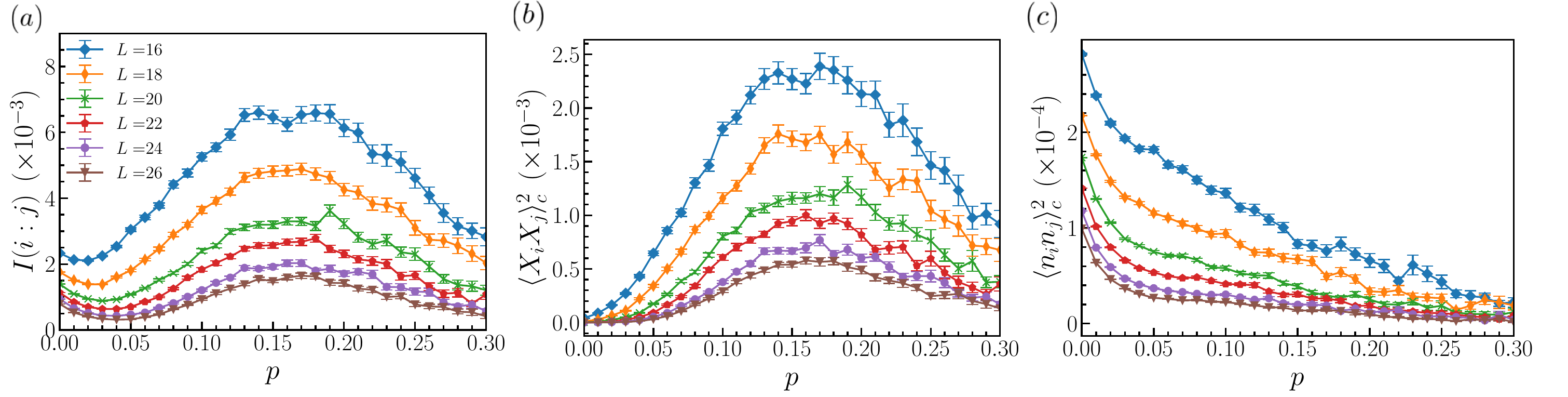}
        \caption{\label{fig:MIXXZZ_L2}
        Steady-state values of (a) the von Neumann mutual information $I(i:j)$, (b) the square of the connected correlation function for Pauli-$X$ operators $\langle X_i X_j\rangle_c^2$, and (c) that for charge operators $\langle n_in_j\rangle_c^2$ between two antipodal sites with $|i-j|=L/2$ for $\bar{n}=1/2$.}
        \end{figure*}

        In Fig.~\ref{fig:MIXXZZ_L2}~(a), we plot the von Neumann mutual information against the measurement rate $p$ for various system sizes.
%====================================================================================
%~~~~~~~~~~~~~~~~~~~~~~~~~~~~~~~~~~~~~~ FIGURE ~~~~~~~~~~~~~~~~~~~~~~~~~~~~~~~~~~~~~~
%====================================================================================
        It exhibits a broad peak around $p \sim 0.15$ as observed in other monitored systems \cite{Li19, Fuji20, Minato22}, indicating the presence of an entanglement transition; correlations are enhanced by critical fluctuation at the transition, whereas they diminish in the volume-law or area-law phase. 
        A similar peak can also be found for the square of the connected correlation function for Pauli-$X$ operators between antipodal sites, $\langle X_i X_j \rangle_c^2$, as shown in Fig.~\ref{fig:MIXXZZ_L2}~(b). 
        On the other hand, the squared correlation function for charge operators, $\langle n_i n_j \rangle_c^2 = \langle Z_i Z_j \rangle_c^2/16$, does not have a peak for finite measurement rate $p$; it takes a maximum at $p=0$ and monotonically decreases with $p$ as seen from Fig.~\ref{fig:MIXXZZ_L2}~(c). 
        
        These qualitatively distinct behaviors of correlation functions depending on the choice of Pauli operators have also been observed in interacting boson systems subject to continuous monitoring with charge conservation \cite{Fuji20}. 
        Such behaviors are not expected for hybrid quantum circuits without symmetry and are indeed peculiar to charge-conserving systems as studied here. 
        In the absence of measurements, a density matrix $\rho(t)$ averaged over random unitary gates reaches an infinite-temperature mixed state for sufficiently long time. 
        In fact, each \textit{pure} state $\ket{\psi(t)}$ evolved by application of random unitary gates is already in a thermal pure state at infinite temperature, meaning that an expectation value $\bra{\psi(t)} O \ket{\psi(t)}$ well approximates the canonical ensemble average of an operator $O$ at infinite temperature. 
        Evaluated with respect to the infinite-temperature mixed state in a fixed charge sector ($N=L/2$ in the present case), the connected correlation function $\langle X_i X_j \rangle_c$ is zero whereas  $\langle n_i n_j \rangle_c$ takes a nonzero value,
        \begin{align}
            \langle n_in_j\rangle_c = -\frac{1}{4(L-1)}.
            \label{eq:nncorr_inf_temp}
        \end{align}
        It turns out that $\langle n_i n_j \rangle_c$ has a nonzero trajectory average even in the presence of measurements, whose scaling behavior is studied in Sec.~\ref{subsec:Charge fluctuation}, whereas $\langle X_i X_j \rangle_c$ remains zero. 
        Thus, the trajectory average for the squared correlation function $\langle n_i n_j \rangle_c^2$ is dominated by a nonzero contribution from the infinite-temperature value of $\langle n_i n_j \rangle_c$, leading to a monotonically decreasing behavior with a peak at $p=0$.
        In contrast, since the trajectory average of $\langle X_i X_j \rangle_c$ is zero, the trajectory average of $\langle X_i X_j \rangle_c^2$ well captures a correlation among trajectories and peaks around the entanglement transition. 
        As detailed in Appendix~\ref{app:infinite-temperature average}, the mutual information computed from correlation functions for the infinite-temperature mixed state with total charge $N=L/2$ also has a finite value:
        \begin{align}
            I(i:j) = \frac{1}{2} \ln \left( 1-\frac{1}{(L-1)^2} \right) -\frac{1}{2(L-1)} \ln \left(1-\frac{2}{L}\right),
            \label{mutualITS}
        \end{align}
        which is in good agreement with a small peak at $p=0$ for the numerically obtained mutual information in Fig.~\ref{fig:MIXXZZ_L2}~(a). 
        
%%%%%%%%%%%%%%%%%%%%%%%%%%%%%%%%%%%%%%%%%%%%%%%%%%%%%%%%%%%%%%%%%%%%%%%%%%%%%%%%%%%%%
%~~~~~~~~~~~~~~~~~~~~~~~~~~~~~~~~~~~~~ SECTION ~~~~~~~~~~~~~~~~~~~~~~~~~~~~~~~~~~~~~~
% Numerical results for half filling system
    % Entanglement transition
        % Bipartite and tripartite mutual
        \subsubsection{Bipartite and tripartite mutual information}\label{subsubsec:Bipartite and tripartite mutual information}
%%%%%%%%%%%%%%%%%%%%%%%%%%%%%%%%%%%%%%%%%%%%%%%%%%%%%%%%%%%%%%%%%%%%%%%%%%%%%%%%%%%%%
        
\begin{figure*}[tbp]
        \includegraphics[width=0.8\textwidth]{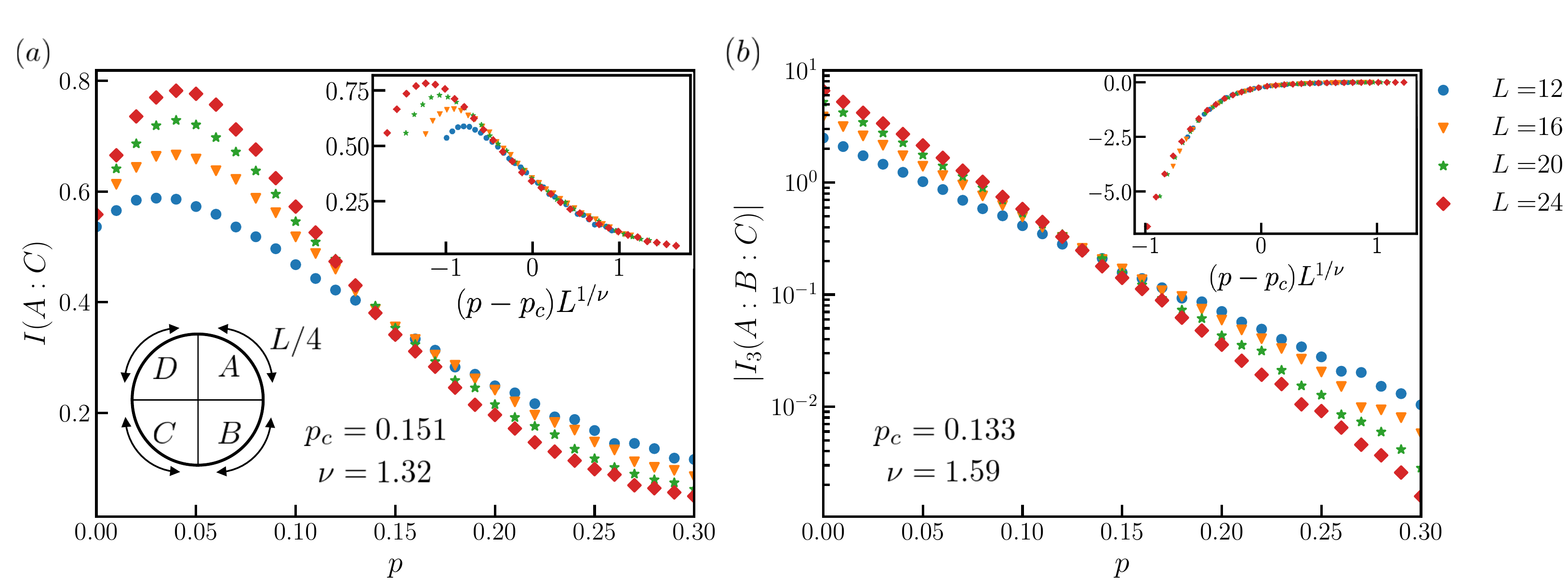}
        \caption{\label{fig:SC_BMITMI_L2} 
        Steady-state values of (a) bipartite mutual information $I(A:C)$ and (b) tripartite mutual information $I_3(A:B:C)$ for $\bar{n}=1/2$ under the partition of the system into four subsystems $A$, $B$, $C$, and $D$ of the length $L/4$.
        The insets show scaling collapses of $I(A:C)$ and $I_3(A:B:C)$ against $(p-p_c) L^{1/\nu}$ for which we have used $(p_c, \nu)=(0.151, 1.32)$ and $(p_c, \nu)=(0.133, 1.59)$, respectively.}
        \end{figure*}

        For a more accurate estimation of the transition point, we can still use the von Neumann mutual information but with a partition different from that used in the previous section. 
        We here divide the system into four contiguous subsystems $A=[i, j)$, $B=[j, k)$, $C=[k, l)$, and $D=[l, i)$. 
        For (1+1)D conformal field theory (CFT), the bipartite mutual information $I(A:C)$ is associated with a four-point correlation function and is a universal function depending only on the cross ratio $\eta$ \cite{Calabrese09b},
        \begin{align}
            I(A:C) = f(\eta), \quad \eta = \frac{x_{ij}x_{kl}}{x_{ik}x_{jl}},
            \label{eq:cross_ratio}
        \end{align}
        where $x_{ij}$ is the chord distance: 
        \begin{align}
            x_{ij} = \frac{L}{\pi} \sin \frac{\pi |i-j|}{L}.
            \label{eq:chord_distance}
        \end{align}
        This implies that, if the entanglement transition has emergent conformal invariance, the bipartite mutual informations for different system sizes but with a fixed ratio of subsystem sizes should coincide with each other at the transition point $p=p_c$. 
        In Fig.~\ref{fig:SC_BMITMI_L2}~(a), we show the bipartite mutual information $I(A:C)$ as a function of the measurement rate $p$ under the partition of the system into four subsystems of the length $L/4$.
%====================================================================================
%~~~~~~~~~~~~~~~~~~~~~~~~~~~~~~~~~~~~~~ FIGURE ~~~~~~~~~~~~~~~~~~~~~~~~~~~~~~~~~~~~~~
%====================================================================================
        Data between different system sizes cross with each other around $p \sim 0.15$. 
        This may indicate that the bipartite mutual information loses the system-size dependence at the transition due to emergent conformal invariance.
        We then perform a scaling collapse with the ansatz
        \begin{align}
            I(A:C) \sim F[(p-p_c)L^{1/\nu}],
            \label{eq:scaling_collpase_I2}
        \end{align}
        where $F(x)$ is a scaling function and $\nu$ is the correlation length exponent. 
        As detailed in Appendix~\ref{app:Scaling analysis}, we obtain $p_c = 0.151(4)$ and $\nu = 1.3(2)$ for the best collapse [see the inset of Fig.~\ref{fig:SC_BMITMI_L2}~(a)].
        
        However, the bipartite mutual information has a nonmonotonic shape for a small $p$ region, which could cause a drift of the crossing point and render the scaling analysis inadequate for small-size systems (see also Refs.~\cite{Zabalo20, Szyniszewski20}). 
        Instead, the tripartite mutual information
        \begin{align}
            I_3(A:B:C) = I(A:B)+I(A:C)-I(A:B\cup C).
        \end{align}
        has been proposed as a more appropriate measure of the entanglement transition \cite{Gullans20a, Zabalo20}, since it is designed to scale with the system size in the volume-law phase, takes a constant value at the conformal invariant transition point, and decays to zero in the area-law phase. 
        As shown in Fig.~\ref{fig:SC_BMITMI_L2}~(b), the tripartite mutual information monotonically decreases with the measurement rate $p$ and crosses at $p \sim 0.13$ between different system sizes. 
        We then assume the same scaling ansatz as in Eq.~\eqref{eq:scaling_collpase_I2} and perform the scaling collapse, which yields $p_c = 0.133(6)$ and $\nu = 1.6(3)$ [see the inset of Fig.~\ref{fig:SC_BMITMI_L2}~(b)]. 
        As detailed in Appendix~\ref{app:Numerical results for N=L/4 system}, we have also performed a similar scaling analysis for filling $\bar{n}=1/4$ and found $p_c = 0.095(7)$ and $\nu = 1.2(3)$ from the bipartite mutual information and $p_c = 0.101(10)$ and $\nu = 1.7(4)$ from the tripartite mutual information. 
        For $\bar{n}=1/6$, we have failed in locating the entanglement transition as we could not find clear crossings of the bipartite or tripartite mutual information from available system sizes.
        
        We note that Ref.~\cite{Agrawal22} has also performed a scaling analysis for the tripartite mutual information in the same hybrid circuit with $\bar{n}=1/2$ and obtained $p_c = 0.105(3)$ and $\nu = 1.32(6)$; 
        the critical exponent $\nu$ agrees with our result within error, but the transition point $p_c$ deviates. 
        We believe that this discrepancy happens not only by actual implementations of the scaling analysis but also by how to collect the original data; 
        we have computed physical quantities just after the projective measurements, but those in Ref.~\cite{Agrawal22} appear to be computed before the measurements. 
        Such a slight difference in the protocol will be negligible in the large-volume limit but still affects quantities for small-size systems.

%%%%%%%%%%%%%%%%%%%%%%%%%%%%%%%%%%%%%%%%%%%%%%%%%%%%%%%%%%%%%%%%%%%%%%%%%%%%%%%%%%%%%
%~~~~~~~~~~~~~~~~~~~~~~~~~~~~~~~~~~~~~ SECTION ~~~~~~~~~~~~~~~~~~~~~~~~~~~~~~~~~~~~~~
% Numerical results for half filling system
    % Entanglement transition
        % Scaling behaviors and CFT
        \subsubsection{Scaling behaviors at entanglement transition}\label{subsubsec:Scaling behaviors and CFT}
%%%%%%%%%%%%%%%%%%%%%%%%%%%%%%%%%%%%%%%%%%%%%%%%%%%%%%%%%%%%%%%%%%%%%%%%%%%%%%%%%%%%%

\begin{figure*}[tbp]
        \includegraphics[width=0.8\textwidth]{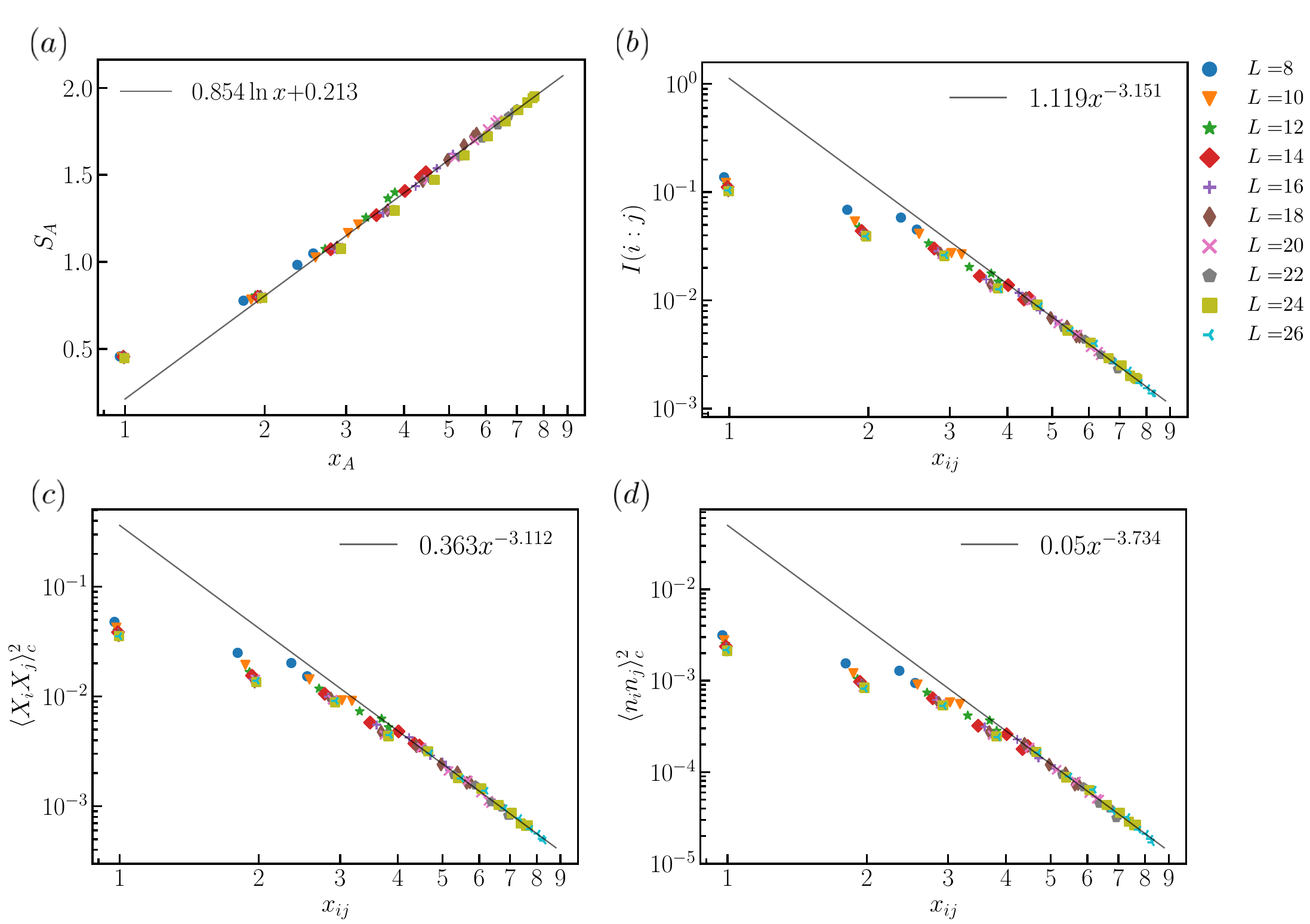}
        \caption{\label{fig:EEMIXXZZ_L2} 
        (a) Steady-state value of von Neumann entanglement entropy is plotted against the chord length $x_A$ of the subsystem $A$ for $L = 8$ to $24$. 
        Steady-state values of (b) two-site mutual information and squared correlation functions (c) $\langle X_iX_j \rangle_c^2$ and (d) $\langle n_in_j \rangle_c^2$ are plotted against the chord distance $x_{ij}$ between two sites for $L = 8$ to $26$. 
        All data are obtained for $\bar{n}=1/2$ at $p=0.13$. 
        The solid lines are fitting functions for data points with $x_{A}, x_{ij} \geq 5$.}
        \end{figure*}

        From the scaling analysis for the tripartite mutual information presented above, we here set $p=0.13$ and study the scaling behaviors of the von Neumann entanglement entropy, two-site mutual information, and squared correlation functions at the entanglement transition. 
        In equilibrium critical systems described by (1+1)D CFT, the von Neumann entanglement entropy under the periodic boundary condition is known to obey the logarithmic scaling \cite{Calabrese04},
        \begin{align}
            S_A = \frac{c}{3}\ln{x_A}+c', 
            \label{eq:CCformula}
        \end{align}
        where $x_A=(L/\pi)\ln(\pi|A|/L)$ is the chord length of the subsystem $A$, $c$ is the central charge, and $c'$ is a non-universal constant. 
        This logarithmic scaling of the entanglement entropy has been numerically observed at various entanglement transitions \cite{Skinner19, Li19, Tang20, Zabalo20, Fuji20, Alberton21, Li21a} and has also been analytically derived through mapping to percolation transitions for some special cases \cite{Vasseur19, Skinner19, Jian20a, Bao20}. 
        In Fig.~\ref{fig:EEMIXXZZ_L2}~(a), we plot the von Neumann entanglement entropy at $p=0.13$ as a function of $x_A$, which is fitted well into Eq.~\eqref{eq:CCformula} with $c = 2.56$ and $c' = 0.21$. 
%====================================================================================
%~~~~~~~~~~~~~~~~~~~~~~~~~~~~~~~~~~~~~~ FIGURE ~~~~~~~~~~~~~~~~~~~~~~~~~~~~~~~~~~~~~~
%====================================================================================
        We remark that, in contrast to equilibrium phase transitions, the coefficient $c$ in Eq.~\eqref{eq:CCformula} is not necessarily interpreted as the central charge for the entanglement transition, as the percolation transitions have a vanishing central charge and $c$ is instead related to a boundary critical exponent. 
        Nevertheless, the logarithmic scaling of the entanglement entropy is a strong indication of emergent conformal invariance at the entanglement transition in our $U(1)$-symmetric monitored circuit. 

        To further substantiate the CFT nature of the entanglement transition, we consider the von Neumann mutual information between two sites $i$ and $j$. 
        As shown in Fig.~\ref{fig:EEMIXXZZ_L2}~(b), the mutual information  decays algebraically at the critical point for large distance. 
        By fitting the numerical result for a large-distance region $x_{ij} \geq 5$ into the scaling form,
        \begin{align}
            I(i:j) \propto x_{ij}^{-2\Delta}, \label{eq:MIscaling}
        \end{align}
        we obtain the critical exponent $\Delta = 1.58$. 
        This behavior is consistent with the (1+1)D CFT \cite{Furukawa09, Calabrese09a, Calabrese11} and has also been observed at entanglement transitions \cite{Skinner19, Li19, Fuji20}. 
        We also consider the squares of connected correlation functions $\langle X_i X_j \rangle_c^2$ and $\langle n_i n_j \rangle_c^2$, which are shown in Figs.~\ref{fig:EEMIXXZZ_L2}(c) and \ref{fig:EEMIXXZZ_L2}(d), respectively. 
        Similarly to the mutual information, they also decay algebraically for large distance as
        \begin{align}
            \langle X_iX_j\rangle_c^2 \propto x_{ij}^{-2\Delta_X} \label{eq:XX2scaling}, \\
            \langle n_in_j\rangle_c^2 \propto x_{ij}^{-2\Delta_n} \label{eq:ZZ2scaling},
        \end{align}
        and we obtain the corresponding critical exponents $\Delta_X = 1.56$ and $\Delta_n = 1.87$ by fitting. 
        The squared correlation function for charged operators $X_i$ has an exponent smaller than that for charge-neutral operators $n_i$ and thus represents a dominant correlation at the entanglement transition. 
        This is also supported by the fact that the exponent $\Delta_X$ for $\langle X_i X_j \rangle_c^2$ is quite close to the exponent $\Delta$ for the mutual information, which provides an upper bound for squared correlation functions as given in Eq.~\eqref{eq:MIbound}. 
        This observation that the squared correlation function for charged operators gives the most dominant correlation at measurement-induced criticality might be a common feature for $U(1)$-symmetric systems as also found in an interacting boson system \cite{Fuji20} and in free fermion systems \cite{Chen20, Alberton21}. 

        As discussed in Sec.~\ref{subsubsec:Mutual information and correlation function}, not only the trajectory average of the squared correlation function $\langle n_i n_j \rangle_c^2$ but also the trajectory average of the connected correlation function $\langle n_i n_j \rangle_c$ does not vanish for general $p$ in our charge-conserving hybrid circuit. 
        As we will see below, the connected correlation function $\langle n_i n_j \rangle_c$ by itself exhibits an algebraic decay, which might indicate a measurement-induced critical phenomenon akin to TLL theory as commonly observed in (1+1)D quantum critical systems with charge conservation.
        
%%%%%%%%%%%%%%%%%%%%%%%%%%%%%%%%%%%%%%%%%%%%%%%%%%%%%%%%%%%%%%%%%%%%%%%%%%%%%%%%%%%%%
%~~~~~~~~~~~~~~~~~~~~~~~~~~~~~~~~~~~~~ SECTION ~~~~~~~~~~~~~~~~~~~~~~~~~~~~~~~~~~~~~~
% Numerical results for half filling system
    % Charge fluctuation
    \subsection{Charge fluctuation}\label{subsec:Charge fluctuation}
        % Bipartite fluctuation
        \subsubsection{Bipartite charge fluctuation}\label{subsubsec:Bipartite charge fluctuation}
%%%%%%%%%%%%%%%%%%%%%%%%%%%%%%%%%%%%%%%%%%%%%%%%%%%%%%%%%%%%%%%%%%%%%%%%%%%%%%%%%%%%%

\begin{figure}[tbp]
        \includegraphics[width=0.45\textwidth]{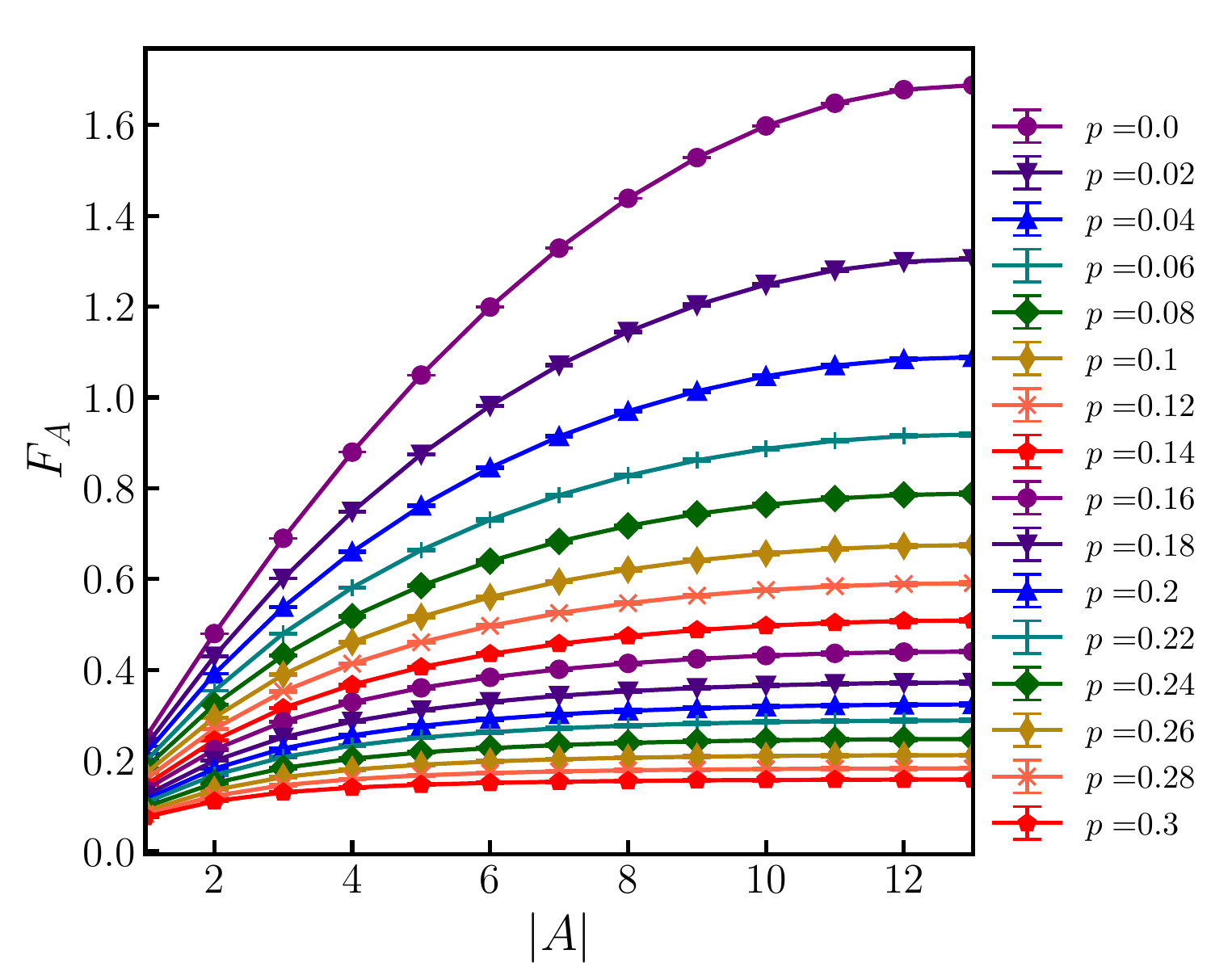}
        \caption{\label{fig:meanBFvsA_L26} 
        Steady-state value of bipartite charge fluctuation $F_A$ as a function of the subsystem size $|A|$ for the $L=26$ and $\bar{n}=1/2$.}
        \end{figure}

        We here discuss that measurement-induced criticality is revealed not only in entanglement, as we have seen above, but also in charge fluctuation for charge conserving systems.
        In analogy with the bipartite entanglement entropy $S_A$, we can introduce a quantity that measures fluctuation of the charge in a subsystem $A$. 
        We are particularly interested in the bipartite charge fluctuation \cite{Klich09, Song10, Song11, Song12, Calabrese12, Rachel12, Frerot15, Crepel21, Estienne22} defined by 
        \begin{align}
            F_A = \textrm{Tr}( \rho n_A^2) - [\textrm{Tr}(\rho n_A)]^2,
        \end{align}
        where $n_A=\sum_{i\in A} n_i$ is the total particle number operator in the subsystem $A$. 
        In Fig.~\ref{fig:meanBFvsA_L26}, we show steady-state values of the bipartite charge fluctuation $F_A$ computed for our hybrid circuit with $L=26$. 
%====================================================================================
%~~~~~~~~~~~~~~~~~~~~~~~~~~~~~~~~~~~~~~ FIGURE ~~~~~~~~~~~~~~~~~~~~~~~~~~~~~~~~~~~~~~
%====================================================================================
        For $p$ small, the bipartite charge fluctuation rapidly increases with the subsystem size $|A|$. 
        This can be readily explained in the absence of measurements where each trajectory is well described by the infinite-temperature mixed state in a fixed charge sector with $N=L/2$ (see Appendix~\ref{app:infinite-temperature average} for detail). 
        We find that the bipartite charge fluctuation is a quadratic function in $|A|$, 
        \begin{align}
            F_A = \frac{|A|(L-|A|)}{4(L-1)},
            \label{eq:BF_inf_temp}
        \end{align}
        in contrast to a simple linear scaling with $|A|$ for the entanglement entropy.
        For a fixed ratio $r_A = |A|/L$, the bipartite charge fluctuation scales extensively with the system size as $F_A \propto r_A(1-r_A) L$.
        On the other hand, it approaches a constant value for large $p$, similarly to the area-law behavior for the entanglement entropy. 
        We then expect that the bipartite charge fluctuation $F_A$ undergoes a phase transition at some finite $p=p_t$, at which its functional form changes from a quadratic one to a constant. 
        We thus call it the measurement-induced charge-fluctuation transition. 

        For (1+1)D critical systems described by TLL theory, it has been shown that the bipartite charge fluctuation $F_A$ for a contiguous subsystem $A$ exhibits a logarithmic scaling \cite{Song10, Song12}, 
        \begin{align}
            F_A \sim \frac{K}{\pi^2}\ln x_A,
            \label{eq:BFTLL}
        \end{align}        
        where $x_A$ is the chord length for the subsystem $A$ and $K$ is the Luttinger parameter. 
        This is reminiscent of the logarithmic scaling of the bipartite entanglement entropy $S_A$.
        Furthermore, if we divide the system into four contiguous subsystems $A=[i, j)$, $B=[j, k)$, $C=[k, l)$, and $D=[l, i)$, the bipartite charge fluctuation for two disjoint intervals $A \cup C$ is given by \cite{Song12},
        \begin{align}
            F_{A\cup C} \sim \frac{K}{\pi^2}\ln\frac{x_{ij}x_{jk}x_{il}x_{kl}}{x_{ik}x_{jl}}.
            \label{eq:BF_disjoint}
        \end{align}
        This leads us to introduce a quantity analogous to the bipartite mutual information in Eq.~\eqref{eq:mutual_inf} by 
        \begin{align}
            \langle n_A n_C\rangle_c = F_A + F_C - F_{A\cup C},
        \end{align}
        which is in fact the connected correlation function between the total charge operator $n_A$ and $n_C$ for the subsystem $A$ and $C$, respectively. 
        This is straightforwardly shown from the fact that the bipartite charge fluctuation $F_A$ can be expressed in terms of connected correlation functions for charge operators $n_i$: 
        \begin{align}
            F_A = \sum_{i,j\in A}\langle n_i n_j \rangle_c \label{eq:BFisSum}.
        \end{align}
        We thus find that the subsystem-charge correlation function $\langle n_A n_C \rangle_c$ becomes a function depending solely on the cross ratio $\eta$ [see Eq.~\eqref{eq:cross_ratio}],
        \begin{align}
            \langle n_A n_C \rangle_c \sim \frac{K}{\pi^2} \ln \frac{x_{il} x_{jk}}{x_{ik} x_{jl}} = \frac{K}{\pi^2} \ln (1-\eta),
        \end{align}
        for critical systems described by TLL theory. 
        Therefore, this quantity can be seen as a charge-correlation counterpart of the bipartite mutual information $I(A:C)$, which is also a function of the cross ratio $\eta$ as discussed in Sec.~\ref{subsubsec:Bipartite and tripartite mutual information}. 
        
\begin{figure}[tbp]
        \includegraphics[width=0.45\textwidth]{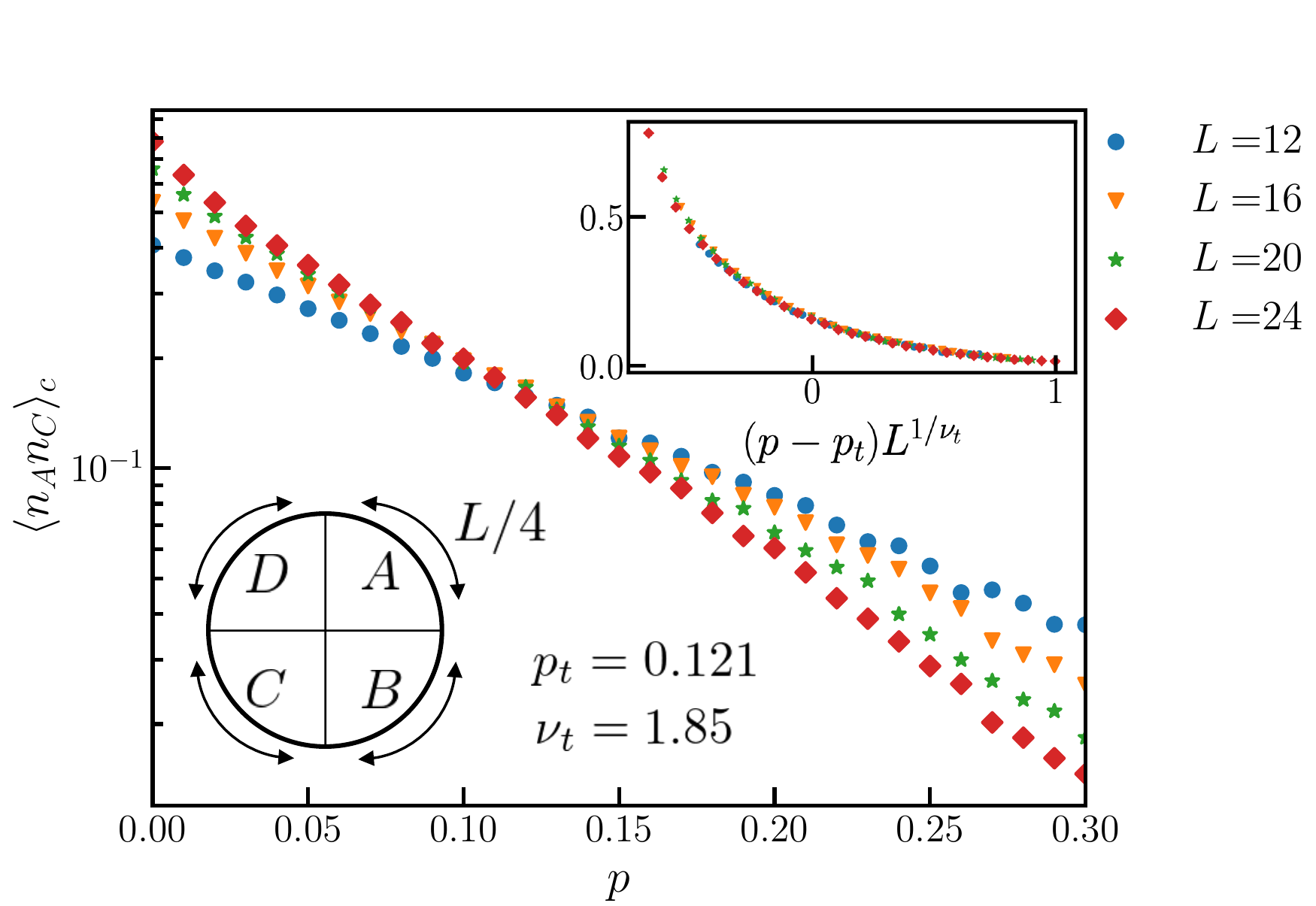}
        \caption{\label{fig:SC_NANC_L2}
        Steady-state value of subsystem-charge correlation function $\langle n_An_C\rangle_c$ between two antipodal regions $A$ and $C$ with $|A|=|C|=L/4$ for $\bar{n}=1/2$. 
        The inset shows scaling collapse of $\langle n_An_C\rangle_c$ against $(p-p_t)L^{1/\nu_t}$ with $(p_t, \nu_t)=(0.121, 1.85)$.
        }
        \end{figure}

        If we suppose that the measurement-induced charge-fluctuation transition for $F_A$ is described by TLL theory, the subsystem-charge correlation function $\langle n_A n_C \rangle_c$ for a fixed ratio of subsystem sizes should not depend on the system size $L$ at the transition point $p=p_t$. 
        For $p<p_t$, we expect that $\langle n_A n_C \rangle_c$ grows with the subsystem size, as we have $\langle n_A n_C \rangle_c = |A||C|/2(L-1)$ at $p=0$ due to the quadratic functional form of $F_A$ in Eq.~\eqref{eq:BF_inf_temp} for the infinite-temperature mixed state. 
        On the other hand, it will decay to zero for $p>p_t$.
        We thus expect that the subsystem-charge correlation functions $\langle n_A n_C \rangle_c$ for different system sizes cross with each other at the charge-fluctuation transition $p=p_t$, similarly to the tripartite mutual information $I(A:B:C)$ at the entanglement transition. 
        This can be clearly seen from Fig.~\ref{fig:SC_NANC_L2}, where we show steady-state values of the subsystem-charge correlation function $\langle n_A n_C \rangle_c$ under the partition of the system into four subsystems with the size $L/4$.
%~~~~~~~~~~~~~~~~~~~~~~~~~~~~~~~~~~~~~~ FIGURE ~~~~~~~~~~~~~~~~~~~~~~~~~~~~~~~~~~~~~~
%====================================================================================
        The curves of $\langle n_A n_C \rangle_c$ monotonically decrease with $p$ and cross between different system sizes around $p \sim 0.12$ as expected. 
        By performing a scaling collapse with the ansatz, 
        \begin{align}
            \langle n_A n_C \rangle_c \sim G[(p-p_t)L^{1/\nu_t}],
            \label{eq:nAnC_scaling}
        \end{align}
        we obtain the critical measurement rate $p_t = 0.121(10)$ and the associated correlation length exponent $\nu_t = 1.9(4)$. 
        The charge-fluctuation transition point $p_t$ is slightly lower than the entanglement transition point $p_c = 0.133(6)$, but they overlap within one standard error.
        
        As detailed in Appendix~\ref{app:Numerical results for N=L/4 system}, we have performed similar scaling analyses for the subsystem-charge correlation function for other filling fractions and found $p_t = 0.071(4)$ and $\nu_t = 1.8(3)$ for $\bar{n}=1/4$ and $p_t = 0.051(10)$ and $\nu_t = 1.8(5)$ for $\bar{n}=1/6$. 
        It is evident that the charge-fluctuation transition, as well as the entanglement transition, depends on the filling fraction $\bar{n}$ and drifts toward $p=0$ as $\bar{n}$ is decreased from $\bar{n}=1/2$. 
        On the other hand, the correlation length exponent $\nu_t$ at the charge-fluctuation transition takes a value close to 2 irrespective of filling. 
        Compared with the $\bar{n}=1/2$ case, difference between the charge-fluctuation and entanglement transition is clearer for $\bar{n}=1/4$, as we found $p_c = 0.095(7)$ for the latter. 
        Thus, it is likely that the two transitions are distinct MIPTs peculiar to monitored circuits with $U(1)$ symmetry.
        
        We now compare our results with the charge-sharpening transition proposed for monitored circuits with $U(1)$ symmetry in Ref.~\cite{Agrawal22}. 
        First of all, we have to notice that the notion of filling is ambiguous for the charge-sharpening transition when it is dynamically characterized by evolution of an initial state that mixes different charge sectors. 
        To be more precise, Ref.~\cite{Agrawal22} has introduced two dynamical characterizations with different initial states: (i) charge variance for a pure state evolved from an equal-weight superposition over all charge basis states and (ii) an ancilla probe that mixes two charge sectors with $N=L/2$ and $N=L/2-1$. 
        As the initial state used in (i) has a maximal weight on the charge sector $N=L/2$ and a small deviation from $N=L/2$ is expected to be immaterial, the charge-sharpening transition in Ref.~\cite{Agrawal22} might correspond to the charge-fluctuation transition at filling fraction $\bar{n}=1/2$ in our case.
        It has been argued that the charge-sharpening transition exists within a volume-law entangled phase and thus the corresponding transition point $p_\sharp$ must be smaller than the entanglement transition point $p_c$. 
        This feature also applies to the charge-fluctuation transition at $p=p_t$.
        While our estimate of the charge-fluctuation transition point $p_t = 0.121(10)$ deviates from that for the charge-sharpening transition $p_\sharp \sim 0.09$, the associated correlation length exponents roughly agree with each other as we have $\nu_t = 1.9(4)$ and $\nu_\sharp = 2.0(3)$. 
        Therefore, we cannot conclusively argue that the two transitions coincide with each other from available data, but they share several common features.

        The next question is whether we have an extended critical phase below the charge-fluctuation transition point $p=p_t$, as predicted in Ref.~\cite{Barratt22a} that the charge-sharpening transition belongs to a BKT universality class and a charge fuzzy phase below the transition exhibits TLL-like critical phenomena. 
        In the following section, we focus on the bipartite charge fluctuation and charge correlation functions below and right at the charge-fluctuation transition and show that this is actually not the case in the monitored Haar-random circuit with $U(1)$ symmetry.
        
%%%%%%%%%%%%%%%%%%%%%%%%%%%%%%%%%%%%%%%%%%%%%%%%%%%%%%%%%%%%%%%%%%%%%%%%%%%%%%%%%%%%%
%~~~~~~~~~~~~~~~~~~~~~~~~~~~~~~~~~~~~~ SECTION ~~~~~~~~~~~~~~~~~~~~~~~~~~~~~~~~~~~~~~
% Numerical results for half filling system
    % Charge fluctuation
        % Scaling behaviors and TLL
        \subsubsection{TLL-like scaling behaviors}\label{subsubsec:Scaling behaviors and TLL}
%%%%%%%%%%%%%%%%%%%%%%%%%%%%%%%%%%%%%%%%%%%%%%%%%%%%%%%%%%%%%%%%%%%%%%%%%%%%%%%%%%%%%
        
\begin{figure}[tbp]
        \includegraphics[width=0.4\textwidth]{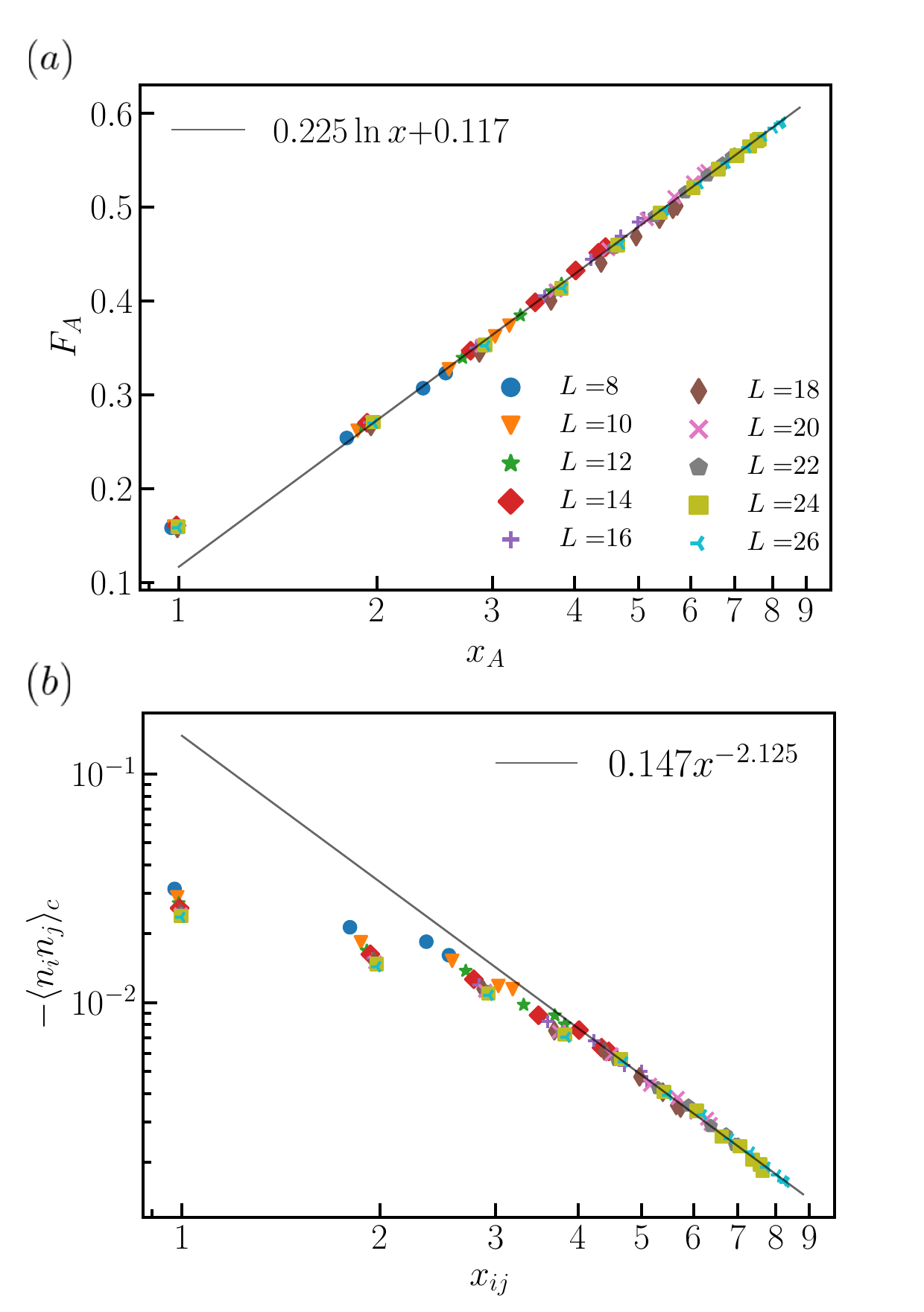}
        \caption{\label{fig:BFZZ_L2} 
        Steady-state values of (a) bipartite charge fluctuation and (b) charge correlation function $-\langle n_i n_j \rangle_c$ are plotted against the chord length $x_A$ of the subsystem $A$ and the chord distance $x_{ij}$ between two sites, respectively. 
        We have used data at $p=0.12$ close to the charge-fluctuation transition for $\bar{n}=1/2$ and $L = 8$ to $26$. 
        The solid lines in both panels are fitting functions obtained by fitting data points with $x_{A}, x_{ij} \geq 5$.}
        \end{figure}

        We first focus on the bipartite charge fluctuation right at the charge-fluctuation transition $p=p_t$. 
        As discussed above, a critical system obeying TLL theory should exhibit a logarithmic scaling for the bipartite charge fluctuation $F_A$ with respect to the subsystem size $|A|$ as given in Eq.~~\eqref{eq:BFTLL}. 
        In Fig.~\ref{fig:BFZZ_L2}~(a), we plot the bipartite charge fluctuation $F_A$ at $p=0.12$ for various systems sizes against the chord length $x_A$ of the subsystem $A$. 
%~~~~~~~~~~~~~~~~~~~~~~~~~~~~~~~~~~~~~~ FIGURE ~~~~~~~~~~~~~~~~~~~~~~~~~~~~~~~~~~~~~~
%====================================================================================
        It is indeed fitted well into the logarithmic scaling form \eqref{eq:BFTLL} from which we can extract the Luttinger parameter $K = 2.22$. 
        This value is close to the universal value $K_\sharp = 2$ predicted at the charge-sharpening transition from a replica field theory for a charge-conserving monitored circuit \cite{Barratt22a}. 
        However, the deviation of the Luttinger parameter $K$ from $K_\sharp = 2$ appears to be stronger for other fillings; we obtain $K=2.57$ and $2.42$ for $\bar{n}=1/4$ and $1/6$, respectively, as shown in Appendix~\ref{app:Numerical results for N=L/4 system}. 
        
        As the bipartite charge fluctuation $F_A$ can be written as a double sum of the charge correlation function $\langle n_i n_j \rangle_c$ [see Eq.~\eqref{eq:BFisSum}], the logarithmic scaling of $F_A$ is related to an algebraic decay of $\langle n_i n_j \rangle_c$ \cite{Song10, Song12}, 
        \begin{align}
            \langle n_i n_j\rangle_c \sim -\frac{K}{2\pi^2}\frac{1}{x_{ij}^2}
            \label{eq:ZZTLL},
        \end{align}
        which is another hallmark of criticality described by TLL theory.
        In Fig.~\ref{fig:BFZZ_L2}~(b), we show the charge correlation function $\langle n_i n_j \rangle_c$ at $p=0.12$ as a function of the chord distance $x_{ij}$. 
        It exhibits an algebraically decay for sufficiently large distance with the exponent $a = 2.13$, which is close to the value $2$ predicted by TLL theory and also by the replica field theory in Ref.~\cite{Barratt22a}. 
        On the other hand, the Luttinger parameter $K$ extracted from the scaling form in Eq.~\eqref{eq:ZZTLL} strongly deviates from the expected universal value $K_\sharp = 2$. 
        We also find $a=2.23$ and $a=2.21$ from similar analyses for $\bar{n}=1/4$ and $\bar{n}=1/6$, respectively (see Appendix~\ref{app:Numerical results for N=L/4 system}).

\begin{figure*}[tbp]
        \includegraphics[width=0.8\textwidth]{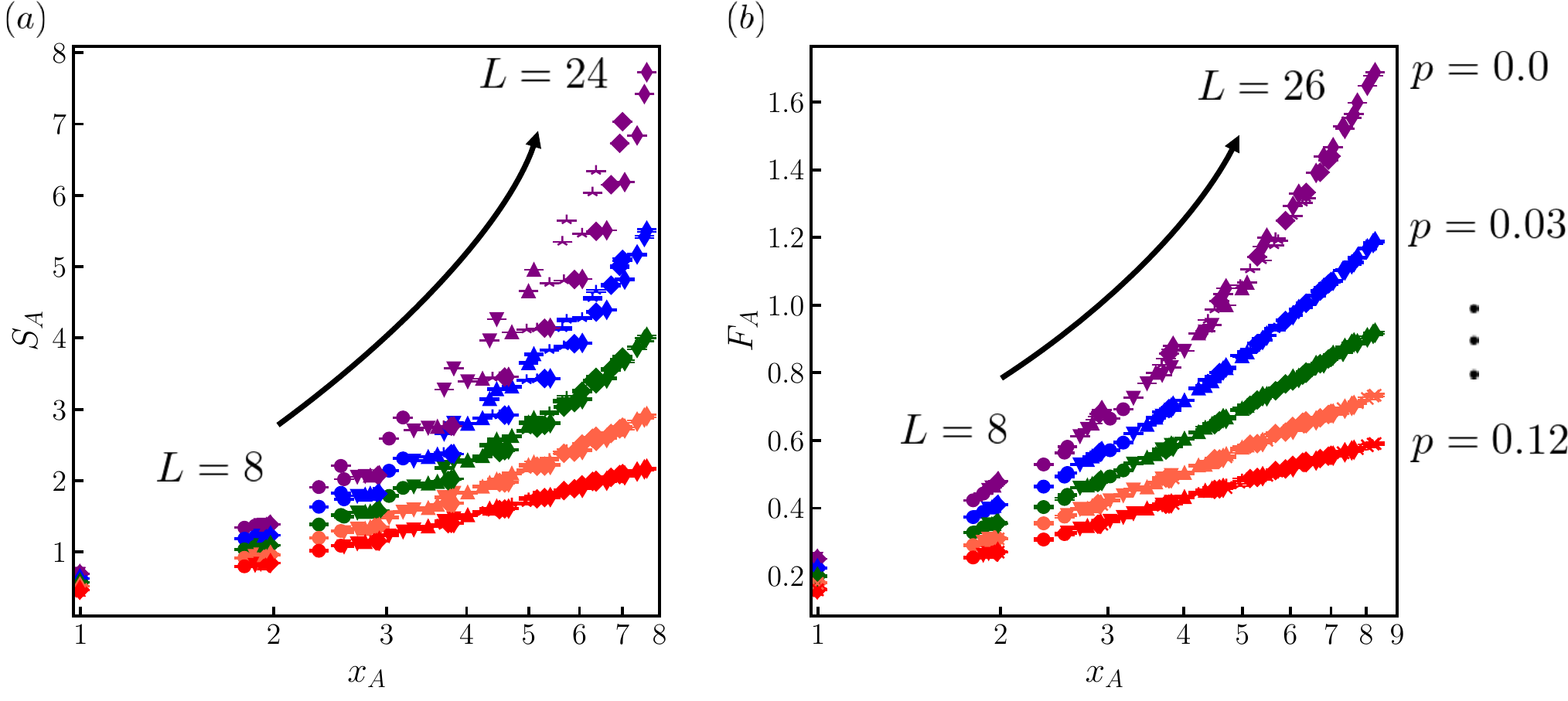}
        \caption{\label{fig:EEBF_L2} 
        Steady-state values of (a) von Neumann entanglement entropy and (b) bipartite charge fluctuation plotted against the chord length $x_A$ of the susbsystem $A$ for $\bar{n}=1/2$ at $p=0.0$, $0.03$, $0.06$, $0.09$, and $0.12$. 
        Data points for different system sizes are distinguished by markers and those with different measurement rates are distinguished by colors.}
        \end{figure*}

        We next examine whether critical properties for charge fluctuation can be seen below the charge-fluctuation transition $p<p_t$. 
        We first compare scaling behaviors of the von Neumann entanglement entropy with those of the bipartite charge fluctuation for $p \leq 0.12$. 
        As shown in Fig.~\ref{fig:EEBF_L2}, the bipartite charge fluctuations for different system sizes collapse well to a single function of the chord length $x_A$, whereas the von Neumann entropies for different system sizes are much more scattered as conformal invariance is not expected in the volume-law phase. 
%~~~~~~~~~~~~~~~~~~~~~~~~~~~~~~~~~~~~~~ FIGURE ~~~~~~~~~~~~~~~~~~~~~~~~~~~~~~~~~~~~~~
%====================================================================================
        We note that, as discussed in Sec.~\ref{subsubsec:Bipartite charge fluctuation}, the bipartite charge fluctuation $F_A$ at $p=0$ is a quadratic function of the subsystem size $|A|$. 
        On the other hand, the bipartite charge fluctuation for finite values of $p$ and for sufficiently large subsystems appears to scale logarithmically, reminiscent of a TLL-like behavior in a charge fuzzy phase below the charge-sharpening transition as predicted in Ref.~\cite{Barratt22a}. 
        
\begin{figure*}[tbp]
        \includegraphics[width=0.9\textwidth]{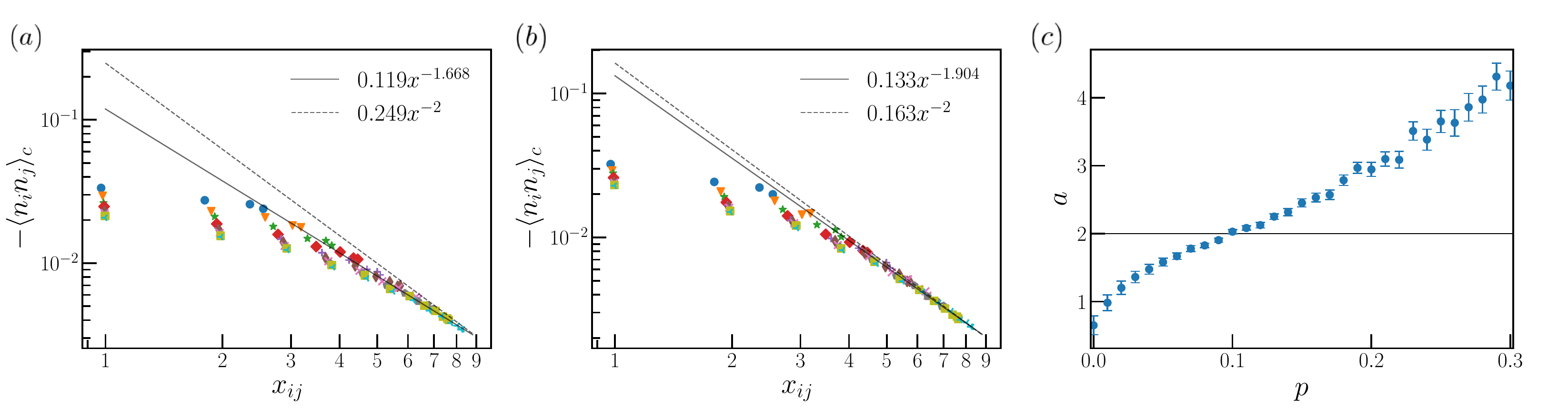}
        \caption{\label{fig:nnDecayExponent}
        Steady-state values of charge correlation function $\langle n_i n_j \rangle_c$ for $\bar{n}=1/2$ at (a) $p=0.06$ and (b) $p=0.09$. 
        The solid lines are fitting functions of the form $C/x_{ij}^a$, whereas the dashed lines are quadratic functions $C'/x_{ij}^2$ shown as reference. 
        (c) Critical exponent $a$ of $\langle n_i n_j \rangle_c$ as a function of the measurement rate $p$ extracted by assuming a power-law form for data points with $x_{ij} \geq 5$. 
        The horizontal solid line shows the exponent $a=2$ predicted by TLL theory. 
        The error bars are least squares uncertainties detailed in the main text.
        }
        \end{figure*}

        In Figs.~\ref{fig:nnDecayExponent}(a) and \ref{fig:nnDecayExponent}(b), we show the charge correlation function $\langle n_i n_j \rangle_c$ for $p=0.06$ and $p=0.09$, respectively. 
%~~~~~~~~~~~~~~~~~~~~~~~~~~~~~~~~~~~~~~ FIGURE ~~~~~~~~~~~~~~~~~~~~~~~~~~~~~~~~~~~~~~
%====================================================================================
        Data points plotted against the chord distance $x_{ij}$ are scattered in a small $x_{ij}$ region but appear to collapse to a single power-law function $1/x_{ij}^a$ for sufficiently large distance.
        Assuming the power-law scaling form, we extract the exponent $a$ by fitting data points of $\langle n_i n_j \rangle_c$ with $x_{ij} \geq 5$, which is shown in Fig.~\ref{fig:nnDecayExponent}(c). 
        Here, the error bars correspond to one standard deviation estimated from the residual sum of squares on least-squares fitting and do not take account of standard errors over trajectories.
        We also note that the power-law fitting would not be reliable for a small and a large $p$ region; since $\langle n_i n_j \rangle_c$ at $p=0$ dose not depend on the subsystem size $|A|$ but does on the system size $L$ as given in Eq.~\eqref{eq:nncorr_inf_temp}, data points for $p \lesssim 0.05$ are still scattered, while those for $p \gtrsim 0.2$ rather decay exponentially with $|A|$. 
        In the intermediate region, the exponent monotonically increases with the measurement rate $p$ and takes a value close to 2 around the charge-fluctuation transition $p \sim 0.12$. 
        However, we cannot clearly observe a plateau of the exponent $2$, which is expected for the charge-fuzzy phase below the charge-sharpening transition \cite{Barratt22a}. 
        This, on the one hand, validates the use of subsystem-charge correlation functions $\langle n_A n_C \rangle_c$ for locating the charge-fluctuation transition as discussed in Sec.~\ref{subsubsec:Bipartite charge fluctuation}; if the transition were of the BKT type and a TLL-like critical phase were extended below the transition, the crossing of $\langle n_A n_C \rangle_c$ at the transition would be obscured and the simple scaling ansatz in Eq.~\eqref{eq:nAnC_scaling} would not hold. 
        As shown in Appendix~\ref{app:Numerical results for N=L/4 system}, similar results have also been obtained for $\bar{n}=1/4$. 
        
\begin{figure}[tbp]
        \includegraphics[width=0.45\textwidth]{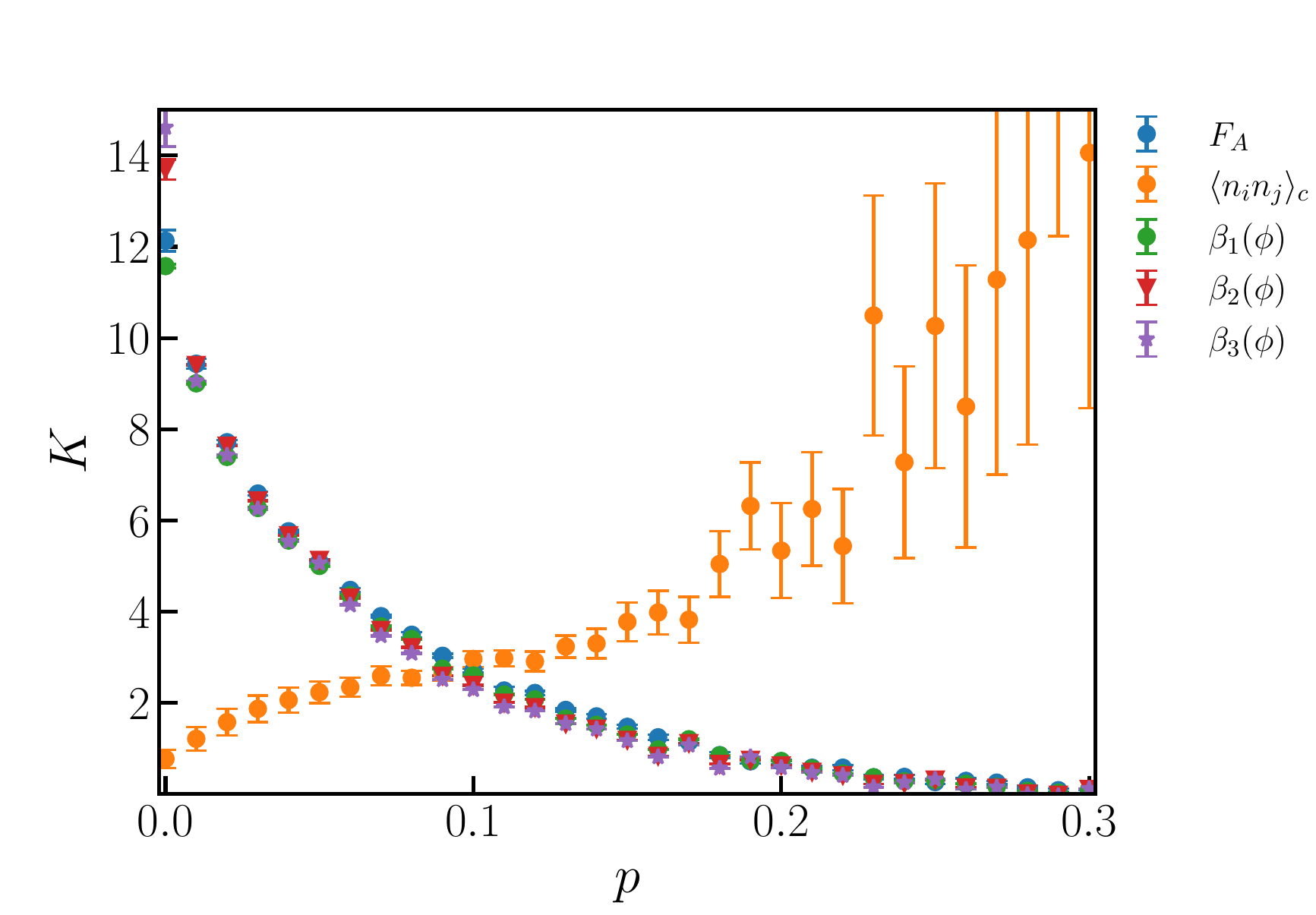}
        \caption{\label{fig:L2_K_co_x5} 
        Luttinger parameter $K$ extracted from the asymptotic behaviors of the bipartite charge fluctuation $F_A$, charge correlation function $\langle n_i n_j \rangle_c$, and $\phi$-dependence of the charged moments $Z_n(\phi)$ with $n=1$, $2$, and $3$ for $\bar{n}=1/2$.}
        \end{figure}

        While the above results for the power-law decay of the charge correlation function $\langle n_i n_j \rangle_c$ do not support the TLL-like critical phase, we here attempt to extract the Luttinger parameter $K$ below the charge-fluctuation transition $p=p_t$ by assuming the logarithmic scaling of the bipartite charge fluctuation $F_A$ in Eq.~\eqref{eq:BFTLL} and the $1/x_{ij}^2$ scaling for $\langle n_i n_j \rangle_c$ in Eq.~\eqref{eq:ZZTLL}. 
        By fitting data points for long-distance regions $x_A, x_{ij} \geq 5$, we obtain the Luttinger parameter $K$ as functions of the measurement rate $p$ as shown in Fig.~\ref{fig:L2_K_co_x5}.
 %~~~~~~~~~~~~~~~~~~~~~~~~~~~~~~~~~~~~~~ FIGURE ~~~~~~~~~~~~~~~~~~~~~~~~~~~~~~~~~~~~~~
%====================================================================================
        In the vicinity of the charge-fluctuation transition $p \sim 0.1$, the estimated Luttinger parameters from $F_A$ and $\langle n_i n_j \rangle_c$ roughly coincide with each other and take values close to 2. 
        However, their deviation is significant away from the transition point, indicating the absence of the TLL-like critical phase below the transition point.

%%%%%%%%%%%%%%%%%%%%%%%%%%%%%%%%%%%%%%%%%%%%%%%%%%%%%%%%%%%%%%%%%%%%%%%%%%%%%%%%%%%%%
%~~~~~~~~~~~~~~~~~~~~~~~~~~~~~~~~~~~~~ SECTION ~~~~~~~~~~~~~~~~~~~~~~~~~~~~~~~~~~~~~~
% Numerical results for half filling system
    % Charge-resolved entanglement
    \subsection{Charge-resolved entanglement}\label{subsec:Charge-resolved entanglement}
%%%%%%%%%%%%%%%%%%%%%%%%%%%%%%%%%%%%%%%%%%%%%%%%%%%%%%%%%%%%%%%%%%%%%%%%%%%%%%%%%%%%%
    
    We here propose symmetry-resolved entanglement \cite{Xavier18, Goldstein18, Feldman19, Bonsignori19, Parez21a, Parez21b, Fraenkel21, Cornfeld18, Parez22} as another diagnostics for measurement-induced criticality in monitored quantum systems with charge conservation. 
    Since the total charge $N$ is conserved in our hybrid circuit, we have $[\rho, n_\textrm{tot}] = 0$ where $n_\textrm{tot}$ is the total charge operator defined in Eq.~\eqref{def:charge}. 
    Then, taking the partial trace over the degrees of freedom supported on the complement of a subsystem $A$ yields $[\rho_A, n_A] = 0$ where $n_A$ is the total charge operator in the subsystem $A$. 
    This implies that the reduced density matrix $\rho_A$ takes a block-diagonal form: 
    \begin{align}
        \rho_A = \bigoplus_{N_A} \Pi_{N_A} \rho_A = \bigoplus_{N_A} p(N_A) \rho_A(N_A),
    \end{align}
    where $N_A$ denotes an eigenvalue of $n_A$, $\Pi_{N_A}$ is the projection operator to the eigenspace associated with the eigenvalue $N_A$, and $p(N_A) = \textrm{Tr}_A (\Pi_{N_A} \rho_A)$ is the probability of finding $N_A$ as the outcome of a measurement of $n_A$. 
    With this definition, the reduced density matrix in each block $\rho_A(N_A)$ is normalized as $\textrm{Tr}_A \rho_A(N_A) = 1$. 
    We can then define the symmetry-resolved R\'enyi entanglement entropy,
    \begin{align}
        S^{(n)}_A(N_A) \equiv \frac{1}{1-n} \ln \textrm{Tr}_A [\rho_A(N_A)]^n.
    \end{align}
    However, direct evaluation of $\rho_A(N_A)$ is often difficult due to the nonlocal nature of the projection operator $\Pi_{N_A}$. 
    Instead, the charged moment defined by 
    \begin{align}
        Z_n(\phi) \equiv \textrm{Tr}_A (\rho_A^n e^{i\phi n_A})
    \end{align}
    has been frequently studied in the literature as it is much easier to analyze. 
    It is related to the symmetry-resolved R\'enyi entropy via 
    \begin{align}
        S^{(n)}_A(N_A) = \frac{1}{1-n} \ln \frac{\mathcal{Z}_n(N_A)}{[\mathcal{Z}_1(N_A)]^n},
    \end{align}
    where $\mathcal{Z}_n(N_A)$ is the Fourier transform of the charged moment $Z_n(\phi)$:
    \begin{align}
        \mathcal{Z}_n(N_A) \equiv \int_{-\pi}^\pi \frac{d\phi}{2\pi} e^{-i\phi N_A} Z_n (\phi) = \textrm{Tr}_A [\Pi_{N_A} \rho_A^n].
    \end{align}
    The parameter $\phi$ introduced here can be seen as a flux for the charged particle.
    
    The charged moment $Z_n(\phi)$ can be seen as a quantity interpolating between entanglement entropy and charge fluctuation under a bipartition. 
    Let us denote the logarithm of the charged moment by $M_n(\phi) \equiv -\ln Z_n(\phi)$. 
    For $\phi=0$, it reduces to the ordinary R\'enyi entanglement entropy up to a factor $n-1$, 
    \begin{align}
        M_n(0) = -\ln \mathrm{Tr}_A \rho_A^n = (n-1) S_A^{(n)} \quad (n \neq 1).
    \end{align}
    On the other hand, its derivative for $n=1$ is related to the bipartite charge fluctuation: 
    \begin{align}
        \partial_{\phi}^2 M_1(0) = (\partial_{\phi}Z_1(0))^2-Z_1(0) \partial_{\phi}^2 Z_1(0) = F_A.
    \end{align}
    Thus, the charged moment is expected to reveal critical properties both at the entanglement transition $p=p_c$ and at the charge-fluctuation transition $p=p_t$. 
    Since in our model $p_c$ and $p_t$ are too close to differentiate these transitions, we here focus on the charged moment near and below the charge-fluctuation transition $p = 0.12$. 

\begin{figure*}[tbp]
    \includegraphics[width=0.95\textwidth]{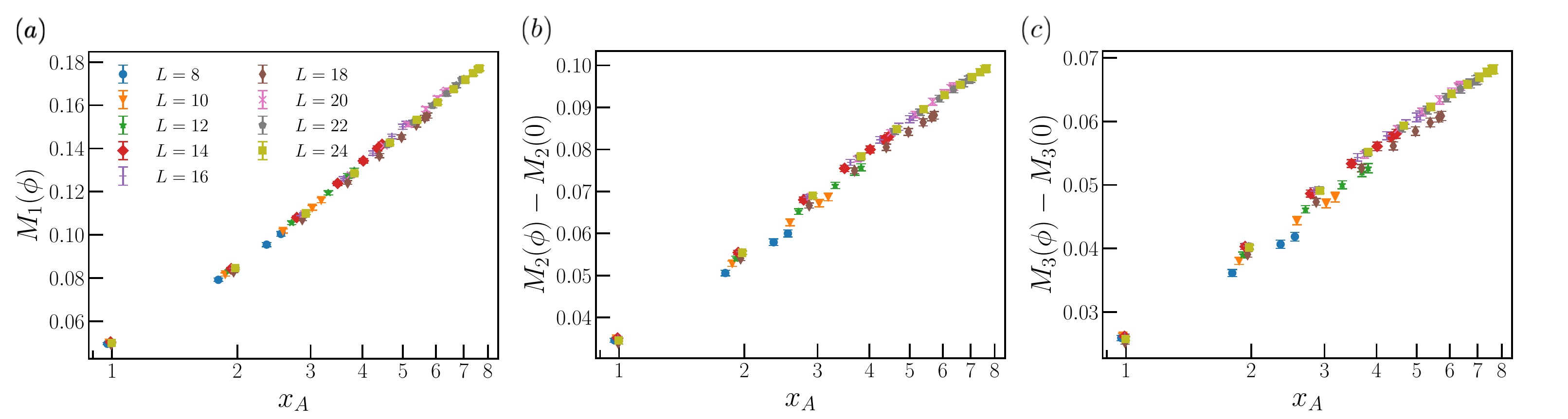}
    \caption{\label{fig:CM_L2} 
    Steady-state values of the logarithms of the charged moments $M_n(\phi)$ with $\phi=\pi/4$ and (a) $n=1$, (b) $n=2$, and (c) $n=3$ are plotted against the chord length $x_A$ of the subsystem $A$ for $\bar{n}=1/2$ and $L = 8$ to $24$. 
    For $M_n(\phi)$ with $n=2$ and $3$, the contribution from $M_n(0)$ is subtracted.}
    \end{figure*}

\begin{figure}[tbp]
    \includegraphics[width=0.45\textwidth]{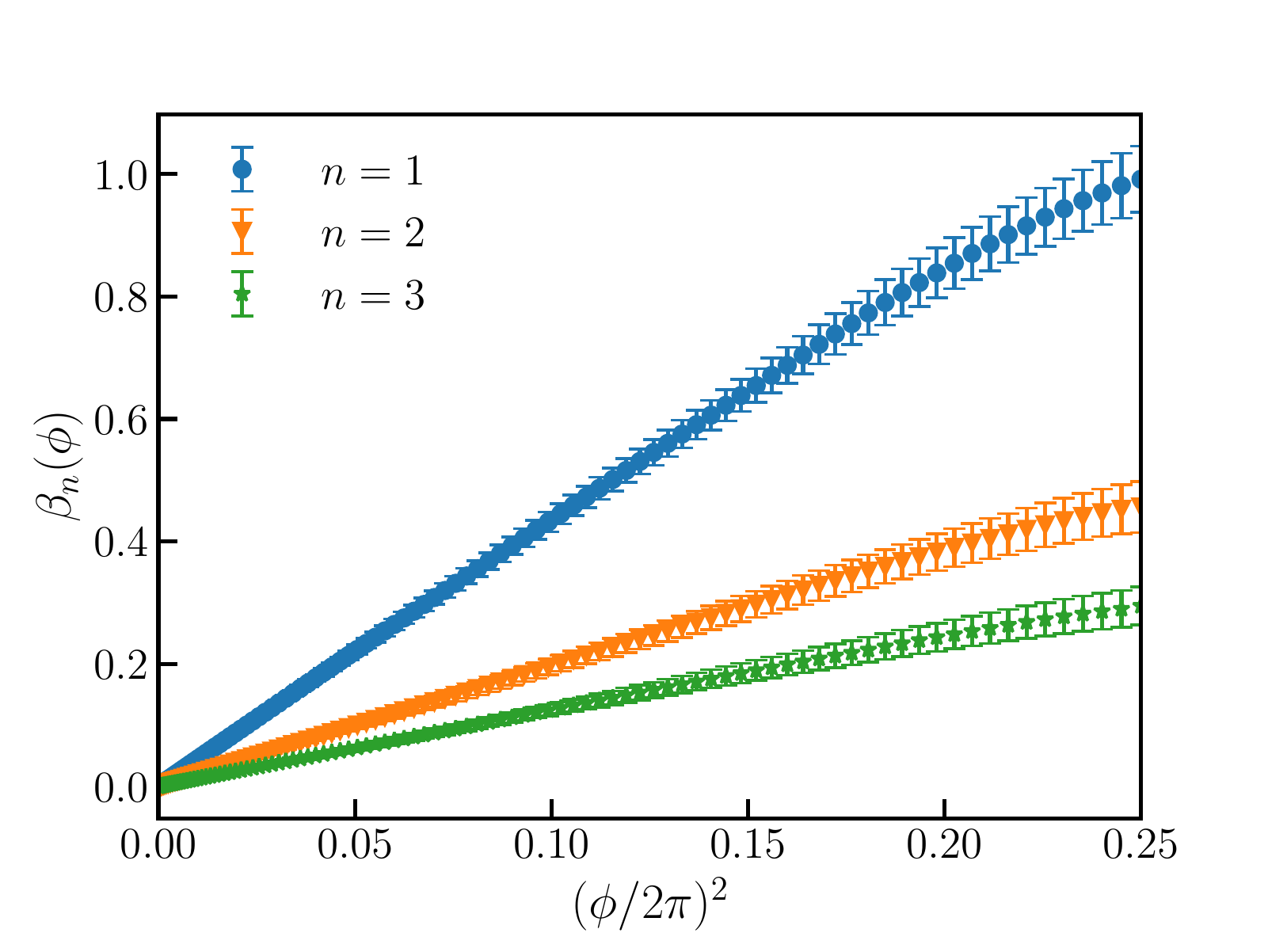}
    \caption{\label{fig:beta_p012}
    Steady-state values of $\beta_n(\phi)$ for $n=1$, $2$, and $3$ are plotted against $(\phi/2\pi)^2$ at $p=0.12$ for $\bar{n}=1/2$, which are estimated from linear fitting of $M_1(\phi)$ and $M_n(\phi)-M_n(0)$ for $n=2,3$ against $\ln x_A$, using data points with $x_A\geq5$. 
    The error-bars are the least squares uncertainties.} 
    \end{figure}

    From TLL theory for (1+1)D critical systems with $U(1)$ charge conservation, the logarithm of the charged moment is expected to scale logarithmically with the chord length of the subsystem size $x_A$ \cite{Xavier18, Goldstein18}, 
    \begin{align}
        M_n(\phi) \sim \gamma_n(\phi)\ln x_A \label{eq:charged moment},
    \end{align}
    where the coefficient $\gamma_n(\phi)$ is a universal quadratic function depending only on the central charge $c$ and the Luttinger parameter $K$, 
    \begin{align}
        \gamma_n(\phi) &= \alpha_n +\beta_n(\phi), \\
        \alpha_n &= \frac{c}{6} \left( n-\frac{1}{n} \right), \\
        \beta_n(\phi) &= \frac{2K}{n} \phi^2. 
        \label{eq:Luttparam}
    \end{align}
    Therefore, we can extract both $c$ and $K$ from the charged moment at criticality described by TLL theory. 
    However, the scaling form in Eq.~\eqref{eq:charged moment} will not hold as it is at the charge-fluctuation transition in our monitored circuit, since the transition occurs within the volume-law phase of entanglement. 
    As $M_n(\phi)$ reduces to the $n$th R\'enyi entanglement entropy for $\phi=0$ and $n \neq 1$, there must be a term linear in the subsystem size $|A|$. 
    It leads us to make the following scaling ansatz:
    \begin{align}
        M_n(\phi) \sim \gamma_n(\phi) \ln x_A + \lambda_n |A| \quad (n \neq 1),
        \label{eq:charged moment corrected}
    \end{align}
    where $\lambda_n$ is a nonuniversal constant independent of $\phi$. 
    Thus, the logarithmic part might be extracted as 
    \begin{align}
        M_n(\phi)-M_n(0) \sim \beta_n(\phi) \ln x_A. 
        \label{eq:beta}
    \end{align}
    We note $M_1(0)=0$ so that $\alpha_1$ and $\lambda_1$ should vanish.
    We examine whether this scaling holds at the charge-fluctuation transition $p=0.12$. 
    In Fig.~\ref{fig:CM_L2}, we show the trajectory averages of $M_1(\phi)$ and $M_n(\phi)-M_n(0)$ for $n=2$ and $3$ as functions of the chord length $x_A$ at $\phi=\pi/4$.
%~~~~~~~~~~~~~~~~~~~~~~~~~~~~~~~~~~~~~~ FIGURE ~~~~~~~~~~~~~~~~~~~~~~~~~~~~~~~~~~~~~~
%====================================================================================
    Their long-length behaviors are well fitted into logarithmic functions. 
    We have also confirmed that the logarithmic scaling generically holds for other values of the flux $\phi$.
    By fitting data points with $x_A \geq 5$ for various values of the flux $\phi$, we obtain the coefficient $\beta_n(\phi)$ in Eq.~\eqref{eq:beta} as a function of $\phi$, which is shown in Fig.~\ref{fig:beta_p012}.
%~~~~~~~~~~~~~~~~~~~~~~~~~~~~~~~~~~~~~~ FIGURE ~~~~~~~~~~~~~~~~~~~~~~~~~~~~~~~~~~~~~~
%====================================================================================
    Here, the error bars are least squares uncertainties as mentioned in Sec.~\ref{subsubsec:Scaling behaviors and TLL}.
    Except for a region around $\phi = \pi$, the coefficient $\beta_n(\phi)$ grows quadratically in $\phi$. 
    This behavior is consistent with one predicted by TLL theory. 
    Supposing that the charged moment obeys the logarithmic scaling form in Eq.~\eqref{eq:beta} and its coefficient $\beta_n(\phi)$ is a quadratic function of the flux $\phi$ as given in Eq.~\eqref{eq:Luttparam}, we have extracted the Luttinger parameter $K$ for general values of the measurement rate $p$. 
    As shown in Fig.~\ref{fig:L2_K_co_x5}, the Luttinger parameters estimated from the charged moments $Z_n(\phi)$ almost coincide with that obtained by the bipartite charge fluctuation $F_A$ but deviate from that obtained by the charge correlation function $\langle n_i n_j \rangle_c$. 
    This again indicates that the TLL-like critical behavior is observed around the charge-fluctuation transition $p \sim 0.1$ but not away from it.

%%%%%%%%%%%%%%%%%%%%%%%%%%%%%%%%%%%%%%%%%%%%%%%%%%%%%%%%%%%%%%%%%%%%%%%%%%%%%%%%%%%%%
%~~~~~~~~~~~~~~~~~~~~~~~~~~~~~~~~~~~~~ SECTION ~~~~~~~~~~~~~~~~~~~~~~~~~~~~~~~~~~~~~~
% Discussion
\section{Discussion}\label{sec:Disucussion}
%%%%%%%%%%%%%%%%%%%%%%%%%%%%%%%%%%%%%%%%%%%%%%%%%%%%%%%%%%%%%%%%%%%%%%%%%%%%%%%%%%%%%

We numerically studied MIPTs in the presence of $U(1)$ symmetry for the (1+1)D monitored Haar-random circuit. 
We particularly focused on steady-state quantities obtained by evolving an initial state with a fixed total charge at a given filling fraction and averaged over different quantum trajectories. 
We first located an entanglement transition between a volume-law and an area-law phase from crossing points in tripartite mutual information and confirmed emergent conformal invariance at the transition from critical scaling properties of various physical quantities, such as logarithmic scaling of entanglement entropy and algebraic decays of squared correlation functions and mutual information. 
We then identified another MIPT, dubbed charge-fluctuation transition, from subsystem-charge correlation functions, which are expected to cross with each other between different system sizes at criticality described by TLL theory. 
The charge-fluctuation transition takes place slightly below the entanglement transition and exhibits critical scaling properties peculiar to TLL criticality, such as logarithmic scaling of the bipartite charge fluctuation and the $1/x^2$ decay of the charge correlation function. 
We also found that the logarithms of charged moments subtracted by the volume-law contribution show logarithmic scaling with universal coefficients quadratic in the flux parameter, which is another characteristic of TLL criticality.

However, it is not very conclusive that the charge-fluctuation transition characterized from static quantities in this study coincides with the charge-sharpening transition characterized from dynamical quantities in Ref.~\cite{Agrawal22}, while the associated correlation length exponents roughly agree with each other.
It also remains unclear that the charge-fluctuation transition exhibits the BKT-type universal nature as expected for the charge-sharpening transition from a replica field theory \cite{Barratt22a}. 
Our numerical results indicate that, although the charge-fluctuation transition in our model exhibits TLL-like critical properties, the corresponding Luttinger parameter $K$ for a general filling fraction does not agree with the universal value $K_\sharp = 2$ expected at the BKT transition. 
We also found that the critical exponent for charge correlation functions below the transition deviates from the universal value $a=2$ expected for TLL criticality, implying the absence of an extended critical phase in the TLL universality class below the charge-fluctuation transition. 
These results suggest that larger system sizes are required to access critical nature extended below the transition or the replica field theory derived in the $d \to \infty$ limit of a monitored circuit with neutral $d$-level qudits in Ref.~\cite{Barratt22a}, which is expected to be valid for large but finite $d$, breaks down or gets modified by a relevant correction for small $d$ ($d=1$ in our case). 
We remark, for the former scenario, that a small-scale numerical simulation is generally hard to observe the BKT transition since the correlation length diverges exponentially in the inverse of the distance from the critical point. 
As another remark, there might be a certain cutoff length below which we cannot observe true scaling behaviors expected for an extended TLL-like critical phase, as found in a monitored free-fermion chain where the entanglement entropy does not exhibit the logarithmic scaling below the BKT transition for a length smaller than a cutoff \cite{Alberton21}.
It will thus be interesting to see how generic the features found in our model for the charge-fluctuation transition hold for larger systems and in other monitored systems with $U(1)$ symmetry, such as a Bose-Hubbard model or a spin chain subject to continuous monitoring \cite{Fuji20, Boorman22, Doggen22}.
It is left for a future work.

There is also an issue regarding experimental feasibility of the MIPT.
Although there are several attempts to experimentally demonstrate MIPTs \cite{Noel22, Koh22}, experimental detection of the MIPT is generally challenging as naive detection schemes, e.g., direct observation of entanglement entropy, require postselection and thereby resources exponentially increasing with both system size and simulation time. 
This issue can be overcome for the Clifford quantum circuit, whose classical simulability allows one to experimentally observe a signature of MIPT \cite{Noel22} with an ancilla probe proposed in Ref.~\cite{Gullans20a}. 
(A similar classical-quantum hybrid protocol for measuring squared quantities, such as the bipartite charge fluctuation, has also been proposed in Ref.~\cite{Garratt22}.)
However, this will not be a cure in the presence of $U(1)$ symmetry; Clifford unitary circuits that conserve a total charge \cite{Richter22}, say, the sum of Pauli $Z$ operators over all sites, cannot generate entanglement from a product state of local basis states with definitive charges, i.e., a product state consisting of eigenstates of the Pauli $Z$ operators. 
Thus, entanglement dies out after a sufficiently long time for any finite measurement rate, and we will not have any steady-state characterizations of MIPTs for Clifford circuits with $U(1)$ symmetry. 

On the other hand, a more general direction for circumventing the postselection problem has recently been proposed by considering adoptive dynamics for which unitary evolution is fed by prior measurement outcomes \cite{Iadecola22, Buchhold22, Friedman22}. 
Therefore, it is also an interesting future task to study whether the nature of the charge-fluctuation transition persists under active feedback by measurement outcomes and whether it can be detected by experimentally accessible quantities in such protocols.

\acknowledgments

We thank Takahiro Sagawa for valuable discussions.
Y.F. is supported by JSPS KAKENHI Grant No. JP20K14402.
The computations have been performed using the facilities of the Supercomputer Center, the Institute for Solid State Physics, the University of Tokyo.

%%%%%%%%%%%%%%%%%%%%%%%%%%%%%%%%%%%%%%%%%%%%%%%%%%%%%%%%%%%%%%%%%%%%%%%%%%%%%%%%%%%%%
%%%%%%%%%%%%%%%%%%%%%%%%%%%%%%%%%%%%%%%%%%%%%%%%%%%%%%%%%%%%%%%%%%%%%%%%%%%%%%%%%%%%%
%~~~~~~~~~~~~~~~~~~~~~~~~~~~~~~~~~~ APPENDIX SECTION ~~~~~~~~~~~~~~~~~~~~~~~~~~~~~~~~
\appendix
% Infinite-temperature average
\section{Infinite-temperature average}\label{app:infinite-temperature average}
%%%%%%%%%%%%%%%%%%%%%%%%%%%%%%%%%%%%%%%%%%%%%%%%%%%%%%%%%%%%%%%%%%%%%%%%%%%%%%%%%%%%%

Here we evaluate the infinite-temperature values of connected correlation functions, bipartite charge fluctuation, and von Neumann mutual information between two sites. 
Since we consider a system with $U(1)$ charge conservation, the canonical ensemble for a given total charge $N$ at infinite temperature is given by the mixed-state density matrix,
\begin{align}
\rho_N = \frac{\Pi_N}{\textrm{Tr} \Pi_N},
\end{align}
where $\Pi_N$ is the projection operator onto the Hilbert subspace with the total charge $N$.
If we assume that each pure state $\ket{\psi(t)}$ evolved by a set of Haar-random unitary gates after sufficiently long time approximates an observable $O$ evaluated at infinite temperature, namely, 
\begin{align}
\langle \psi(t) | O | \psi(t) \rangle \approx \textrm{Tr} (\rho_N O) \quad (t \to \infty), 
\end{align}
the steady-state value of the corresponding trajectory average will also be given by the infinite-temperature average $\textrm{Tr}(\rho_N O)$ in the absence of measurements. 
In this appendix, we consider a system of $L$ qubits and denote the infinite-temperature average of an operator $O$ in a fixed charge sector with total charge $N$ by $\langle O \rangle \equiv \textrm{Tr}(\rho_N O)$.

Let us express a local charge operator in the computational basis as
\begin{align}
    n_i = \ket{1}\bra{1}.
\end{align}
We also define a creation and annihilation operator by
\begin{align}
    b_i^\dagger = \ket{1}\bra{0}, \quad b_i=\ket{0}\bra{1},
\end{align}
respectively. 
We first examine the infinite-temperature average of $n_i$, 
\begin{align}
    \langle n_i \rangle = \frac{\mathrm{Tr}(\Pi_N n_i)}{\mathrm{Tr}\Pi_N}.
\end{align}
The denominator is nothing but the dimension of the subspace with the total charge $N$ and is given by the number of cases to create $N$ particles among $L$ sites,
\begin{align}
    \mathrm{Tr}\Pi_N = \mqty(L\\N) = \frac{L!}{N!(L-N)!}.
\end{align}
The numerator is given by the number of particle configurations with $i$th site occupied among all possible configurations with $N$ particles in $L$ sites.
Thus, we find
\begin{align}
    \langle n_i \rangle = \frac{\mqty(L-1\\N-1)}{\mqty(L\\N)}=\frac{N}{L}
\end{align}
Similarly, $\textrm{Tr} (\Pi_N n_i n_j)$ for $i \neq j$ is given by the number of configurations with $i$th and $j$th sites occupied among all possible configurations with $N$ particles in $L$ sites. 
We then find the infinite-temperature average of a charge correlation function,
\begin{align}
    \langle n_i n_j \rangle = \frac{\mqty(L-2\\N-2)}{\mqty(L\\N)} = \frac{N(N-1)}{L(L-1)} \quad (i \neq j)
\end{align}
On the other hand, the average of any off-diagonal correlation function vanishes:
\begin{align}
    \langle b_i^\dagger b_j \rangle = 0 \quad (i \neq j)
\end{align}

It is now straightforward to evaluate the connected correlation function for local charge operators: 
\begin{align}
    \langle n_i n_j \rangle_c 
    &= \langle n_i n_j \rangle - \langle n_i \rangle \langle n_j \rangle \nonumber \\
    &= \frac{N(N-L)}{L^2(L-1)} \quad (i \neq j)
\end{align}
Since the local charge variance is computed as
\begin{align}
    \langle n_i n_i\rangle_c 
    = \langle n_i^2 \rangle -\langle n_i \rangle^2
    = \frac{N(L-N)}{L^2},
\end{align}
we can evaluate the bipartite charge fluctuation in a subsystem $A$ as
\begin{align}
    F_A &= \sum_{i \in A} \langle n_i n_i \rangle_c +\sum_{i \in A} \sum_{j \neq i} \langle n_i n_j \rangle_c \nonumber \\
    &= \frac{N(L-N)|A|(L-|A|)}{L^2(L-1)}.
\end{align}

We then evaluate the von Neumann mutual information between two sites, $I(i:j) = S_i + S_j - S_{i\cup j}$. 
Due to charge conservation, the single-site reduced density matrix $\rho_i$ takes a diagonal form
\begin{align}
    \rho_i = \langle 1-n_i \rangle\ket{0}\bra{0}+\langle n_i \rangle\ket{1}\bra{1}.
\end{align}
The two-site reduced density matrix $\rho_{i\cup j}$ also takes the block-diagonal form
\begin{align}
    \rho_{i\cup j} &= \langle (1-n_i) (1-n_j) \rangle \ket{00} \bra{00} \nonumber \\
    &+ \langle (1-n_i) n_j \rangle \ket{01} \bra{01} \nonumber \\
    &+ \langle b_i b_j^\dagger \rangle \ket{01} \bra{10}
    +\langle b_i^\dagger b_j \rangle \ket{10} \bra{01} \nonumber \\
    &+ \langle n_i (1-n_j) \rangle \ket{10} \bra{10} + \langle n_i n_j \rangle \ket{11} \bra{11}.
\end{align}
Plugging the infinite-temperature averages of $n_i$, $n_i n_j$, and $b^\dagger_i b_j$ into these expressions, we find the von Neumann entanglement entropy for a single site and two sites, 
\begin{align}
    S_i &= -\frac{N}{L}\ln\frac{N}{L}-\qty(1-\frac{N}{L})\ln\qty(1-\frac{N}{L}), \\
    S_{i\cup j} &= -\qty(1-\frac{2N}{L}+\frac{N(N-1)}{L(L-1)}) \nonumber \\
    &\times\ln\qty(1-\frac{2N}{L}+\frac{N(N-1)}{L(L-1)}) \nonumber\\
    &- \frac{N(N-1)}{L(L-1)}\ln\frac{N(N-1)}{L(L-1)} \nonumber\\
    &- 2\qty(\frac{N}{L}-\frac{N(N-1)}{L(L-1)})\ln\qty(\frac{N}{L}-\frac{N(N-1)}{L(L-1)}),
\end{align}
from which we can compute the two-site mutual information $I(i:j)$.
This gives the expression of $I(i:j)$ for half filling $N=L/2$ in Eq.~\eqref{mutualITS}.
%~~~~~~~~~~~~~~~~~~~~~~~~~~~~~~~~~~~~~~ FIGURE ~~~~~~~~~~~~~~~~~~~~~~~~~~~~~~~~~~~~~~
    \begin{figure*}[tbp]
    \includegraphics[width=0.9\textwidth]{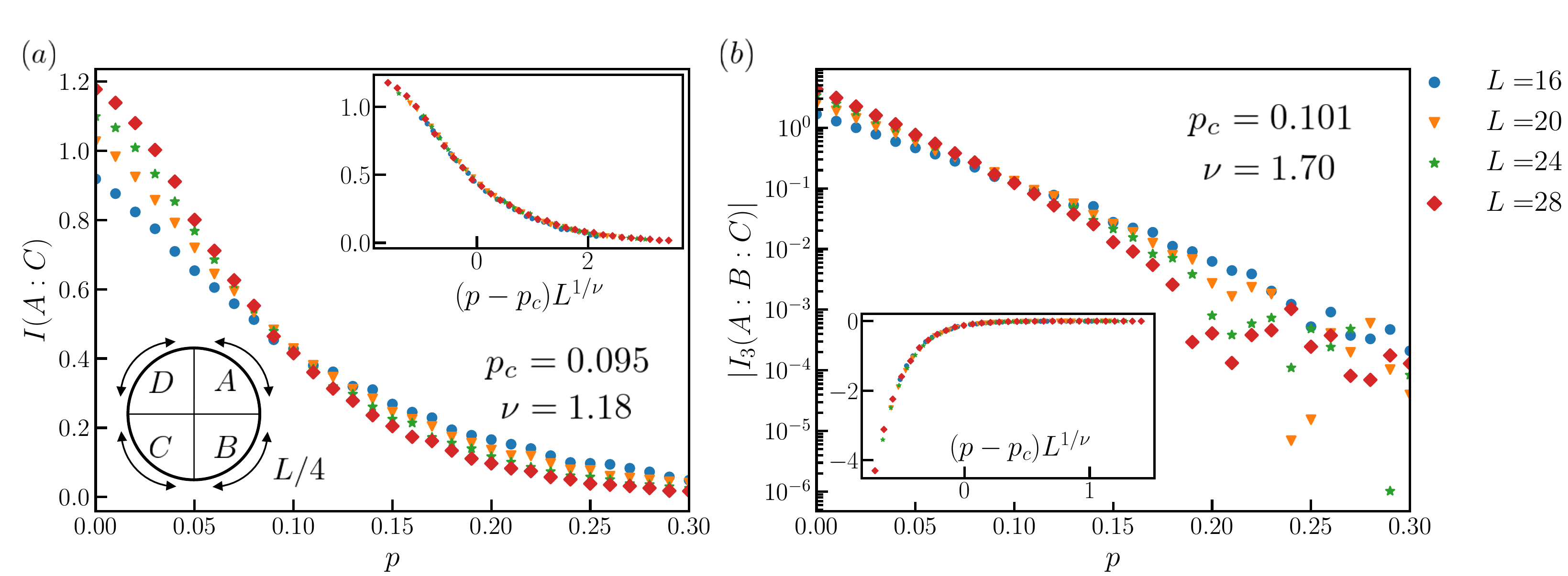}
    \caption{\label{fig:SC_BMITMI_L4} 
    Steady-state values of (a) bipartite mutual information $I(A:C)$ and (b) tripartite mutual information $I_3(A:B:C)$ for $\bar{n}=1/4$ under partition of the system into four contiguous subsystems with $|A|=|B|=|C|=|D|=L/4$. 
    The insets show their scaling collapses against $(p-p_c)L^{1/\nu}$. 
    The critical point $p_c$ and the critical exponent $\nu$ of the entanglement transition is determined as $(p_c, \nu)=(0.095, 1.18)$ from $I(A:C)$ and $(p_c, \nu)=(0.101, 1.70)$ from $I_3(A:B:C)$.
    }
    \end{figure*}
%====================================================================================

%%%%%%%%%%%%%%%%%%%%%%%%%%%%%%%%%%%%%%%%%%%%%%%%%%%%%%%%%%%%%%%%%%%%%%%%%%%%%%%%%%%%%
%~~~~~~~~~~~~~~~~~~~~~~~~~~~~~~~~~~ APPENDIX SECTION ~~~~~~~~~~~~~~~~~~~~~~~~~~~~~~~~
% Numerical results for other fillings
\section{Numerical results for other fillings}\label{app:Numerical results for N=L/4 system}
%%%%%%%%%%%%%%%%%%%%%%%%%%%%%%%%%%%%%%%%%%%%%%%%%%%%%%%%%%%%%%%%%%%%%%%%%%%%%%%%%%%%%

In the main text, we have focused on the half-filling case with $\bar{n}=N/L=1/2$. 
Here we show numerical results for other filling fractions $\bar{n}=1/4$ and $\bar{n}=1/6$. 
For $\bar{n}=1/4$, we have obtained critical properties qualitatively similar to those of the $\bar{n}=1/2$ case for both entanglement transition and charge-fluctuation transition. 
On the other hand, the entanglement transition is obscured for $\bar{n}=1/6$ and only the charge-fluctuation transition has been identified.

%~~~~~~~~~~~~~~~~~~~~~~~~~~~~~~~~~~~~~ SECTION ~~~~~~~~~~~~~~~~~~~~~~~~~~~~~~~~~~~~~~
% Numerical results for N=L/4 system
    % n=1/4 filling
    \subsection{$\bar{n}=1/4$ filling}
%%%%%%%%%%%%%%%%%%%%%%%%%%%%%%%%%%%%%%%%%%%%%%%%%%%%%%%%%%%%%%%%%%%%%%%%%%%%%%%%%%%%%

    We here set the filling fraction to $\bar{n}=1/4$ and use the system sizes from $L=8$ to $L=28$ for entanglement quantities and from $L=8$ to $L=32$ for charge fluctuation and correlation functions. 
    Any steady-state quantities are computed at $t=2L$ and averaged over 1000 trajectories.
    
\begin{figure*}[tbp]
    \includegraphics[width=0.9\textwidth]{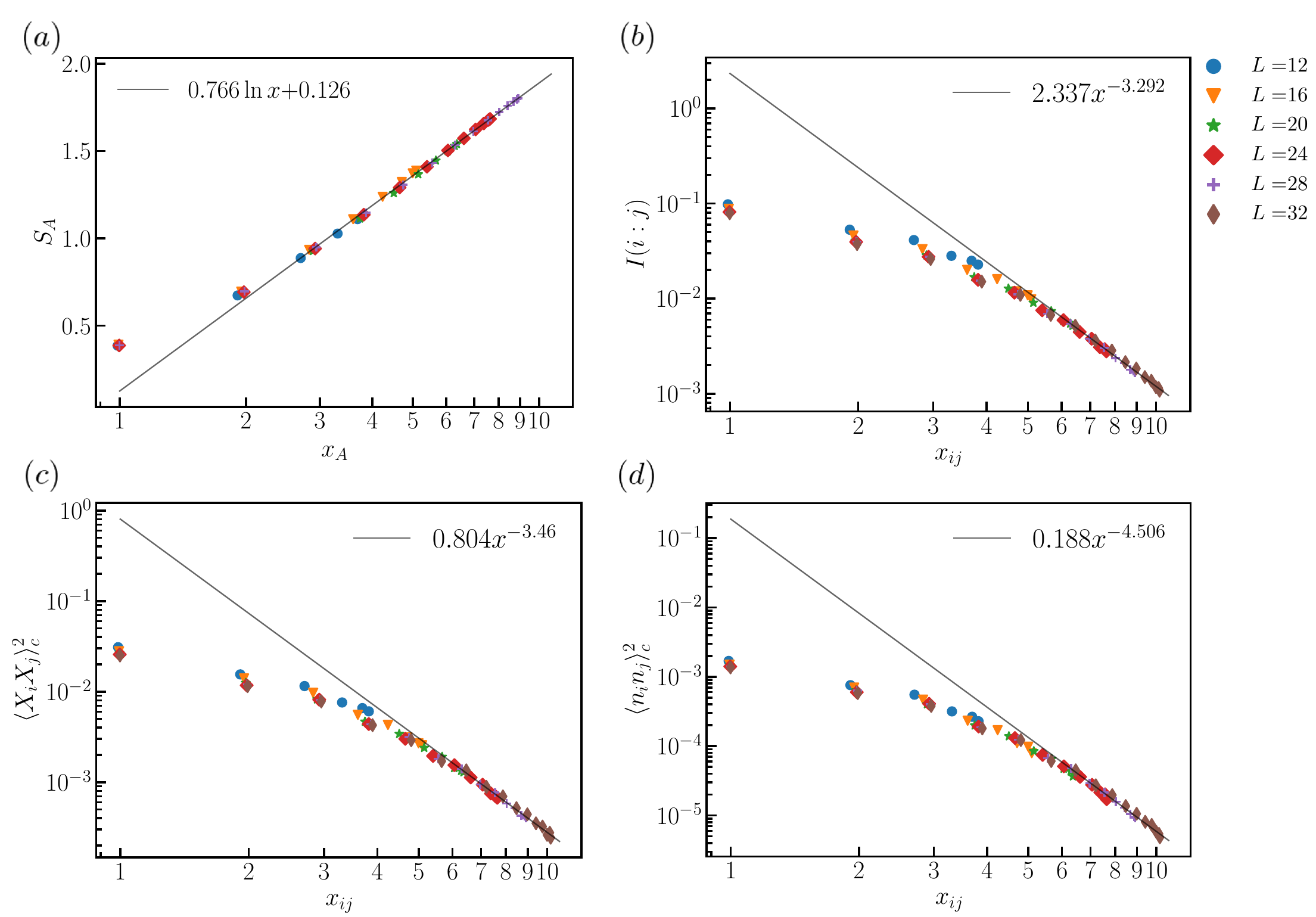}
    \caption{\label{fig:EEMIXXZZ_L4} 
    (a) Steady-state values of von Neumann entanglement entropy are plotted against the chord length $x_A$ of the subsystem $A$ for $L = 12$ to $28$. 
    Steady-state values of (b) mutual information and squared correlation functions (c) $\langle X_iX_j \rangle_c^2$ and (d) $\langle n_in_j \rangle_c^2$ are plotted against the chord distance $x_{ij}$ between two sites for $L = 12$ to $32$. 
    All data are obtained for $\bar{n}=1/4$ at $p=0.1$.
    The solid lines are fitting functions for data points with $x_{A}, x_{ij} \geq 7$.}
    \end{figure*}

    We first locate the entanglement transition between a volume-law and an area-law phase by the bipartite mutual information $I(A:B)$ and the tripartite mutual information $I_3(A:B:C)$ under partition of the system into four contiguous subsystems with $|A|=|B|=|C|=|D|=L/4$. 
    As in the $\bar{n}=1/2$ case, those quantities with different system sizes are expected to cross with each other at the entanglement transition. 
    In Fig.~\ref{fig:SC_BMITMI_L4}, we plot the bipartite and tripartite  mutual information against the measurement rate $p$, which clearly exhibit crossings around $p \sim 0.1$. 
    We then perform scaling collapse by assuming the ansatz of the form \eqref{eq:scaling_collpase_I2}, which yields $p_c=0.095(7)$ and $\nu=1.2(3)$ from the bipartite mutual information and $p_c=0.101(10)$ and $\nu=1.7(4)$ from the tripartite mutual information. 
    Compared with the case of filling $\bar{n}=1/2$, both estimates for the transition point $p_c$ are in good agreement, reflecting the monotonic behavior of the bipartite mutual information. 
    By setting the measurement rate to $p=0.1$, we study the scaling behaviors of the von Neumann entanglement entropy $S_A$, the von Neumann mutual information between two sites $I(i:j)$, and squared correlation functions $\langle X_i X_j \rangle_c^2$ and $\langle n_i n_j \rangle_c^2$ at the entanglement transition, which are shown in Fig.~\ref{fig:EEMIXXZZ_L4}.
%~~~~~~~~~~~~~~~~~~~~~~~~~~~~~~~~~~~~~~ FIGURE ~~~~~~~~~~~~~~~~~~~~~~~~~~~~~~~~~~~~~~
%====================================================================================
    The entanglement entropy exhibits the logarithmic scaling form in Eq.~\eqref{eq:CCformula} with the coefficient $c=2.30$, which is extracted by fitting data points with $x_A \geq 7$. 
    The mutual information $I(i:j)$ and the squared correlation functions $\langle X_i X_j \rangle_c^2$ and $\langle n_i n_j \rangle_c^2$ show algebraic decays for large distance with exponents $\Delta = 1.65$, $\Delta_X = 1.73$, and $\Delta_n = 2.25$, respectively, which are again obtained by fitting data points with $x_{ij} \geq 7$. 
    These observations indicate emergent conformal invariance at the entanglement transition $p=0.1$ for $\bar{n}=1/4$, similarly to the $\bar{n}=1/2$ case discussed in the main text.

\begin{figure*}[tbp]
    \includegraphics[width=1.0\textwidth]{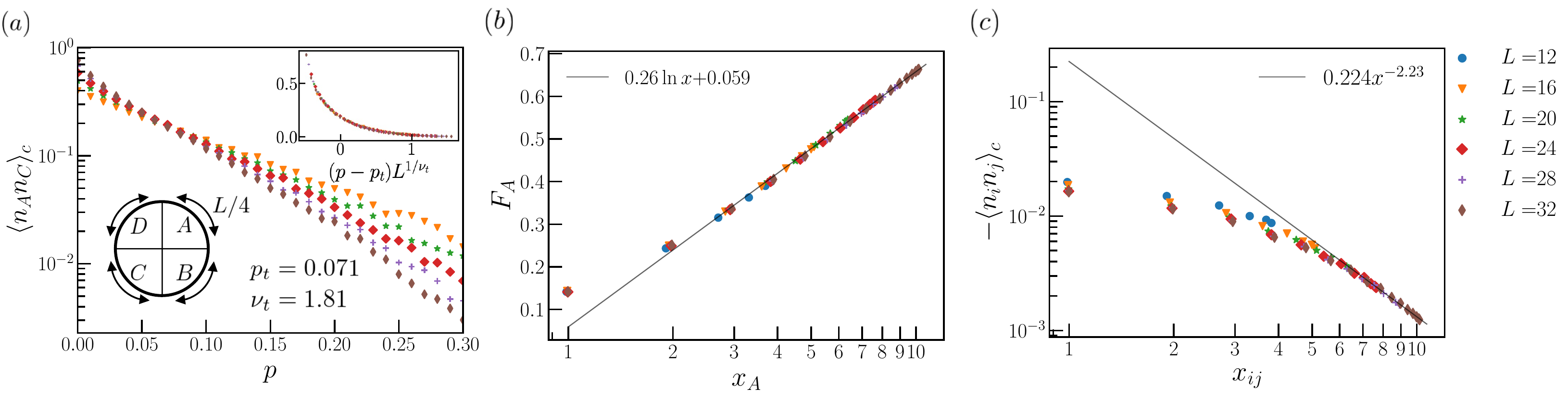}
    \caption{\label{fig:BFZZ_L4} 
    (a) Steady-state values of subsystem-charge correlation function $\langle n_A n_C \rangle_c$ between two antipodal regions $A$ and $C$ with $|A|=|C|=L/4$ for $\bar{n}=1/4$. 
    The inset shows scaling collapse with $p_t = 0.071$ and $\nu_t = 1.81$. 
    Steady-state values of (b) bipartite charge fluctuation and (c) charge correlation function are plotted against $x_A$ and $x_{ij}$, respectively, for $\bar{n}=1/4$ at $p=0.07$. 
    The solid lines are fitting functions obtained from data points with $x_{A}, x_{ij} \geq 7$.
    }
    \end{figure*}

\begin{figure}[tbp]
    \includegraphics[width=0.45\textwidth]{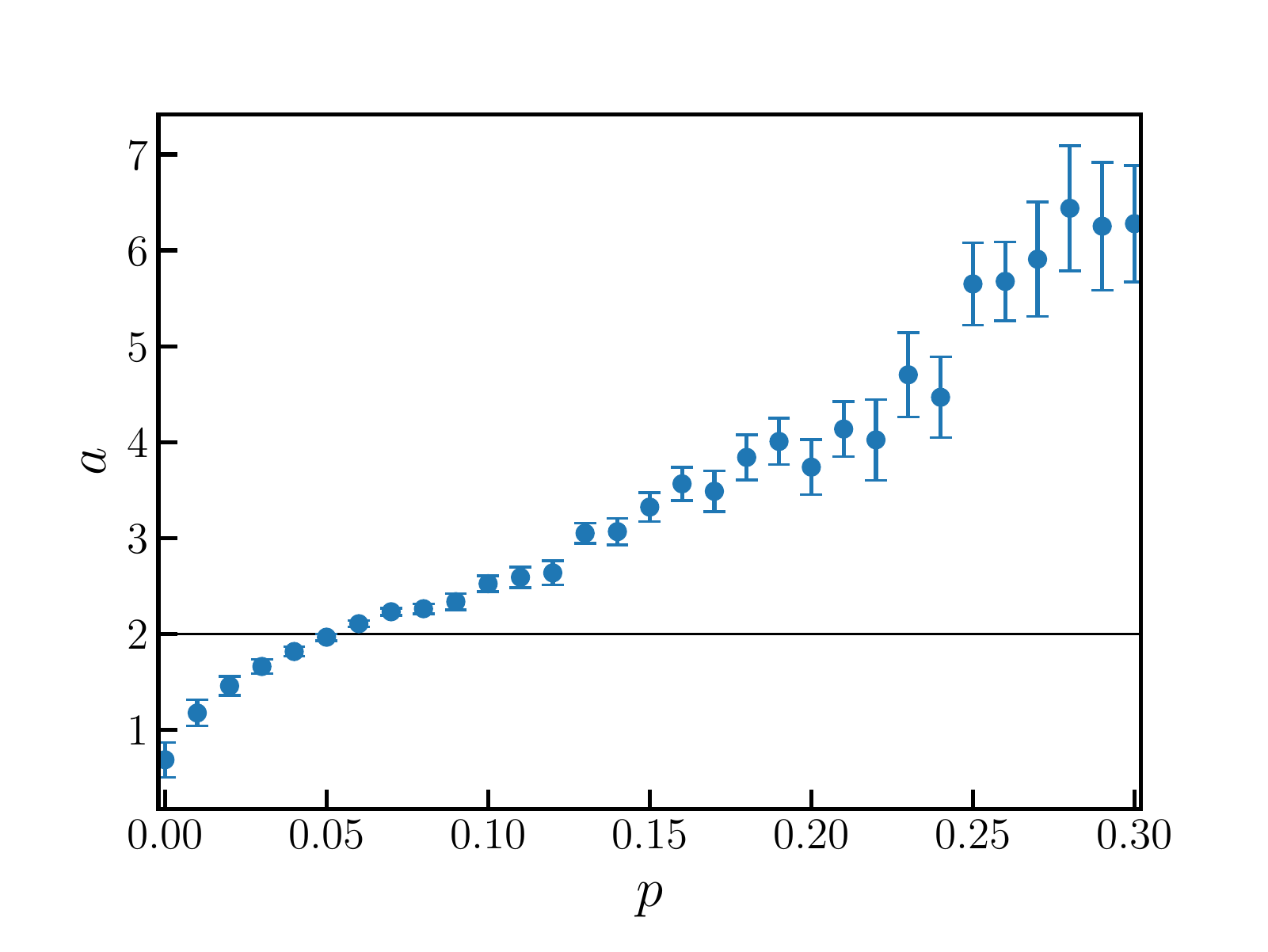}
    \caption{\label{fig:nnDecayExponent_L4} 
    Critical exponent $a$ of $\langle n_i n_j \rangle_c$ for $\bar{n}=1/4$ as a function of the measurement rate $p$ extracted by assuming the power-law form $C/x_{ij}^a$ for data points with $x_{ij} \geq 7$. 
    The horizontal solid line shows the exponent $a=2$ predicted by TLL theory. 
    The error bars correspond to one standard deviation computed from least squares uncertainty. 
    }
    \end{figure}

    We next locate the charge-fluctuation transition by inspecting the bipartite charge fluctuation $F_A$. 
    In Fig.~\ref{fig:BFZZ_L4}~(a), we show the subsystem-charge correlation function $\langle n_A n_C \rangle_c$ as functions of $p$ under the same partition as used above.
%~~~~~~~~~~~~~~~~~~~~~~~~~~~~~~~~~~~~~~ FIGURE ~~~~~~~~~~~~~~~~~~~~~~~~~~~~~~~~~~~~~~
    The correlation functions between different system sizes cross with each other around $p \sim 0.07$, signaling the size-independence of $\langle n_A n_C \rangle_c$ as observed for $\bar{n}=1/2$. 
    By performing scaling collapse with the ansatz in Eq.~\eqref{eq:nAnC_scaling}, we find the charge-fluctuation transition point $p_t = 0.071(4)$ and the associated correlation-length exponent $\nu_t = 1.8(3)$ for the best collapse. 
    Compared with the $\bar{n}=1/2$ case, discrepancy between the entanglement transition and the charge-fluctuation transition is clearer for $\bar{n}=1/4$. 
    The latter transition exists within the volume-law phase of entanglement and may correspond to the charge-sharpening transition predicted in Refs.~\cite{Barratt22a}. 
    We then study the scaling behaviors of the bipartite charge fluctuation $F_A$ and charge correlation function $\langle n_i n_j \rangle_c$ at $p=0.07$, which are shown in Figs.~\ref{fig:BFZZ_L4}(b) and \ref{fig:BFZZ_L4}(c), respectively. 
    The bipartite charge fluctuation clearly exhibits the logarithmic scaling form in Eq.~\eqref{eq:BFTLL} with the Luttinger parameter $K=2.57$, which has been extracted by fitting data points with $x_A \geq 7$. 
    The charge correlation function decays in the power-law form of Eq.~\eqref{eq:ZZTLL} for large distance and we obtain the associated exponent $a=2.23$ by fitting data points with $x_{ij} \geq 7$. 
    Compared with the $\bar{n}=1/2$ case, the Luttinger parameter $K$ at the charge-fluctuation transition deviates from the universal value $K_\sharp=2$ predicted at the charge-sharpening transition in Ref.~\cite{Barratt22a}. 
    On the other hand, the exponent of $\langle n_i n_j \rangle_c$ remains close to the universal value $2$ expected from the TLL theory. 
    These results indicate emergent TLL-like criticality at the charge-fluctuation transition $p=0.07$, but do not strongly support the BKT transition where the Luttinger parameter $K$ takes a universal value.
    To see the possibility of an extended critical phase below the charge-fluctuation transition, we assume the power-law scaling form $1/x^a_{ij}$ for $\langle n_i n_j \rangle_c$ and extract the exponent $a$ as a function of the measurement rate $p$ by fitting data points with $x_{ij} \geq 7$. 
    The result is given in Fig.~\ref{fig:nnDecayExponent_L4}. 
%~~~~~~~~~~~~~~~~~~~~~~~~~~~~~~~~~~~~~~ FIGURE ~~~~~~~~~~~~~~~~~~~~~~~~~~~~~~~~~~~~~~
%====================================================================================
    Similarly to the $\bar{n}=1/2$ case, we cannot clearly find a plateau of the exponent $a=2$ as expected for a critical phase described by TLL theory.

%~~~~~~~~~~~~~~~~~~~~~~~~~~~~~~~~~~~~~ SECTION ~~~~~~~~~~~~~~~~~~~~~~~~~~~~~~~~~~~~~~
% Numerical results for other fillings
    % n=1/6 filling
    \subsection{$\bar{n}=1/6$ filling}
%%%%%%%%%%%%%%%%%%%%%%%%%%%%%%%%%%%%%%%%%%%%%%%%%%%%%%%%%%%%%%%%%%%%%%%%%%%%%%%%%%%%%
    
    We here present several results for the filling fraction $\bar{n}=1/6$.
    We use the system sizes from $L=12$ to $L=30$ for entanglement quantities and from $L=12$ to $L=36$ for charge fluctuation and correlation functions.
    As we have done for the other fillings, any steady-state quantities are computed at $t=2L$ and averaged over 1000 trajectories.
    
\begin{figure*}[tbp]
    \includegraphics[width=0.9\textwidth]{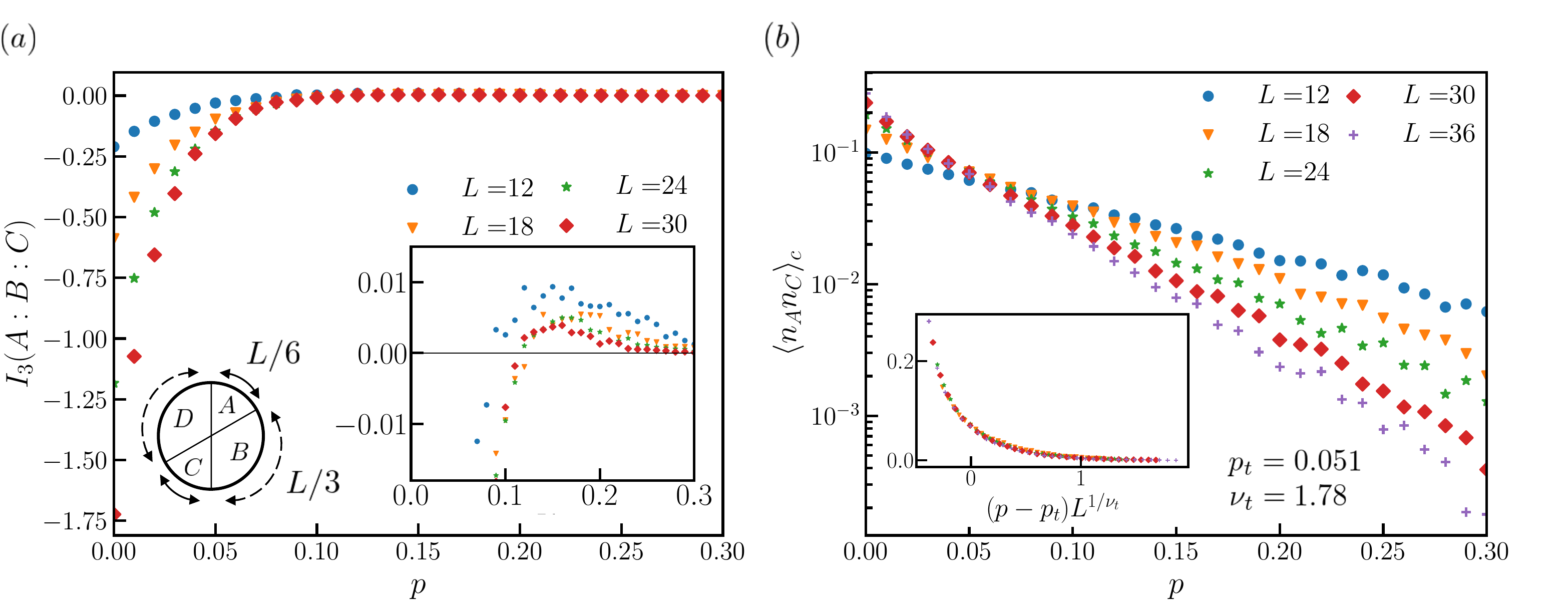}
    \caption{\label{fig:TMI_NANC_L6}
    Steady-state values of (a) tripartite mutual information $I_3(A:B:C)$ and (b) subsystem-charge correlation function for $\bar{n}=1/6$ under the partition of the system into four subsystems $A$, $B$, $C$, and $D$ with $|A|=|C|=L/6$ and $|B|=|D|=L/3$.
    The inset in (a) shows a magnified view in the vicinity of $I_3(A:B:C)=0$ (horizontal solid line), and that in (b) shows scaling collapse of $\langle n_A n_C\rangle_c$ with $p_t=0.051$ and $\nu_t=1.78$.
    }
    \end{figure*}

    We first try to locate the entanglement transition from the bipartite and tripartite mutual information under partition of the system into four contiguous subsystems with sizes $|A|=|C|=L/6$ and $|B|=|D|=L/3$.
    However, both of them do not exhibit a clear crossing for available system sizes.
    As shown in Fig.~\ref{fig:TMI_NANC_L6}~(a), the tripartite mutual information $I_3(A:B:C)$ can even take a positive value by increasing the measurement rate $p$ and becomes a nonmonotonic function of $p$; due to low density of particles in small-size systems, typical trajectories subject to measurements cannot acquire enough entanglement to scramble quantum information encoded in the system, resulting in a positive tripartite mutual information, in contrast to a negative one expected for scrambling states \cite{Iyoda18}. 
%~~~~~~~~~~~~~~~~~~~~~~~~~~~~~~~~~~~~~~ FIGURE ~~~~~~~~~~~~~~~~~~~~~~~~~~~~~~~~~~~~~~
%====================================================================================
    These features make scaling collapse for both bipartite and tripartite mutual information inadequate and do not allow us to reliably estimate the entanglement transition point.

\begin{figure*}[tbp]
    \includegraphics[width=0.9\textwidth]{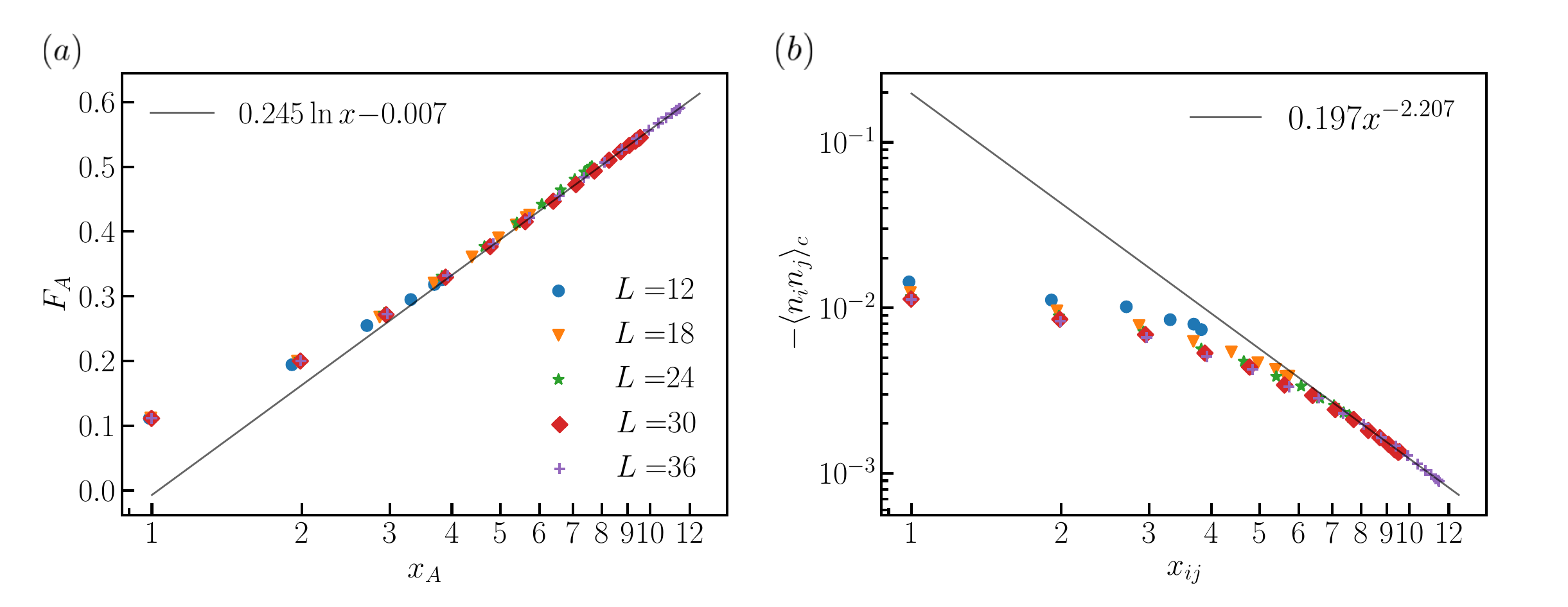}
    \caption{\label{fig:BFZZ_L6} 
    Steady-state values of (a) bipartite charge fluctuation and (b) charge correlation function $\langle n_i n_j \rangle_c$ are plotted against the chord length $x_A$ of the subsystem $A$ and the chord distance $x_{ij}$ between two sites, respectively. 
    We have used data at $p=0.05$ close to the charge-fluctuation transition for $\bar{n}=1/6$ and $L = 12$ to $36$. 
    The solid lines in both panels are fitting functions obtained by using data points with $x_{A}, x_{ij} \geq 8$.
    }
    \end{figure*}

    Nevertheless, the subsystem-charge correlation function $\langle n_A n_C \rangle_c$ computed under the same partition behaves in a similar fashion as observed for the other fillings. 
    As shown in Fig.~\ref{fig:TMI_NANC_L6}~(b), the subsystem-charge correlation functions exhibit a crossing around $p=0.05$ between different system sizes. 
    By performing scaling analysis, we obtain the charge-fluctuation transition point $p_t = 0.051(10)$ and the correlation-length exponent $\nu_t = 1.8(5)$ for the optimal collapse. 
    We then consider the bipartite charge fluctuation $F_A$ and charge correlation function $\langle n_i n_j \rangle_c$ at $p=0.05$, which are shown in Fig.~\ref{fig:BFZZ_L6}. 
%~~~~~~~~~~~~~~~~~~~~~~~~~~~~~~~~~~~~~~ FIGURE ~~~~~~~~~~~~~~~~~~~~~~~~~~~~~~~~~~~~~~
%====================================================================================
    The bipartite charge fluctuation for large subsystem size takes the logarithmic scaling form in Eq.~\eqref{eq:BFTLL} with the Luttinger parameter $K=2.42$, which is extracted by fitting data points with $x_A \geq 8$. 
    The charge correlation function seems to decay in the power-law form in Eq.~\eqref{eq:ZZTLL} for large distance with the exponent $a=2.21$, which is also extracted by fitting data points with $x_{ij} \geq 8$. 
    Similarly to the $\bar{n}=1/4$ case discussed above, the exponent $a$ takes a value close to $2$ expected from TLL theory, but the Luttinger parameter $K$ appears to deviate from the universal value $K_\sharp =2$ predicted at the charge-sharpening transition.

%%%%%%%%%%%%%%%%%%%%%%%%%%%%%%%%%%%%%%%%%%%%%%%%%%%%%%%%%%%%%%%%%%%%%%%%%%%%%%%%%%%%%
%~~~~~~~~~~~~~~~~~~~~~~~~~~~~~~~~~~ APPENDIX SECTION ~~~~~~~~~~~~~~~~~~~~~~~~~~~~~~~~
% Scaling analysis
\section{Scaling analysis}\label{app:Scaling analysis}
%%%%%%%%%%%%%%%%%%%%%%%%%%%%%%%%%%%%%%%%%%%%%%%%%%%%%%%%%%%%%%%%%%%%%%%%%%%%%%%%%%%%%

In Secs.~\ref{subsubsec:Bipartite and tripartite mutual information} and \ref{subsubsec:Bipartite charge fluctuation}, we have performed scaling analyses for the steady-state values of the bipartite mutual information $I(A:C)$, the tripartite mutual information $I_3(A:B:C)$, and the subsystem-charge correlation function $\langle n_A n_C\rangle_c$ for $\bar{n}=1/2$. 
In Appendix~\ref{app:Numerical results for N=L/4 system}, we have performed similar scaling analyses for $\bar{n}=1/4$ and $\bar{n}=1/6$. 
In this appendix, we provide some details about those scaling analyses. 
We follow the method proposed in Ref.~\cite{Houdayer04}. 
The original data is a set of $I(A:C)$, $I_3(A:B:C)$, or $\langle n_A n_C \rangle_c$, which we denote by $y_{ij}$, for various values of the system size $L_i$ and the measurement rate $p_j$.
For a given value of the critical measurement rate $p_c$ and the correlation length exponent $\nu$, we generate a set of data points $(x_{ij}, y_{ij}, dy_{ij})$ where $x_{ij} = (p_j-p_c) L_i^{1/\nu}$ and $dy_{ij}$ is the standard error of $y_{ij}$. 
We then define an objective function
\begin{align}
Q = \frac{1}{\mathcal{N}} \sum_{i,j} \frac{(y_{ij}-Y_{ij})^2}{dy_{ij}^2+dY_{ij}^2}, 
\end{align}
where $\mathcal{N}$ is the number of terms involved in the sum and $Y_{ij}$ and $dY_{ij}$ are the scaling function and its standard error, respectively, estimated from the data points as detailed in Ref.~\cite{Houdayer04}. 
For a given data set, the best estimate for $p_c$ and $\nu$ is obtained by numerically minimizing the objective function $Q$.

\begin{figure*}[tbp]
    \includegraphics[width=0.95\textwidth]{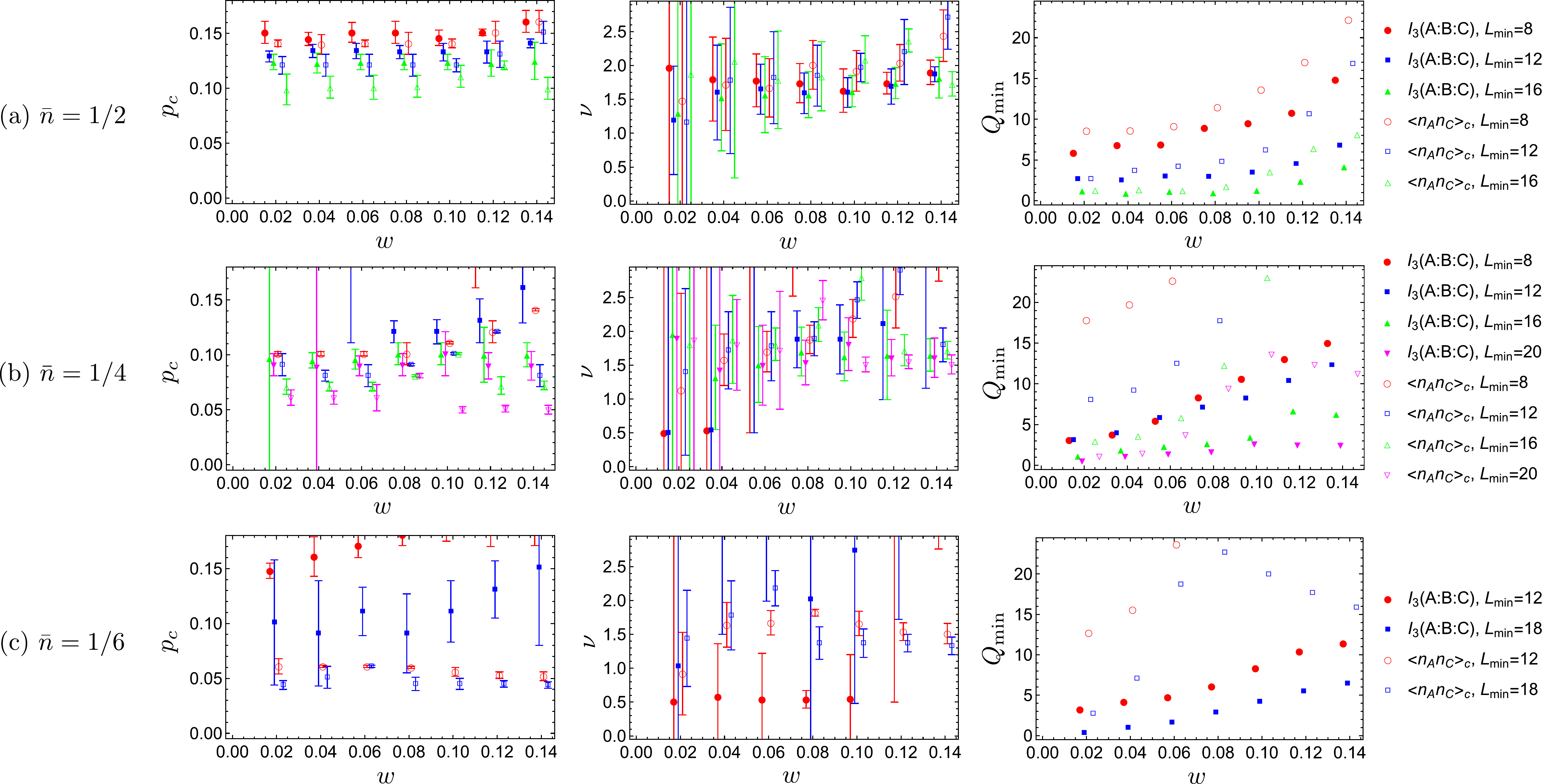}
    \caption{\label{fig:SC_bootstrap}
    Estimated value of the critical measurement rate $p_c$ and the correlation length exponent $\nu$ and corresponding minimum of the objective function $Q_\textrm{min}$ for the tripartite mutual information $I_3(A:B:C)$ and the subsystem-charge correlation function $\langle n_A n_C \rangle_c$ and for each data set specified by the data range $w$ and minimal system length $L_\textrm{min}$. 
    While we have used only $w=0.02, 0.04, 0.06, 0.08, 0.1, 0.12$, and $0.14$, the data points are slightly shifted from these values for visibility.
    }
    \end{figure*}

As the scaling ansatz holds only in the vicinity of the transition and for sufficiently large system sizes, we have to carefully choose a data set used for the scaling analysis for reliable estimate of $p_c$ and $\nu$. 
Specifically, following Ref.~\cite{Luitz15}, we collect data points for measurement rates $p_j$ centered around $p_c$ with a width $2w$, i.e., $p_j \in [p_c-w, p_c+w]$, and for system lengths $L_i \geq L_\textrm{min}$. 
By varying the width $2w$ and the minimal system length $L_\textrm{min}$, we have performed the minimization of the objective function $Q$ within $p_c \in [0.02, 0.25]$ and $\nu \in [0.5, 5]$. 
The minimal value of the objective function $Q_\textrm{min}$ and the corresponding value of $p_c$ and $\nu$ are obtained as functions of $w$ and $L_\textrm{min}$ as shown in Fig.~\ref{fig:SC_bootstrap} for the tripartite mutual information $I_3(A:B:C)$ and the subsystem-charge correlation function $\langle n_A n_C \rangle_c$. 
%~~~~~~~~~~~~~~~~~~~~~~~~~~~~~~~~~~~~~~ FIGURE ~~~~~~~~~~~~~~~~~~~~~~~~~~~~~~~~~~~~~~
%====================================================================================
We then select a data set such that the objective function attains a minimum $1 \lesssim Q_\textrm{min} \lesssim 5$ and the associated value of $w$ ($L_\textrm{min}$) is as large (small) as possible to ensure that enough data points are included in the analysis. 
For $\bar{n}=1/2$, we have chosen $w=0.08$ and $L_\textrm{min}=12$ for all three quantities. 
For $\bar{n}=1/4$, we have chosen $w=0.08$ and $L_\textrm{min}=16$ for $I(A:B)$ and $I_3(A:B:C)$ and $w=0.06$ and $L_\textrm{min}=16$ for $\langle n_A n_C \rangle_c$. 
For $\bar{n}=1/6$, the scaling analysis for $I(A:B)$ and $I_3(A:B:C)$ is unsuccessful as the optimal value of $\nu$ generically flows towards the outside of the range under consideration. 
At this filling, we have chosen $w=0.04$ and $L_\textrm{min}=18$ for $\langle n_A n_C \rangle_c$. 
For each data set, the error for the estimated value of $p_c$ and $\nu$ is obtained as a minimal value of $dp_c$ and $d\nu$ such that the rectangular region with corners $(p_c \pm dp_c, \nu \pm d\nu)$ in the parameter space of $(p_c, \nu)$ encloses a region in which the objective function takes $Q \leq Q_\textrm{min}+1$.

One might be worried about that using different sets of the system size for estimating the entanglement and charge-fluctuation transition causes a superficial discrepancy between the two transition points for $\bar{n}=1/4$. 
By restricting data points to those with $16 \leq L \leq 28$ for $\bar{n}=1/4$, the scaling analysis for the subsystem-charge correlation function $\langle n_A n_C \rangle_c$ yields $p_t = 0.073(9)$ and $\nu_t = 1.6(5)$ for $w=0.06$ and $p_t = 0.081(3)$ and $\nu_t = 1.8(2)$ for $w=0.08$. 
Thus, it does not significantly affect the estimation of the charge-fluctuation transition point.

%%%%%%%%%%%%%%%%%%%%%%%%%%%%%%%%%%%%%%%%%%%%%%%%%%%%%%%%%%%%%%%%%%%%%%%%%%%%%%%%%%%%%
%~~~~~~~~~~~~~~~~~~~~~~~~~~~~~~~~~~ BIBLIOGRAPHY ~~~~~~~~~~~~~~~~~~~~~~~~~~~~~~~~~~~~
\bibliography{MIPTU1MonitCircuit}

\begin{thebibliography}{94}%
\makeatletter
\providecommand \@ifxundefined [1]{%
 \@ifx{#1\undefined}
}%
\providecommand \@ifnum [1]{%
 \ifnum #1\expandafter \@firstoftwo
 \else \expandafter \@secondoftwo
 \fi
}%
\providecommand \@ifx [1]{%
 \ifx #1\expandafter \@firstoftwo
 \else \expandafter \@secondoftwo
 \fi
}%
\providecommand \natexlab [1]{#1}%
\providecommand \enquote  [1]{``#1''}%
\providecommand \bibnamefont  [1]{#1}%
\providecommand \bibfnamefont [1]{#1}%
\providecommand \citenamefont [1]{#1}%
\providecommand \href@noop [0]{\@secondoftwo}%
\providecommand \href [0]{\begingroup \@sanitize@url \@href}%
\providecommand \@href[1]{\@@startlink{#1}\@@href}%
\providecommand \@@href[1]{\endgroup#1\@@endlink}%
\providecommand \@sanitize@url [0]{\catcode `\\12\catcode `\$12\catcode
  `\&12\catcode `\#12\catcode `\^12\catcode `\_12\catcode `\%12\relax}%
\providecommand \@@startlink[1]{}%
\providecommand \@@endlink[0]{}%
\providecommand \url  [0]{\begingroup\@sanitize@url \@url }%
\providecommand \@url [1]{\endgroup\@href {#1}{\urlprefix }}%
\providecommand \urlprefix  [0]{URL }%
\providecommand \Eprint [0]{\href }%
\providecommand \doibase [0]{https://doi.org/}%
\providecommand \selectlanguage [0]{\@gobble}%
\providecommand \bibinfo  [0]{\@secondoftwo}%
\providecommand \bibfield  [0]{\@secondoftwo}%
\providecommand \translation [1]{[#1]}%
\providecommand \BibitemOpen [0]{}%
\providecommand \bibitemStop [0]{}%
\providecommand \bibitemNoStop [0]{.\EOS\space}%
\providecommand \EOS [0]{\spacefactor3000\relax}%
\providecommand \BibitemShut  [1]{\csname bibitem#1\endcsname}%
\let\auto@bib@innerbib\@empty
%</preamble>
\bibitem [{\citenamefont {Potter}\ and\ \citenamefont
  {Vasseur}(2022)}]{Potter22}%
  \BibitemOpen
  \bibfield  {author} {\bibinfo {author} {\bibfnamefont {A.~C.}\ \bibnamefont
  {Potter}}\ and\ \bibinfo {author} {\bibfnamefont {R.}~\bibnamefont
  {Vasseur}},\ }\bibfield  {title} {\bibinfo {title} {{Entanglement Dynamics in
  Hybrid Quantum Circuits}},\ }in\ \href
  {https://doi.org/10.1007/978-3-031-03998-0_9} {\emph {\bibinfo {booktitle}
  {Entanglement in Spin Chains: From Theory to Quantum Technology
  Applications}}},\ \bibinfo {editor} {edited by\ \bibinfo {editor}
  {\bibfnamefont {A.}~\bibnamefont {Bayat}}, \bibinfo {editor} {\bibfnamefont
  {S.}~\bibnamefont {Bose}},\ and\ \bibinfo {editor} {\bibfnamefont
  {H.}~\bibnamefont {Johannesson}}}\ (\bibinfo  {publisher} {Springer
  International Publishing},\ \bibinfo {address} {Cham},\ \bibinfo {year}
  {2022})\ pp.\ \bibinfo {pages} {211--249}\BibitemShut {NoStop}%
\bibitem [{\citenamefont {Fisher}\ \emph {et~al.}(2022)\citenamefont {Fisher},
  \citenamefont {Khemani}, \citenamefont {Nahum},\ and\ \citenamefont
  {Vijay}}]{Fisher22}%
  \BibitemOpen
  \bibfield  {author} {\bibinfo {author} {\bibfnamefont {M.~P.~A.}\
  \bibnamefont {Fisher}}, \bibinfo {author} {\bibfnamefont {V.}~\bibnamefont
  {Khemani}}, \bibinfo {author} {\bibfnamefont {A.}~\bibnamefont {Nahum}},\
  and\ \bibinfo {author} {\bibfnamefont {S.}~\bibnamefont {Vijay}},\ }\bibfield
   {title} {\bibinfo {title} {{Random Quantum Circuits}},\ }\Eprint
  {https://arxiv.org/abs/2207.14280} {arXiv:2207.14280}  (\bibinfo {year}
  {2022})\BibitemShut {NoStop}%
\bibitem [{\citenamefont {Li}\ \emph {et~al.}(2018)\citenamefont {Li},
  \citenamefont {Chen},\ and\ \citenamefont {Fisher}}]{Li18}%
  \BibitemOpen
  \bibfield  {author} {\bibinfo {author} {\bibfnamefont {Y.}~\bibnamefont
  {Li}}, \bibinfo {author} {\bibfnamefont {X.}~\bibnamefont {Chen}},\ and\
  \bibinfo {author} {\bibfnamefont {M.~P.~A.}\ \bibnamefont {Fisher}},\
  }\bibfield  {title} {\bibinfo {title} {{Quantum Zeno effect and the many-body
  entanglement transition}},\ }\href
  {https://doi.org/10.1103/PhysRevB.98.205136} {\bibfield  {journal} {\bibinfo
  {journal} {Phys. Rev. B}\ }\textbf {\bibinfo {volume} {98}},\ \bibinfo
  {pages} {205136} (\bibinfo {year} {2018})}\BibitemShut {NoStop}%
\bibitem [{\citenamefont {Skinner}\ \emph {et~al.}(2019)\citenamefont
  {Skinner}, \citenamefont {Ruhman},\ and\ \citenamefont {Nahum}}]{Skinner19}%
  \BibitemOpen
  \bibfield  {author} {\bibinfo {author} {\bibfnamefont {B.}~\bibnamefont
  {Skinner}}, \bibinfo {author} {\bibfnamefont {J.}~\bibnamefont {Ruhman}},\
  and\ \bibinfo {author} {\bibfnamefont {A.}~\bibnamefont {Nahum}},\ }\bibfield
   {title} {\bibinfo {title} {{Measurement-Induced Phase Transitions in the
  Dynamics of Entanglement}},\ }\href
  {https://doi.org/10.1103/PhysRevX.9.031009} {\bibfield  {journal} {\bibinfo
  {journal} {Phys. Rev. X}\ }\textbf {\bibinfo {volume} {9}},\ \bibinfo {pages}
  {031009} (\bibinfo {year} {2019})}\BibitemShut {NoStop}%
\bibitem [{\citenamefont {Chan}\ \emph {et~al.}(2019)\citenamefont {Chan},
  \citenamefont {Nandkishore}, \citenamefont {Pretko},\ and\ \citenamefont
  {Smith}}]{Chan19}%
  \BibitemOpen
  \bibfield  {author} {\bibinfo {author} {\bibfnamefont {A.}~\bibnamefont
  {Chan}}, \bibinfo {author} {\bibfnamefont {R.~M.}\ \bibnamefont
  {Nandkishore}}, \bibinfo {author} {\bibfnamefont {M.}~\bibnamefont
  {Pretko}},\ and\ \bibinfo {author} {\bibfnamefont {G.}~\bibnamefont
  {Smith}},\ }\bibfield  {title} {\bibinfo {title} {{Unitary-projective
  entanglement dynamics}},\ }\href {https://doi.org/10.1103/PhysRevB.99.224307}
  {\bibfield  {journal} {\bibinfo  {journal} {Phys. Rev. B}\ }\textbf {\bibinfo
  {volume} {99}},\ \bibinfo {pages} {224307} (\bibinfo {year}
  {2019})}\BibitemShut {NoStop}%
\bibitem [{\citenamefont {Li}\ \emph {et~al.}(2019)\citenamefont {Li},
  \citenamefont {Chen},\ and\ \citenamefont {Fisher}}]{Li19}%
  \BibitemOpen
  \bibfield  {author} {\bibinfo {author} {\bibfnamefont {Y.}~\bibnamefont
  {Li}}, \bibinfo {author} {\bibfnamefont {X.}~\bibnamefont {Chen}},\ and\
  \bibinfo {author} {\bibfnamefont {M.~P.~A.}\ \bibnamefont {Fisher}},\
  }\bibfield  {title} {\bibinfo {title} {{Measurement-driven entanglement
  transition in hybrid quantum circuits}},\ }\href
  {https://doi.org/10.1103/PhysRevB.100.134306} {\bibfield  {journal} {\bibinfo
   {journal} {Phys. Rev. B}\ }\textbf {\bibinfo {volume} {100}},\ \bibinfo
  {pages} {134306} (\bibinfo {year} {2019})}\BibitemShut {NoStop}%
\bibitem [{\citenamefont {Szyniszewski}\ \emph {et~al.}(2019)\citenamefont
  {Szyniszewski}, \citenamefont {Romito},\ and\ \citenamefont
  {Schomerus}}]{Szyniszewski19}%
  \BibitemOpen
  \bibfield  {author} {\bibinfo {author} {\bibfnamefont {M.}~\bibnamefont
  {Szyniszewski}}, \bibinfo {author} {\bibfnamefont {A.}~\bibnamefont
  {Romito}},\ and\ \bibinfo {author} {\bibfnamefont {H.}~\bibnamefont
  {Schomerus}},\ }\bibfield  {title} {\bibinfo {title} {{Entanglement
  transition from variable-strength weak measurements}},\ }\href
  {https://doi.org/10.1103/PhysRevB.100.064204} {\bibfield  {journal} {\bibinfo
   {journal} {Phys. Rev. B}\ }\textbf {\bibinfo {volume} {100}},\ \bibinfo
  {pages} {064204} (\bibinfo {year} {2019})}\BibitemShut {NoStop}%
\bibitem [{\citenamefont {Choi}\ \emph {et~al.}(2020)\citenamefont {Choi},
  \citenamefont {Bao}, \citenamefont {Qi},\ and\ \citenamefont
  {Altman}}]{Choi20}%
  \BibitemOpen
  \bibfield  {author} {\bibinfo {author} {\bibfnamefont {S.}~\bibnamefont
  {Choi}}, \bibinfo {author} {\bibfnamefont {Y.}~\bibnamefont {Bao}}, \bibinfo
  {author} {\bibfnamefont {X.-L.}\ \bibnamefont {Qi}},\ and\ \bibinfo {author}
  {\bibfnamefont {E.}~\bibnamefont {Altman}},\ }\bibfield  {title} {\bibinfo
  {title} {{Quantum Error Correction in Scrambling Dynamics and
  Measurement-Induced Phase Transition}},\ }\href
  {https://doi.org/10.1103/PhysRevLett.125.030505} {\bibfield  {journal}
  {\bibinfo  {journal} {Phys. Rev. Lett.}\ }\textbf {\bibinfo {volume} {125}},\
  \bibinfo {pages} {030505} (\bibinfo {year} {2020})}\BibitemShut {NoStop}%
\bibitem [{\citenamefont {Iaconis}\ \emph {et~al.}(2020)\citenamefont
  {Iaconis}, \citenamefont {Lucas},\ and\ \citenamefont {Chen}}]{Iaconis20}%
  \BibitemOpen
  \bibfield  {author} {\bibinfo {author} {\bibfnamefont {J.}~\bibnamefont
  {Iaconis}}, \bibinfo {author} {\bibfnamefont {A.}~\bibnamefont {Lucas}},\
  and\ \bibinfo {author} {\bibfnamefont {X.}~\bibnamefont {Chen}},\ }\bibfield
  {title} {\bibinfo {title} {{Measurement-induced phase transitions in quantum
  automaton circuits}},\ }\href {https://doi.org/10.1103/PhysRevB.102.224311}
  {\bibfield  {journal} {\bibinfo  {journal} {Phys. Rev. B}\ }\textbf {\bibinfo
  {volume} {102}},\ \bibinfo {pages} {224311} (\bibinfo {year}
  {2020})}\BibitemShut {NoStop}%
\bibitem [{\citenamefont {Tang}\ and\ \citenamefont {Zhu}(2020)}]{Tang20}%
  \BibitemOpen
  \bibfield  {author} {\bibinfo {author} {\bibfnamefont {Q.}~\bibnamefont
  {Tang}}\ and\ \bibinfo {author} {\bibfnamefont {W.}~\bibnamefont {Zhu}},\
  }\bibfield  {title} {\bibinfo {title} {{Measurement-induced phase transition:
  A case study in the nonintegrable model by density-matrix renormalization
  group calculations}},\ }\href
  {https://doi.org/10.1103/PhysRevResearch.2.013022} {\bibfield  {journal}
  {\bibinfo  {journal} {Phys. Rev. Research}\ }\textbf {\bibinfo {volume}
  {2}},\ \bibinfo {pages} {013022} (\bibinfo {year} {2020})}\BibitemShut
  {NoStop}%
\bibitem [{\citenamefont {Turkeshi}\ \emph {et~al.}(2020)\citenamefont
  {Turkeshi}, \citenamefont {Fazio},\ and\ \citenamefont
  {Dalmonte}}]{Turkeshi20}%
  \BibitemOpen
  \bibfield  {author} {\bibinfo {author} {\bibfnamefont {X.}~\bibnamefont
  {Turkeshi}}, \bibinfo {author} {\bibfnamefont {R.}~\bibnamefont {Fazio}},\
  and\ \bibinfo {author} {\bibfnamefont {M.}~\bibnamefont {Dalmonte}},\
  }\bibfield  {title} {\bibinfo {title} {{Measurement-induced criticality in
  $(2+1)$-dimensional hybrid quantum circuits}},\ }\href
  {https://doi.org/10.1103/PhysRevB.102.014315} {\bibfield  {journal} {\bibinfo
   {journal} {Phys. Rev. B}\ }\textbf {\bibinfo {volume} {102}},\ \bibinfo
  {pages} {014315} (\bibinfo {year} {2020})}\BibitemShut {NoStop}%
\bibitem [{\citenamefont {Zabalo}\ \emph {et~al.}(2020)\citenamefont {Zabalo},
  \citenamefont {Gullans}, \citenamefont {Wilson}, \citenamefont
  {Gopalakrishnan}, \citenamefont {Huse},\ and\ \citenamefont
  {Pixley}}]{Zabalo20}%
  \BibitemOpen
  \bibfield  {author} {\bibinfo {author} {\bibfnamefont {A.}~\bibnamefont
  {Zabalo}}, \bibinfo {author} {\bibfnamefont {M.~J.}\ \bibnamefont {Gullans}},
  \bibinfo {author} {\bibfnamefont {J.~H.}\ \bibnamefont {Wilson}}, \bibinfo
  {author} {\bibfnamefont {S.}~\bibnamefont {Gopalakrishnan}}, \bibinfo
  {author} {\bibfnamefont {D.~A.}\ \bibnamefont {Huse}},\ and\ \bibinfo
  {author} {\bibfnamefont {J.~H.}\ \bibnamefont {Pixley}},\ }\bibfield  {title}
  {\bibinfo {title} {{Critical properties of the measurement-induced transition
  in random quantum circuits}},\ }\href
  {https://doi.org/10.1103/PhysRevB.101.060301} {\bibfield  {journal} {\bibinfo
   {journal} {Phys. Rev. B}\ }\textbf {\bibinfo {volume} {101}},\ \bibinfo
  {pages} {060301} (\bibinfo {year} {2020})}\BibitemShut {NoStop}%
\bibitem [{\citenamefont {Fan}\ \emph {et~al.}(2021)\citenamefont {Fan},
  \citenamefont {Vijay}, \citenamefont {Vishwanath},\ and\ \citenamefont
  {You}}]{Fan21}%
  \BibitemOpen
  \bibfield  {author} {\bibinfo {author} {\bibfnamefont {R.}~\bibnamefont
  {Fan}}, \bibinfo {author} {\bibfnamefont {S.}~\bibnamefont {Vijay}}, \bibinfo
  {author} {\bibfnamefont {A.}~\bibnamefont {Vishwanath}},\ and\ \bibinfo
  {author} {\bibfnamefont {Y.-Z.}\ \bibnamefont {You}},\ }\bibfield  {title}
  {\bibinfo {title} {{Self-organized error correction in random unitary
  circuits with measurement}},\ }\href
  {https://doi.org/10.1103/PhysRevB.103.174309} {\bibfield  {journal} {\bibinfo
   {journal} {Phys. Rev. B}\ }\textbf {\bibinfo {volume} {103}},\ \bibinfo
  {pages} {174309} (\bibinfo {year} {2021})}\BibitemShut {NoStop}%
\bibitem [{\citenamefont {Lavasani}\ \emph
  {et~al.}(2021{\natexlab{a}})\citenamefont {Lavasani}, \citenamefont
  {Alavirad},\ and\ \citenamefont {Barkeshli}}]{Lavasani21a}%
  \BibitemOpen
  \bibfield  {author} {\bibinfo {author} {\bibfnamefont {A.}~\bibnamefont
  {Lavasani}}, \bibinfo {author} {\bibfnamefont {Y.}~\bibnamefont {Alavirad}},\
  and\ \bibinfo {author} {\bibfnamefont {M.}~\bibnamefont {Barkeshli}},\
  }\bibfield  {title} {\bibinfo {title} {{Measurement-induced topological
  entanglement transitions in symmetric random quantum circuits}},\ }\href
  {https://doi.org/10.1038/s41567-020-01112-z} {\bibfield  {journal} {\bibinfo
  {journal} {Nat. Phys.}\ }\textbf {\bibinfo {volume} {17}},\ \bibinfo {pages}
  {342} (\bibinfo {year} {2021}{\natexlab{a}})}\BibitemShut {NoStop}%
\bibitem [{\citenamefont {Lavasani}\ \emph
  {et~al.}(2021{\natexlab{b}})\citenamefont {Lavasani}, \citenamefont
  {Alavirad},\ and\ \citenamefont {Barkeshli}}]{Lavasani21b}%
  \BibitemOpen
  \bibfield  {author} {\bibinfo {author} {\bibfnamefont {A.}~\bibnamefont
  {Lavasani}}, \bibinfo {author} {\bibfnamefont {Y.}~\bibnamefont {Alavirad}},\
  and\ \bibinfo {author} {\bibfnamefont {M.}~\bibnamefont {Barkeshli}},\
  }\bibfield  {title} {\bibinfo {title} {{Topological Order and Criticality in
  $(2+1)\mathrm{D}$ Monitored Random Quantum Circuits}},\ }\href
  {https://doi.org/10.1103/PhysRevLett.127.235701} {\bibfield  {journal}
  {\bibinfo  {journal} {Phys. Rev. Lett.}\ }\textbf {\bibinfo {volume} {127}},\
  \bibinfo {pages} {235701} (\bibinfo {year} {2021}{\natexlab{b}})}\BibitemShut
  {NoStop}%
\bibitem [{\citenamefont {Li}\ \emph {et~al.}(2021)\citenamefont {Li},
  \citenamefont {Chen}, \citenamefont {Ludwig},\ and\ \citenamefont
  {Fisher}}]{Li21a}%
  \BibitemOpen
  \bibfield  {author} {\bibinfo {author} {\bibfnamefont {Y.}~\bibnamefont
  {Li}}, \bibinfo {author} {\bibfnamefont {X.}~\bibnamefont {Chen}}, \bibinfo
  {author} {\bibfnamefont {A.~W.~W.}\ \bibnamefont {Ludwig}},\ and\ \bibinfo
  {author} {\bibfnamefont {M.~P.~A.}\ \bibnamefont {Fisher}},\ }\bibfield
  {title} {\bibinfo {title} {{Conformal invariance and quantum nonlocality in
  critical hybrid circuits}},\ }\href
  {https://doi.org/10.1103/PhysRevB.104.104305} {\bibfield  {journal} {\bibinfo
   {journal} {Phys. Rev. B}\ }\textbf {\bibinfo {volume} {104}},\ \bibinfo
  {pages} {104305} (\bibinfo {year} {2021})}\BibitemShut {NoStop}%
\bibitem [{\citenamefont {Lu}\ and\ \citenamefont {Grover}(2021)}]{Lu21}%
  \BibitemOpen
  \bibfield  {author} {\bibinfo {author} {\bibfnamefont {T.-C.}\ \bibnamefont
  {Lu}}\ and\ \bibinfo {author} {\bibfnamefont {T.}~\bibnamefont {Grover}},\
  }\bibfield  {title} {\bibinfo {title} {{Spacetime duality between
  localization transitions and measurement-induced transitions}},\ }\href
  {https://doi.org/10.1103/PRXQuantum.2.040319} {\bibfield  {journal} {\bibinfo
   {journal} {PRX Quantum}\ }\textbf {\bibinfo {volume} {2}},\ \bibinfo {pages}
  {040319} (\bibinfo {year} {2021})}\BibitemShut {NoStop}%
\bibitem [{\citenamefont {Lunt}\ \emph {et~al.}(2021)\citenamefont {Lunt},
  \citenamefont {Szyniszewski},\ and\ \citenamefont {Pal}}]{Lunt21}%
  \BibitemOpen
  \bibfield  {author} {\bibinfo {author} {\bibfnamefont {O.}~\bibnamefont
  {Lunt}}, \bibinfo {author} {\bibfnamefont {M.}~\bibnamefont {Szyniszewski}},\
  and\ \bibinfo {author} {\bibfnamefont {A.}~\bibnamefont {Pal}},\ }\bibfield
  {title} {\bibinfo {title} {{Measurement-induced criticality and entanglement
  clusters: A study of one-dimensional and two-dimensional Clifford
  circuits}},\ }\href {https://doi.org/10.1103/PhysRevB.104.155111} {\bibfield
  {journal} {\bibinfo  {journal} {Phys. Rev. B}\ }\textbf {\bibinfo {volume}
  {104}},\ \bibinfo {pages} {155111} (\bibinfo {year} {2021})}\BibitemShut
  {NoStop}%
\bibitem [{\citenamefont {Sang}\ and\ \citenamefont {Hsieh}(2021)}]{Sang21}%
  \BibitemOpen
  \bibfield  {author} {\bibinfo {author} {\bibfnamefont {S.}~\bibnamefont
  {Sang}}\ and\ \bibinfo {author} {\bibfnamefont {T.~H.}\ \bibnamefont
  {Hsieh}},\ }\bibfield  {title} {\bibinfo {title} {{Measurement-protected
  quantum phases}},\ }\href {https://doi.org/10.1103/PhysRevResearch.3.023200}
  {\bibfield  {journal} {\bibinfo  {journal} {Phys. Rev. Research}\ }\textbf
  {\bibinfo {volume} {3}},\ \bibinfo {pages} {023200} (\bibinfo {year}
  {2021})}\BibitemShut {NoStop}%
\bibitem [{\citenamefont {C\^ot\'e}\ and\ \citenamefont
  {Kourtis}(2022)}]{Cote22}%
  \BibitemOpen
  \bibfield  {author} {\bibinfo {author} {\bibfnamefont {J.}~\bibnamefont
  {C\^ot\'e}}\ and\ \bibinfo {author} {\bibfnamefont {S.}~\bibnamefont
  {Kourtis}},\ }\bibfield  {title} {\bibinfo {title} {{Entanglement Phase
  Transition with Spin Glass Criticality}},\ }\href
  {https://doi.org/10.1103/PhysRevLett.128.240601} {\bibfield  {journal}
  {\bibinfo  {journal} {Phys. Rev. Lett.}\ }\textbf {\bibinfo {volume} {128}},\
  \bibinfo {pages} {240601} (\bibinfo {year} {2022})}\BibitemShut {NoStop}%
\bibitem [{\citenamefont {Sierant}\ and\ \citenamefont
  {Turkeshi}(2022)}]{Sierant22b}%
  \BibitemOpen
  \bibfield  {author} {\bibinfo {author} {\bibfnamefont {P.}~\bibnamefont
  {Sierant}}\ and\ \bibinfo {author} {\bibfnamefont {X.}~\bibnamefont
  {Turkeshi}},\ }\bibfield  {title} {\bibinfo {title} {{Universal Behavior
  beyond Multifractality of Wave Functions at Measurement-Induced Phase
  Transitions}},\ }\href {https://doi.org/10.1103/PhysRevLett.128.130605}
  {\bibfield  {journal} {\bibinfo  {journal} {Phys. Rev. Lett.}\ }\textbf
  {\bibinfo {volume} {128}},\ \bibinfo {pages} {130605} (\bibinfo {year}
  {2022})}\BibitemShut {NoStop}%
\bibitem [{\citenamefont {Zabalo}\ \emph {et~al.}(2022)\citenamefont {Zabalo},
  \citenamefont {Gullans}, \citenamefont {Wilson}, \citenamefont {Vasseur},
  \citenamefont {Ludwig}, \citenamefont {Gopalakrishnan}, \citenamefont
  {Huse},\ and\ \citenamefont {Pixley}}]{Zabalo22}%
  \BibitemOpen
  \bibfield  {author} {\bibinfo {author} {\bibfnamefont {A.}~\bibnamefont
  {Zabalo}}, \bibinfo {author} {\bibfnamefont {M.~J.}\ \bibnamefont {Gullans}},
  \bibinfo {author} {\bibfnamefont {J.~H.}\ \bibnamefont {Wilson}}, \bibinfo
  {author} {\bibfnamefont {R.}~\bibnamefont {Vasseur}}, \bibinfo {author}
  {\bibfnamefont {A.~W.~W.}\ \bibnamefont {Ludwig}}, \bibinfo {author}
  {\bibfnamefont {S.}~\bibnamefont {Gopalakrishnan}}, \bibinfo {author}
  {\bibfnamefont {D.~A.}\ \bibnamefont {Huse}},\ and\ \bibinfo {author}
  {\bibfnamefont {J.~H.}\ \bibnamefont {Pixley}},\ }\bibfield  {title}
  {\bibinfo {title} {{Operator Scaling Dimensions and Multifractality at
  Measurement-Induced Transitions}},\ }\href
  {https://doi.org/10.1103/PhysRevLett.128.050602} {\bibfield  {journal}
  {\bibinfo  {journal} {Phys. Rev. Lett.}\ }\textbf {\bibinfo {volume} {128}},\
  \bibinfo {pages} {050602} (\bibinfo {year} {2022})}\BibitemShut {NoStop}%
\bibitem [{\citenamefont {Lang}\ and\ \citenamefont
  {B\"uchler}(2020)}]{Lang20}%
  \BibitemOpen
  \bibfield  {author} {\bibinfo {author} {\bibfnamefont {N.}~\bibnamefont
  {Lang}}\ and\ \bibinfo {author} {\bibfnamefont {H.~P.}\ \bibnamefont
  {B\"uchler}},\ }\bibfield  {title} {\bibinfo {title} {{Entanglement
  transition in the projective transverse field Ising model}},\ }\href
  {https://doi.org/10.1103/PhysRevB.102.094204} {\bibfield  {journal} {\bibinfo
   {journal} {Phys. Rev. B}\ }\textbf {\bibinfo {volume} {102}},\ \bibinfo
  {pages} {094204} (\bibinfo {year} {2020})}\BibitemShut {NoStop}%
\bibitem [{\citenamefont {Ippoliti}\ \emph {et~al.}(2021)\citenamefont
  {Ippoliti}, \citenamefont {Gullans}, \citenamefont {Gopalakrishnan},
  \citenamefont {Huse},\ and\ \citenamefont {Khemani}}]{Ippoliti21}%
  \BibitemOpen
  \bibfield  {author} {\bibinfo {author} {\bibfnamefont {M.}~\bibnamefont
  {Ippoliti}}, \bibinfo {author} {\bibfnamefont {M.~J.}\ \bibnamefont
  {Gullans}}, \bibinfo {author} {\bibfnamefont {S.}~\bibnamefont
  {Gopalakrishnan}}, \bibinfo {author} {\bibfnamefont {D.~A.}\ \bibnamefont
  {Huse}},\ and\ \bibinfo {author} {\bibfnamefont {V.}~\bibnamefont
  {Khemani}},\ }\bibfield  {title} {\bibinfo {title} {{Entanglement Phase
  Transitions in Measurement-Only Dynamics}},\ }\href
  {https://doi.org/10.1103/PhysRevX.11.011030} {\bibfield  {journal} {\bibinfo
  {journal} {Phys. Rev. X}\ }\textbf {\bibinfo {volume} {11}},\ \bibinfo
  {pages} {011030} (\bibinfo {year} {2021})}\BibitemShut {NoStop}%
\bibitem [{\citenamefont {Klocke}\ and\ \citenamefont
  {Buchhold}(2022)}]{Klocke22}%
  \BibitemOpen
  \bibfield  {author} {\bibinfo {author} {\bibfnamefont {K.}~\bibnamefont
  {Klocke}}\ and\ \bibinfo {author} {\bibfnamefont {M.}~\bibnamefont
  {Buchhold}},\ }\bibfield  {title} {\bibinfo {title} {{Topological order and
  entanglement dynamics in the measurement-only XZZX quantum code}},\ }\href
  {https://doi.org/10.1103/PhysRevB.106.104307} {\bibfield  {journal} {\bibinfo
   {journal} {Phys. Rev. B}\ }\textbf {\bibinfo {volume} {106}},\ \bibinfo
  {pages} {104307} (\bibinfo {year} {2022})}\BibitemShut {NoStop}%
\bibitem [{\citenamefont {Lavasani}\ \emph {et~al.}(2022)\citenamefont
  {Lavasani}, \citenamefont {Luo},\ and\ \citenamefont {Vijay}}]{Lavasani22}%
  \BibitemOpen
  \bibfield  {author} {\bibinfo {author} {\bibfnamefont {A.}~\bibnamefont
  {Lavasani}}, \bibinfo {author} {\bibfnamefont {Z.-X.}\ \bibnamefont {Luo}},\
  and\ \bibinfo {author} {\bibfnamefont {S.}~\bibnamefont {Vijay}},\ }\bibfield
   {title} {\bibinfo {title} {{Monitored Quantum Dynamics and the Kitaev Spin
  Liquid}},\ }\Eprint {https://arxiv.org/abs/2207.02877} {arXiv:2207.02877}
  (\bibinfo {year} {2022})\BibitemShut {NoStop}%
\bibitem [{\citenamefont {Sriram}\ \emph {et~al.}(2022)\citenamefont {Sriram},
  \citenamefont {Rakovszky}, \citenamefont {Khemani},\ and\ \citenamefont
  {Ippoliti}}]{Sriam22}%
  \BibitemOpen
  \bibfield  {author} {\bibinfo {author} {\bibfnamefont {A.}~\bibnamefont
  {Sriram}}, \bibinfo {author} {\bibfnamefont {T.}~\bibnamefont {Rakovszky}},
  \bibinfo {author} {\bibfnamefont {V.}~\bibnamefont {Khemani}},\ and\ \bibinfo
  {author} {\bibfnamefont {M.}~\bibnamefont {Ippoliti}},\ }\bibfield  {title}
  {\bibinfo {title} {{Topology, criticality, and dynamically generated qubits
  in a stochastic measurement-only Kitaev model}},\ }\Eprint
  {https://arxiv.org/abs/2207.07096} {arXiv:2207.07096}  (\bibinfo {year}
  {2022})\BibitemShut {NoStop}%
\bibitem [{\citenamefont {Alberton}\ \emph {et~al.}(2021)\citenamefont
  {Alberton}, \citenamefont {Buchhold},\ and\ \citenamefont
  {Diehl}}]{Alberton21}%
  \BibitemOpen
  \bibfield  {author} {\bibinfo {author} {\bibfnamefont {O.}~\bibnamefont
  {Alberton}}, \bibinfo {author} {\bibfnamefont {M.}~\bibnamefont {Buchhold}},\
  and\ \bibinfo {author} {\bibfnamefont {S.}~\bibnamefont {Diehl}},\ }\bibfield
   {title} {\bibinfo {title} {{Entanglement Transition in a Monitored
  Free-Fermion Chain: From Extended Criticality to Area Law}},\ }\href
  {https://doi.org/10.1103/PhysRevLett.126.170602} {\bibfield  {journal}
  {\bibinfo  {journal} {Phys. Rev. Lett.}\ }\textbf {\bibinfo {volume} {126}},\
  \bibinfo {pages} {170602} (\bibinfo {year} {2021})}\BibitemShut {NoStop}%
\bibitem [{\citenamefont {Buchhold}\ \emph {et~al.}(2021)\citenamefont
  {Buchhold}, \citenamefont {Minoguchi}, \citenamefont {Altland},\ and\
  \citenamefont {Diehl}}]{Buchhold21}%
  \BibitemOpen
  \bibfield  {author} {\bibinfo {author} {\bibfnamefont {M.}~\bibnamefont
  {Buchhold}}, \bibinfo {author} {\bibfnamefont {Y.}~\bibnamefont {Minoguchi}},
  \bibinfo {author} {\bibfnamefont {A.}~\bibnamefont {Altland}},\ and\ \bibinfo
  {author} {\bibfnamefont {S.}~\bibnamefont {Diehl}},\ }\bibfield  {title}
  {\bibinfo {title} {{Effective Theory for the Measurement-Induced Phase
  Transition of Dirac Fermions}},\ }\href
  {https://doi.org/10.1103/PhysRevX.11.041004} {\bibfield  {journal} {\bibinfo
  {journal} {Phys. Rev. X}\ }\textbf {\bibinfo {volume} {11}},\ \bibinfo
  {pages} {041004} (\bibinfo {year} {2021})}\BibitemShut {NoStop}%
\bibitem [{\citenamefont {Kells}\ \emph {et~al.}(2021)\citenamefont {Kells},
  \citenamefont {Meidan},\ and\ \citenamefont {Romito}}]{Kells21}%
  \BibitemOpen
  \bibfield  {author} {\bibinfo {author} {\bibfnamefont {G.}~\bibnamefont
  {Kells}}, \bibinfo {author} {\bibfnamefont {D.}~\bibnamefont {Meidan}},\ and\
  \bibinfo {author} {\bibfnamefont {A.}~\bibnamefont {Romito}},\ }\bibfield
  {title} {\bibinfo {title} {{Topological transitions with continuously
  monitored free fermions}},\ }\Eprint {https://arxiv.org/abs/2112.09787}
  {arXiv:2112.09787}  (\bibinfo {year} {2021})\BibitemShut {NoStop}%
\bibitem [{\citenamefont {Turkeshi}\ \emph {et~al.}(2021)\citenamefont
  {Turkeshi}, \citenamefont {Biella}, \citenamefont {Fazio}, \citenamefont
  {Dalmonte},\ and\ \citenamefont {Schir\'o}}]{Turkeshi21}%
  \BibitemOpen
  \bibfield  {author} {\bibinfo {author} {\bibfnamefont {X.}~\bibnamefont
  {Turkeshi}}, \bibinfo {author} {\bibfnamefont {A.}~\bibnamefont {Biella}},
  \bibinfo {author} {\bibfnamefont {R.}~\bibnamefont {Fazio}}, \bibinfo
  {author} {\bibfnamefont {M.}~\bibnamefont {Dalmonte}},\ and\ \bibinfo
  {author} {\bibfnamefont {M.}~\bibnamefont {Schir\'o}},\ }\bibfield  {title}
  {\bibinfo {title} {{Measurement-induced entanglement transitions in the
  quantum Ising chain: From infinite to zero clicks}},\ }\href
  {https://doi.org/10.1103/PhysRevB.103.224210} {\bibfield  {journal} {\bibinfo
   {journal} {Phys. Rev. B}\ }\textbf {\bibinfo {volume} {103}},\ \bibinfo
  {pages} {224210} (\bibinfo {year} {2021})}\BibitemShut {NoStop}%
\bibitem [{\citenamefont {Piccitto}\ \emph {et~al.}(2022)\citenamefont
  {Piccitto}, \citenamefont {Russomanno},\ and\ \citenamefont
  {Rossini}}]{Piccitto22}%
  \BibitemOpen
  \bibfield  {author} {\bibinfo {author} {\bibfnamefont {G.}~\bibnamefont
  {Piccitto}}, \bibinfo {author} {\bibfnamefont {A.}~\bibnamefont
  {Russomanno}},\ and\ \bibinfo {author} {\bibfnamefont {D.}~\bibnamefont
  {Rossini}},\ }\bibfield  {title} {\bibinfo {title} {{Entanglement transitions
  in the quantum Ising chain: A comparison between different unravelings of the
  same Lindbladian}},\ }\href {https://doi.org/10.1103/PhysRevB.105.064305}
  {\bibfield  {journal} {\bibinfo  {journal} {Phys. Rev. B}\ }\textbf {\bibinfo
  {volume} {105}},\ \bibinfo {pages} {064305} (\bibinfo {year}
  {2022})}\BibitemShut {NoStop}%
\bibitem [{\citenamefont {Turkeshi}\ \emph {et~al.}(2022)\citenamefont
  {Turkeshi}, \citenamefont {Dalmonte}, \citenamefont {Fazio},\ and\
  \citenamefont {Schir\`o}}]{Turkeshi22}%
  \BibitemOpen
  \bibfield  {author} {\bibinfo {author} {\bibfnamefont {X.}~\bibnamefont
  {Turkeshi}}, \bibinfo {author} {\bibfnamefont {M.}~\bibnamefont {Dalmonte}},
  \bibinfo {author} {\bibfnamefont {R.}~\bibnamefont {Fazio}},\ and\ \bibinfo
  {author} {\bibfnamefont {M.}~\bibnamefont {Schir\`o}},\ }\bibfield  {title}
  {\bibinfo {title} {{Entanglement transitions from stochastic resetting of
  non-Hermitian quasiparticles}},\ }\href
  {https://doi.org/10.1103/PhysRevB.105.L241114} {\bibfield  {journal}
  {\bibinfo  {journal} {Phys. Rev. B}\ }\textbf {\bibinfo {volume} {105}},\
  \bibinfo {pages} {L241114} (\bibinfo {year} {2022})}\BibitemShut {NoStop}%
\bibitem [{\citenamefont {Fuji}\ and\ \citenamefont {Ashida}(2020)}]{Fuji20}%
  \BibitemOpen
  \bibfield  {author} {\bibinfo {author} {\bibfnamefont {Y.}~\bibnamefont
  {Fuji}}\ and\ \bibinfo {author} {\bibfnamefont {Y.}~\bibnamefont {Ashida}},\
  }\bibfield  {title} {\bibinfo {title} {{Measurement-induced quantum
  criticality under continuous monitoring}},\ }\href
  {https://doi.org/10.1103/PhysRevB.102.054302} {\bibfield  {journal} {\bibinfo
   {journal} {Phys. Rev. B}\ }\textbf {\bibinfo {volume} {102}},\ \bibinfo
  {pages} {054302} (\bibinfo {year} {2020})}\BibitemShut {NoStop}%
\bibitem [{\citenamefont {Szyniszewski}\ \emph {et~al.}(2020)\citenamefont
  {Szyniszewski}, \citenamefont {Romito},\ and\ \citenamefont
  {Schomerus}}]{Szyniszewski20}%
  \BibitemOpen
  \bibfield  {author} {\bibinfo {author} {\bibfnamefont {M.}~\bibnamefont
  {Szyniszewski}}, \bibinfo {author} {\bibfnamefont {A.}~\bibnamefont
  {Romito}},\ and\ \bibinfo {author} {\bibfnamefont {H.}~\bibnamefont
  {Schomerus}},\ }\bibfield  {title} {\bibinfo {title} {{Universality of
  Entanglement Transitions from Stroboscopic to Continuous Measurements}},\
  }\href {https://doi.org/10.1103/PhysRevLett.125.210602} {\bibfield  {journal}
  {\bibinfo  {journal} {Phys. Rev. Lett.}\ }\textbf {\bibinfo {volume} {125}},\
  \bibinfo {pages} {210602} (\bibinfo {year} {2020})}\BibitemShut {NoStop}%
\bibitem [{\citenamefont {Jian}\ \emph {et~al.}(2021)\citenamefont {Jian},
  \citenamefont {Liu}, \citenamefont {Chen}, \citenamefont {Swingle},\ and\
  \citenamefont {Zhang}}]{SKJian21}%
  \BibitemOpen
  \bibfield  {author} {\bibinfo {author} {\bibfnamefont {S.-K.}\ \bibnamefont
  {Jian}}, \bibinfo {author} {\bibfnamefont {C.}~\bibnamefont {Liu}}, \bibinfo
  {author} {\bibfnamefont {X.}~\bibnamefont {Chen}}, \bibinfo {author}
  {\bibfnamefont {B.}~\bibnamefont {Swingle}},\ and\ \bibinfo {author}
  {\bibfnamefont {P.}~\bibnamefont {Zhang}},\ }\bibfield  {title} {\bibinfo
  {title} {{Measurement-Induced Phase Transition in the Monitored
  Sachdev-Ye-Kitaev Model}},\ }\href
  {https://doi.org/10.1103/PhysRevLett.127.140601} {\bibfield  {journal}
  {\bibinfo  {journal} {Phys. Rev. Lett.}\ }\textbf {\bibinfo {volume} {127}},\
  \bibinfo {pages} {140601} (\bibinfo {year} {2021})}\BibitemShut {NoStop}%
\bibitem [{\citenamefont {Van~Regemortel}\ \emph {et~al.}(2021)\citenamefont
  {Van~Regemortel}, \citenamefont {Cian}, \citenamefont {Seif}, \citenamefont
  {Dehghani},\ and\ \citenamefont {Hafezi}}]{VanRegemortel21}%
  \BibitemOpen
  \bibfield  {author} {\bibinfo {author} {\bibfnamefont {M.}~\bibnamefont
  {Van~Regemortel}}, \bibinfo {author} {\bibfnamefont {Z.-P.}\ \bibnamefont
  {Cian}}, \bibinfo {author} {\bibfnamefont {A.}~\bibnamefont {Seif}}, \bibinfo
  {author} {\bibfnamefont {H.}~\bibnamefont {Dehghani}},\ and\ \bibinfo
  {author} {\bibfnamefont {M.}~\bibnamefont {Hafezi}},\ }\bibfield  {title}
  {\bibinfo {title} {{Entanglement Entropy Scaling Transition under Competing
  Monitoring Protocols}},\ }\href
  {https://doi.org/10.1103/PhysRevLett.126.123604} {\bibfield  {journal}
  {\bibinfo  {journal} {Phys. Rev. Lett.}\ }\textbf {\bibinfo {volume} {126}},\
  \bibinfo {pages} {123604} (\bibinfo {year} {2021})}\BibitemShut {NoStop}%
\bibitem [{\citenamefont {Boorman}\ \emph {et~al.}(2022)\citenamefont
  {Boorman}, \citenamefont {Szyniszewski}, \citenamefont {Schomerus},\ and\
  \citenamefont {Romito}}]{Boorman22}%
  \BibitemOpen
  \bibfield  {author} {\bibinfo {author} {\bibfnamefont {T.}~\bibnamefont
  {Boorman}}, \bibinfo {author} {\bibfnamefont {M.}~\bibnamefont
  {Szyniszewski}}, \bibinfo {author} {\bibfnamefont {H.}~\bibnamefont
  {Schomerus}},\ and\ \bibinfo {author} {\bibfnamefont {A.}~\bibnamefont
  {Romito}},\ }\bibfield  {title} {\bibinfo {title} {{Diagnostics of
  entanglement dynamics in noisy and disordered spin chains via the
  measurement-induced steady-state entanglement transition}},\ }\href
  {https://doi.org/10.1103/PhysRevB.105.144202} {\bibfield  {journal} {\bibinfo
   {journal} {Phys. Rev. B}\ }\textbf {\bibinfo {volume} {105}},\ \bibinfo
  {pages} {144202} (\bibinfo {year} {2022})}\BibitemShut {NoStop}%
\bibitem [{\citenamefont {Doggen}\ \emph {et~al.}(2022)\citenamefont {Doggen},
  \citenamefont {Gefen}, \citenamefont {Gornyi}, \citenamefont {Mirlin},\ and\
  \citenamefont {Polyakov}}]{Doggen22}%
  \BibitemOpen
  \bibfield  {author} {\bibinfo {author} {\bibfnamefont {E.~V.~H.}\
  \bibnamefont {Doggen}}, \bibinfo {author} {\bibfnamefont {Y.}~\bibnamefont
  {Gefen}}, \bibinfo {author} {\bibfnamefont {I.~V.}\ \bibnamefont {Gornyi}},
  \bibinfo {author} {\bibfnamefont {A.~D.}\ \bibnamefont {Mirlin}},\ and\
  \bibinfo {author} {\bibfnamefont {D.~G.}\ \bibnamefont {Polyakov}},\
  }\bibfield  {title} {\bibinfo {title} {{Generalized quantum measurements with
  matrix product states: Entanglement phase transition and clusterization}},\
  }\href {https://doi.org/10.1103/PhysRevResearch.4.023146} {\bibfield
  {journal} {\bibinfo  {journal} {Phys. Rev. Research}\ }\textbf {\bibinfo
  {volume} {4}},\ \bibinfo {pages} {023146} (\bibinfo {year}
  {2022})}\BibitemShut {NoStop}%
\bibitem [{\citenamefont {Sahu}\ \emph {et~al.}(2021)\citenamefont {Sahu},
  \citenamefont {Jian}, \citenamefont {Bentsen},\ and\ \citenamefont
  {Swingle}}]{Sahu21}%
  \BibitemOpen
  \bibfield  {author} {\bibinfo {author} {\bibfnamefont {S.}~\bibnamefont
  {Sahu}}, \bibinfo {author} {\bibfnamefont {S.-K.}\ \bibnamefont {Jian}},
  \bibinfo {author} {\bibfnamefont {G.}~\bibnamefont {Bentsen}},\ and\ \bibinfo
  {author} {\bibfnamefont {B.}~\bibnamefont {Swingle}},\ }\bibfield  {title}
  {\bibinfo {title} {{Entanglement Phases in large-N hybrid Brownian circuits
  with long-range couplings}},\ }\Eprint {https://arxiv.org/abs/2109.00013}
  {arXiv:2109.00013}  (\bibinfo {year} {2021})\BibitemShut {NoStop}%
\bibitem [{\citenamefont {Block}\ \emph {et~al.}(2022)\citenamefont {Block},
  \citenamefont {Bao}, \citenamefont {Choi}, \citenamefont {Altman},\ and\
  \citenamefont {Yao}}]{Block22}%
  \BibitemOpen
  \bibfield  {author} {\bibinfo {author} {\bibfnamefont {M.}~\bibnamefont
  {Block}}, \bibinfo {author} {\bibfnamefont {Y.}~\bibnamefont {Bao}}, \bibinfo
  {author} {\bibfnamefont {S.}~\bibnamefont {Choi}}, \bibinfo {author}
  {\bibfnamefont {E.}~\bibnamefont {Altman}},\ and\ \bibinfo {author}
  {\bibfnamefont {N.~Y.}\ \bibnamefont {Yao}},\ }\bibfield  {title} {\bibinfo
  {title} {{Measurement-Induced Transition in Long-Range Interacting Quantum
  Circuits}},\ }\href {https://doi.org/10.1103/PhysRevLett.128.010604}
  {\bibfield  {journal} {\bibinfo  {journal} {Phys. Rev. Lett.}\ }\textbf
  {\bibinfo {volume} {128}},\ \bibinfo {pages} {010604} (\bibinfo {year}
  {2022})}\BibitemShut {NoStop}%
\bibitem [{\citenamefont {Hashizume}\ \emph {et~al.}(2022)\citenamefont
  {Hashizume}, \citenamefont {Bentsen},\ and\ \citenamefont
  {Daley}}]{Hashizume22}%
  \BibitemOpen
  \bibfield  {author} {\bibinfo {author} {\bibfnamefont {T.}~\bibnamefont
  {Hashizume}}, \bibinfo {author} {\bibfnamefont {G.}~\bibnamefont {Bentsen}},\
  and\ \bibinfo {author} {\bibfnamefont {A.~J.}\ \bibnamefont {Daley}},\
  }\bibfield  {title} {\bibinfo {title} {{Measurement-induced phase transitions
  in sparse nonlocal scramblers}},\ }\href
  {https://doi.org/10.1103/PhysRevResearch.4.013174} {\bibfield  {journal}
  {\bibinfo  {journal} {Phys. Rev. Research}\ }\textbf {\bibinfo {volume}
  {4}},\ \bibinfo {pages} {013174} (\bibinfo {year} {2022})}\BibitemShut
  {NoStop}%
\bibitem [{\citenamefont {Minato}\ \emph {et~al.}(2022)\citenamefont {Minato},
  \citenamefont {Sugimoto}, \citenamefont {Kuwahara},\ and\ \citenamefont
  {Saito}}]{Minato22}%
  \BibitemOpen
  \bibfield  {author} {\bibinfo {author} {\bibfnamefont {T.}~\bibnamefont
  {Minato}}, \bibinfo {author} {\bibfnamefont {K.}~\bibnamefont {Sugimoto}},
  \bibinfo {author} {\bibfnamefont {T.}~\bibnamefont {Kuwahara}},\ and\
  \bibinfo {author} {\bibfnamefont {K.}~\bibnamefont {Saito}},\ }\bibfield
  {title} {\bibinfo {title} {{Fate of Measurement-Induced Phase Transition in
  Long-Range Interactions}},\ }\href
  {https://doi.org/10.1103/PhysRevLett.128.010603} {\bibfield  {journal}
  {\bibinfo  {journal} {Phys. Rev. Lett.}\ }\textbf {\bibinfo {volume} {128}},\
  \bibinfo {pages} {010603} (\bibinfo {year} {2022})}\BibitemShut {NoStop}%
\bibitem [{\citenamefont {M\"uller}\ \emph {et~al.}(2022)\citenamefont
  {M\"uller}, \citenamefont {Diehl},\ and\ \citenamefont
  {Buchhold}}]{Muller22}%
  \BibitemOpen
  \bibfield  {author} {\bibinfo {author} {\bibfnamefont {T.}~\bibnamefont
  {M\"uller}}, \bibinfo {author} {\bibfnamefont {S.}~\bibnamefont {Diehl}},\
  and\ \bibinfo {author} {\bibfnamefont {M.}~\bibnamefont {Buchhold}},\
  }\bibfield  {title} {\bibinfo {title} {{Measurement-Induced Dark State Phase
  Transitions in Long-Ranged Fermion Systems}},\ }\href
  {https://doi.org/10.1103/PhysRevLett.128.010605} {\bibfield  {journal}
  {\bibinfo  {journal} {Phys. Rev. Lett.}\ }\textbf {\bibinfo {volume} {128}},\
  \bibinfo {pages} {010605} (\bibinfo {year} {2022})}\BibitemShut {NoStop}%
\bibitem [{\citenamefont {Sierant}\ \emph {et~al.}(2022)\citenamefont
  {Sierant}, \citenamefont {Chiriac{\`{o}}}, \citenamefont {Surace},
  \citenamefont {Sharma}, \citenamefont {Turkeshi}, \citenamefont {Dalmonte},
  \citenamefont {Fazio},\ and\ \citenamefont {Pagano}}]{Sierant22a}%
  \BibitemOpen
  \bibfield  {author} {\bibinfo {author} {\bibfnamefont {P.}~\bibnamefont
  {Sierant}}, \bibinfo {author} {\bibfnamefont {G.}~\bibnamefont
  {Chiriac{\`{o}}}}, \bibinfo {author} {\bibfnamefont {F.~M.}\ \bibnamefont
  {Surace}}, \bibinfo {author} {\bibfnamefont {S.}~\bibnamefont {Sharma}},
  \bibinfo {author} {\bibfnamefont {X.}~\bibnamefont {Turkeshi}}, \bibinfo
  {author} {\bibfnamefont {M.}~\bibnamefont {Dalmonte}}, \bibinfo {author}
  {\bibfnamefont {R.}~\bibnamefont {Fazio}},\ and\ \bibinfo {author}
  {\bibfnamefont {G.}~\bibnamefont {Pagano}},\ }\bibfield  {title} {\bibinfo
  {title} {Dissipative {F}loquet {D}ynamics: from {S}teady {S}tate to
  {M}easurement {I}nduced {C}riticality in {T}rapped-ion {C}hains},\ }\href
  {https://doi.org/10.22331/q-2022-02-02-638} {\bibfield  {journal} {\bibinfo
  {journal} {{Quantum}}\ }\textbf {\bibinfo {volume} {6}},\ \bibinfo {pages}
  {638} (\bibinfo {year} {2022})}\BibitemShut {NoStop}%
\bibitem [{\citenamefont {Sharma}\ \emph {et~al.}(2022)\citenamefont {Sharma},
  \citenamefont {Turkeshi}, \citenamefont {Fazio},\ and\ \citenamefont
  {Dalmonte}}]{Sharma22}%
  \BibitemOpen
  \bibfield  {author} {\bibinfo {author} {\bibfnamefont {S.}~\bibnamefont
  {Sharma}}, \bibinfo {author} {\bibfnamefont {X.}~\bibnamefont {Turkeshi}},
  \bibinfo {author} {\bibfnamefont {R.}~\bibnamefont {Fazio}},\ and\ \bibinfo
  {author} {\bibfnamefont {M.}~\bibnamefont {Dalmonte}},\ }\bibfield  {title}
  {\bibinfo {title} {{Measurement-induced criticality in extended and
  long-range unitary circuits}},\ }\href
  {https://doi.org/10.21468/SciPostPhysCore.5.2.023} {\bibfield  {journal}
  {\bibinfo  {journal} {SciPost Phys. Core}\ }\textbf {\bibinfo {volume} {5}},\
  \bibinfo {pages} {023} (\bibinfo {year} {2022})}\BibitemShut {NoStop}%
\bibitem [{\citenamefont {Bao}\ \emph {et~al.}(2020)\citenamefont {Bao},
  \citenamefont {Choi},\ and\ \citenamefont {Altman}}]{Bao20}%
  \BibitemOpen
  \bibfield  {author} {\bibinfo {author} {\bibfnamefont {Y.}~\bibnamefont
  {Bao}}, \bibinfo {author} {\bibfnamefont {S.}~\bibnamefont {Choi}},\ and\
  \bibinfo {author} {\bibfnamefont {E.}~\bibnamefont {Altman}},\ }\bibfield
  {title} {\bibinfo {title} {{Theory of the phase transition in random unitary
  circuits with measurements}},\ }\href
  {https://doi.org/10.1103/PhysRevB.101.104301} {\bibfield  {journal} {\bibinfo
   {journal} {Phys. Rev. B}\ }\textbf {\bibinfo {volume} {101}},\ \bibinfo
  {pages} {104301} (\bibinfo {year} {2020})}\BibitemShut {NoStop}%
\bibitem [{\citenamefont {Jian}\ \emph {et~al.}(2020)\citenamefont {Jian},
  \citenamefont {You}, \citenamefont {Vasseur},\ and\ \citenamefont
  {Ludwig}}]{Jian20a}%
  \BibitemOpen
  \bibfield  {author} {\bibinfo {author} {\bibfnamefont {C.-M.}\ \bibnamefont
  {Jian}}, \bibinfo {author} {\bibfnamefont {Y.-Z.}\ \bibnamefont {You}},
  \bibinfo {author} {\bibfnamefont {R.}~\bibnamefont {Vasseur}},\ and\ \bibinfo
  {author} {\bibfnamefont {A.~W.~W.}\ \bibnamefont {Ludwig}},\ }\bibfield
  {title} {\bibinfo {title} {{Measurement-induced criticality in random quantum
  circuits}},\ }\href {https://doi.org/10.1103/PhysRevB.101.104302} {\bibfield
  {journal} {\bibinfo  {journal} {Phys. Rev. B}\ }\textbf {\bibinfo {volume}
  {101}},\ \bibinfo {pages} {104302} (\bibinfo {year} {2020})}\BibitemShut
  {NoStop}%
\bibitem [{\citenamefont {Nahum}\ and\ \citenamefont
  {Skinner}(2020)}]{Nahum20}%
  \BibitemOpen
  \bibfield  {author} {\bibinfo {author} {\bibfnamefont {A.}~\bibnamefont
  {Nahum}}\ and\ \bibinfo {author} {\bibfnamefont {B.}~\bibnamefont
  {Skinner}},\ }\bibfield  {title} {\bibinfo {title} {{Entanglement and
  dynamics of diffusion-annihilation processes with Majorana defects}},\ }\href
  {https://doi.org/10.1103/PhysRevResearch.2.023288} {\bibfield  {journal}
  {\bibinfo  {journal} {Phys. Rev. Research}\ }\textbf {\bibinfo {volume}
  {2}},\ \bibinfo {pages} {023288} (\bibinfo {year} {2020})}\BibitemShut
  {NoStop}%
\bibitem [{\citenamefont {Li}\ and\ \citenamefont {Fisher}(2021)}]{Li21b}%
  \BibitemOpen
  \bibfield  {author} {\bibinfo {author} {\bibfnamefont {Y.}~\bibnamefont
  {Li}}\ and\ \bibinfo {author} {\bibfnamefont {M.~P.~A.}\ \bibnamefont
  {Fisher}},\ }\bibfield  {title} {\bibinfo {title} {{Robust decoding in
  monitored dynamics of open quantum systems with $\mathbb{Z}_2$ symmetry}},\
  }\Eprint {https://arxiv.org/abs/2108.04274} {arXiv:2108.04274}  (\bibinfo
  {year} {2021})\BibitemShut {NoStop}%
\bibitem [{\citenamefont {Han}\ and\ \citenamefont {Chen}(2022)}]{Han22}%
  \BibitemOpen
  \bibfield  {author} {\bibinfo {author} {\bibfnamefont {Y.}~\bibnamefont
  {Han}}\ and\ \bibinfo {author} {\bibfnamefont {X.}~\bibnamefont {Chen}},\
  }\bibfield  {title} {\bibinfo {title} {{Measurement-induced criticality in
  $\mathbb{Z}_{2}$-symmetric quantum automaton circuits}},\ }\href
  {https://doi.org/10.1103/PhysRevB.105.064306} {\bibfield  {journal} {\bibinfo
   {journal} {Phys. Rev. B}\ }\textbf {\bibinfo {volume} {105}},\ \bibinfo
  {pages} {064306} (\bibinfo {year} {2022})}\BibitemShut {NoStop}%
\bibitem [{\citenamefont {Tasaki}(2022)}]{Tasaki22}%
  \BibitemOpen
  \bibfield  {author} {\bibinfo {author} {\bibfnamefont {H.}~\bibnamefont
  {Tasaki}},\ }\bibfield  {title} {\bibinfo {title} {{The Lieb-Schultz-Mattis
  Theorem: A Topological Point of View}},\ }in\ \href
  {https://doi.org/10.4171/90} {\emph {\bibinfo {booktitle} {The Physics and
  Mathematics of Elliott Lieb}}},\ Vol.~\bibinfo {volume} {2},\ \bibinfo
  {editor} {edited by\ \bibinfo {editor} {\bibfnamefont {R.~L.}\ \bibnamefont
  {Frank}}, \bibinfo {editor} {\bibfnamefont {A.}~\bibnamefont {Laptev}},
  \bibinfo {editor} {\bibfnamefont {M.}~\bibnamefont {Lewin}},\ and\ \bibinfo
  {editor} {\bibfnamefont {R.}~\bibnamefont {Seiringer.}}}\ (\bibinfo
  {publisher} {European Mathematical Society Press},\ \bibinfo {year} {2022})\
  pp.\ \bibinfo {pages} {405--446}\BibitemShut {NoStop}%
\bibitem [{\citenamefont {Bao}\ \emph {et~al.}(2021)\citenamefont {Bao},
  \citenamefont {Choi},\ and\ \citenamefont {Altman}}]{Bao21}%
  \BibitemOpen
  \bibfield  {author} {\bibinfo {author} {\bibfnamefont {Y.}~\bibnamefont
  {Bao}}, \bibinfo {author} {\bibfnamefont {S.}~\bibnamefont {Choi}},\ and\
  \bibinfo {author} {\bibfnamefont {E.}~\bibnamefont {Altman}},\ }\bibfield
  {title} {\bibinfo {title} {{Symmetry enriched phases of quantum circuits}},\
  }\href {https://doi.org/10.1016/j.aop.2021.168618} {\bibfield  {journal}
  {\bibinfo  {journal} {Ann. Phys.}\ }\textbf {\bibinfo {volume} {435}},\
  \bibinfo {pages} {168618} (\bibinfo {year} {2021})}\BibitemShut {NoStop}%
\bibitem [{\citenamefont {Agrawal}\ \emph {et~al.}(2022)\citenamefont
  {Agrawal}, \citenamefont {Zabalo}, \citenamefont {Chen}, \citenamefont
  {Wilson}, \citenamefont {Potter}, \citenamefont {Pixley}, \citenamefont
  {Gopalakrishnan},\ and\ \citenamefont {Vasseur}}]{Agrawal22}%
  \BibitemOpen
  \bibfield  {author} {\bibinfo {author} {\bibfnamefont {U.}~\bibnamefont
  {Agrawal}}, \bibinfo {author} {\bibfnamefont {A.}~\bibnamefont {Zabalo}},
  \bibinfo {author} {\bibfnamefont {K.}~\bibnamefont {Chen}}, \bibinfo {author}
  {\bibfnamefont {J.~H.}\ \bibnamefont {Wilson}}, \bibinfo {author}
  {\bibfnamefont {A.~C.}\ \bibnamefont {Potter}}, \bibinfo {author}
  {\bibfnamefont {J.~H.}\ \bibnamefont {Pixley}}, \bibinfo {author}
  {\bibfnamefont {S.}~\bibnamefont {Gopalakrishnan}},\ and\ \bibinfo {author}
  {\bibfnamefont {R.}~\bibnamefont {Vasseur}},\ }\bibfield  {title} {\bibinfo
  {title} {{Entanglement and Charge-Sharpening Transitions in U(1) Symmetric
  Monitored Quantum Circuits}},\ }\href
  {https://doi.org/10.1103/PhysRevX.12.041002} {\bibfield  {journal} {\bibinfo
  {journal} {Phys. Rev. X}\ }\textbf {\bibinfo {volume} {12}},\ \bibinfo
  {pages} {041002} (\bibinfo {year} {2022})}\BibitemShut {NoStop}%
\bibitem [{\citenamefont {Barratt}\ \emph
  {et~al.}(2022{\natexlab{a}})\citenamefont {Barratt}, \citenamefont {Agrawal},
  \citenamefont {Gopalakrishnan}, \citenamefont {Huse}, \citenamefont
  {Vasseur},\ and\ \citenamefont {Potter}}]{Barratt22a}%
  \BibitemOpen
  \bibfield  {author} {\bibinfo {author} {\bibfnamefont {F.}~\bibnamefont
  {Barratt}}, \bibinfo {author} {\bibfnamefont {U.}~\bibnamefont {Agrawal}},
  \bibinfo {author} {\bibfnamefont {S.}~\bibnamefont {Gopalakrishnan}},
  \bibinfo {author} {\bibfnamefont {D.~A.}\ \bibnamefont {Huse}}, \bibinfo
  {author} {\bibfnamefont {R.}~\bibnamefont {Vasseur}},\ and\ \bibinfo {author}
  {\bibfnamefont {A.~C.}\ \bibnamefont {Potter}},\ }\bibfield  {title}
  {\bibinfo {title} {{Field Theory of Charge Sharpening in Symmetric Monitored
  Quantum Circuits}},\ }\href {https://doi.org/10.1103/PhysRevLett.129.120604}
  {\bibfield  {journal} {\bibinfo  {journal} {Phys. Rev. Lett.}\ }\textbf
  {\bibinfo {volume} {129}},\ \bibinfo {pages} {120604} (\bibinfo {year}
  {2022}{\natexlab{a}})}\BibitemShut {NoStop}%
\bibitem [{\citenamefont {Barratt}\ \emph
  {et~al.}(2022{\natexlab{b}})\citenamefont {Barratt}, \citenamefont {Agrawal},
  \citenamefont {Potter}, \citenamefont {Gopalakrishnan},\ and\ \citenamefont
  {Vasseur}}]{Barratt22b}%
  \BibitemOpen
  \bibfield  {author} {\bibinfo {author} {\bibfnamefont {F.}~\bibnamefont
  {Barratt}}, \bibinfo {author} {\bibfnamefont {U.}~\bibnamefont {Agrawal}},
  \bibinfo {author} {\bibfnamefont {A.~C.}\ \bibnamefont {Potter}}, \bibinfo
  {author} {\bibfnamefont {S.}~\bibnamefont {Gopalakrishnan}},\ and\ \bibinfo
  {author} {\bibfnamefont {R.}~\bibnamefont {Vasseur}},\ }\bibfield  {title}
  {\bibinfo {title} {{Transitions in the learnability of global charges from
  local measurements}},\ }\Eprint {https://arxiv.org/abs/2206.12429}
  {arXiv:2206.12429}  (\bibinfo {year} {2022}{\natexlab{b}})\BibitemShut
  {NoStop}%
\bibitem [{\citenamefont {Mezzadri}(2006)}]{Mezzadri06}%
  \BibitemOpen
  \bibfield  {author} {\bibinfo {author} {\bibfnamefont {F.}~\bibnamefont
  {Mezzadri}},\ }\bibfield  {title} {\bibinfo {title} {{How to generate random
  matrices from the classical compact groups}},\ }\Eprint
  {https://arxiv.org/abs/math-ph/0609050} {arXiv:math-ph/0609050}  (\bibinfo
  {year} {2006})\BibitemShut {NoStop}%
\bibitem [{\citenamefont {Wolf}\ \emph {et~al.}(2008)\citenamefont {Wolf},
  \citenamefont {Verstraete}, \citenamefont {Hastings},\ and\ \citenamefont
  {Cirac}}]{Wolf08}%
  \BibitemOpen
  \bibfield  {author} {\bibinfo {author} {\bibfnamefont {M.~M.}\ \bibnamefont
  {Wolf}}, \bibinfo {author} {\bibfnamefont {F.}~\bibnamefont {Verstraete}},
  \bibinfo {author} {\bibfnamefont {M.~B.}\ \bibnamefont {Hastings}},\ and\
  \bibinfo {author} {\bibfnamefont {J.~I.}\ \bibnamefont {Cirac}},\ }\bibfield
  {title} {\bibinfo {title} {{Area Laws in Quantum Systems: Mutual Information
  and Correlations}},\ }\href {https://doi.org/10.1103/PhysRevLett.100.070502}
  {\bibfield  {journal} {\bibinfo  {journal} {Phys. Rev. Lett.}\ }\textbf
  {\bibinfo {volume} {100}},\ \bibinfo {pages} {070502} (\bibinfo {year}
  {2008})}\BibitemShut {NoStop}%
\bibitem [{\citenamefont {Calabrese}\ and\ \citenamefont
  {Cardy}(2009)}]{Calabrese09b}%
  \BibitemOpen
  \bibfield  {author} {\bibinfo {author} {\bibfnamefont {P.}~\bibnamefont
  {Calabrese}}\ and\ \bibinfo {author} {\bibfnamefont {J.}~\bibnamefont
  {Cardy}},\ }\bibfield  {title} {\bibinfo {title} {{Entanglement entropy and
  conformal field theory}},\ }\href
  {https://doi.org/10.1088/1751-8113/42/50/504005} {\bibfield  {journal}
  {\bibinfo  {journal} {J. Phys. A: Math. Theor.}\ }\textbf {\bibinfo {volume}
  {42}},\ \bibinfo {pages} {504005} (\bibinfo {year} {2009})}\BibitemShut
  {NoStop}%
\bibitem [{\citenamefont {Gullans}\ and\ \citenamefont
  {Huse}(2020)}]{Gullans20a}%
  \BibitemOpen
  \bibfield  {author} {\bibinfo {author} {\bibfnamefont {M.~J.}\ \bibnamefont
  {Gullans}}\ and\ \bibinfo {author} {\bibfnamefont {D.~A.}\ \bibnamefont
  {Huse}},\ }\bibfield  {title} {\bibinfo {title} {{Dynamical Purification
  Phase Transition Induced by Quantum Measurements}},\ }\href
  {https://doi.org/10.1103/PhysRevX.10.041020} {\bibfield  {journal} {\bibinfo
  {journal} {Phys. Rev. X}\ }\textbf {\bibinfo {volume} {10}},\ \bibinfo
  {pages} {041020} (\bibinfo {year} {2020})}\BibitemShut {NoStop}%
\bibitem [{\citenamefont {Calabrese}\ and\ \citenamefont
  {Cardy}(2004)}]{Calabrese04}%
  \BibitemOpen
  \bibfield  {author} {\bibinfo {author} {\bibfnamefont {P.}~\bibnamefont
  {Calabrese}}\ and\ \bibinfo {author} {\bibfnamefont {J.}~\bibnamefont
  {Cardy}},\ }\bibfield  {title} {\bibinfo {title} {{Entanglement entropy and
  quantum field theory}},\ }\href
  {https://doi.org/10.1088/1742-5468/2004/06/p06002} {\bibfield  {journal}
  {\bibinfo  {journal} {J. Stat. Mech.}\ }\textbf {\bibinfo {volume} {2004}},\
  \bibinfo {pages} {P06002} (\bibinfo {year} {2004})}\BibitemShut {NoStop}%
\bibitem [{\citenamefont {Vasseur}\ \emph {et~al.}(2019)\citenamefont
  {Vasseur}, \citenamefont {Potter}, \citenamefont {You},\ and\ \citenamefont
  {Ludwig}}]{Vasseur19}%
  \BibitemOpen
  \bibfield  {author} {\bibinfo {author} {\bibfnamefont {R.}~\bibnamefont
  {Vasseur}}, \bibinfo {author} {\bibfnamefont {A.~C.}\ \bibnamefont {Potter}},
  \bibinfo {author} {\bibfnamefont {Y.-Z.}\ \bibnamefont {You}},\ and\ \bibinfo
  {author} {\bibfnamefont {A.~W.~W.}\ \bibnamefont {Ludwig}},\ }\bibfield
  {title} {\bibinfo {title} {{Entanglement transitions from holographic random
  tensor networks}},\ }\href {https://doi.org/10.1103/PhysRevB.100.134203}
  {\bibfield  {journal} {\bibinfo  {journal} {Phys. Rev. B}\ }\textbf {\bibinfo
  {volume} {100}},\ \bibinfo {pages} {134203} (\bibinfo {year}
  {2019})}\BibitemShut {NoStop}%
\bibitem [{\citenamefont {Furukawa}\ \emph {et~al.}(2009)\citenamefont
  {Furukawa}, \citenamefont {Pasquier},\ and\ \citenamefont
  {Shiraishi}}]{Furukawa09}%
  \BibitemOpen
  \bibfield  {author} {\bibinfo {author} {\bibfnamefont {S.}~\bibnamefont
  {Furukawa}}, \bibinfo {author} {\bibfnamefont {V.}~\bibnamefont {Pasquier}},\
  and\ \bibinfo {author} {\bibfnamefont {J.}~\bibnamefont {Shiraishi}},\
  }\bibfield  {title} {\bibinfo {title} {{Mutual Information and Boson Radius
  in a $c=1$ Critical System in One Dimension}},\ }\href
  {https://doi.org/10.1103/PhysRevLett.102.170602} {\bibfield  {journal}
  {\bibinfo  {journal} {Phys. Rev. Lett.}\ }\textbf {\bibinfo {volume} {102}},\
  \bibinfo {pages} {170602} (\bibinfo {year} {2009})}\BibitemShut {NoStop}%
\bibitem [{\citenamefont {Calabrese}\ \emph {et~al.}(2009)\citenamefont
  {Calabrese}, \citenamefont {Cardy},\ and\ \citenamefont
  {Tonni}}]{Calabrese09a}%
  \BibitemOpen
  \bibfield  {author} {\bibinfo {author} {\bibfnamefont {P.}~\bibnamefont
  {Calabrese}}, \bibinfo {author} {\bibfnamefont {J.}~\bibnamefont {Cardy}},\
  and\ \bibinfo {author} {\bibfnamefont {E.}~\bibnamefont {Tonni}},\ }\bibfield
   {title} {\bibinfo {title} {{Entanglement entropy of two disjoint intervals
  in conformal field theory}},\ }\href
  {https://doi.org/10.1088/1742-5468/2009/11/p11001} {\bibfield  {journal}
  {\bibinfo  {journal} {J. Stat. Mech.}\ }\textbf {\bibinfo {volume} {2009}},\
  \bibinfo {pages} {P11001} (\bibinfo {year} {2009})}\BibitemShut {NoStop}%
\bibitem [{\citenamefont {Calabrese}\ \emph {et~al.}(2011)\citenamefont
  {Calabrese}, \citenamefont {Cardy},\ and\ \citenamefont
  {Tonni}}]{Calabrese11}%
  \BibitemOpen
  \bibfield  {author} {\bibinfo {author} {\bibfnamefont {P.}~\bibnamefont
  {Calabrese}}, \bibinfo {author} {\bibfnamefont {J.}~\bibnamefont {Cardy}},\
  and\ \bibinfo {author} {\bibfnamefont {E.}~\bibnamefont {Tonni}},\ }\bibfield
   {title} {\bibinfo {title} {{Entanglement entropy of two disjoint intervals
  in conformal field theory: {II}}},\ }\href
  {https://doi.org/10.1088/1742-5468/2011/01/p01021} {\bibfield  {journal}
  {\bibinfo  {journal} {J. Stat. Mech.}\ }\textbf {\bibinfo {volume} {2011}},\
  \bibinfo {pages} {P01021} (\bibinfo {year} {2011})}\BibitemShut {NoStop}%
\bibitem [{\citenamefont {Chen}\ \emph {et~al.}(2020)\citenamefont {Chen},
  \citenamefont {Li}, \citenamefont {Fisher},\ and\ \citenamefont
  {Lucas}}]{Chen20}%
  \BibitemOpen
  \bibfield  {author} {\bibinfo {author} {\bibfnamefont {X.}~\bibnamefont
  {Chen}}, \bibinfo {author} {\bibfnamefont {Y.}~\bibnamefont {Li}}, \bibinfo
  {author} {\bibfnamefont {M.~P.~A.}\ \bibnamefont {Fisher}},\ and\ \bibinfo
  {author} {\bibfnamefont {A.}~\bibnamefont {Lucas}},\ }\bibfield  {title}
  {\bibinfo {title} {{Emergent conformal symmetry in nonunitary random dynamics
  of free fermions}},\ }\href
  {https://doi.org/10.1103/PhysRevResearch.2.033017} {\bibfield  {journal}
  {\bibinfo  {journal} {Phys. Rev. Research}\ }\textbf {\bibinfo {volume}
  {2}},\ \bibinfo {pages} {033017} (\bibinfo {year} {2020})}\BibitemShut
  {NoStop}%
\bibitem [{\citenamefont {Klich}\ and\ \citenamefont
  {Levitov}(2009)}]{Klich09}%
  \BibitemOpen
  \bibfield  {author} {\bibinfo {author} {\bibfnamefont {I.}~\bibnamefont
  {Klich}}\ and\ \bibinfo {author} {\bibfnamefont {L.}~\bibnamefont
  {Levitov}},\ }\bibfield  {title} {\bibinfo {title} {{Quantum Noise as an
  Entanglement Meter}},\ }\href
  {https://doi.org/10.1103/PhysRevLett.102.100502} {\bibfield  {journal}
  {\bibinfo  {journal} {Phys. Rev. Lett.}\ }\textbf {\bibinfo {volume} {102}},\
  \bibinfo {pages} {100502} (\bibinfo {year} {2009})}\BibitemShut {NoStop}%
\bibitem [{\citenamefont {Song}\ \emph {et~al.}(2010)\citenamefont {Song},
  \citenamefont {Rachel},\ and\ \citenamefont {Le~Hur}}]{Song10}%
  \BibitemOpen
  \bibfield  {author} {\bibinfo {author} {\bibfnamefont {H.~F.}\ \bibnamefont
  {Song}}, \bibinfo {author} {\bibfnamefont {S.}~\bibnamefont {Rachel}},\ and\
  \bibinfo {author} {\bibfnamefont {K.}~\bibnamefont {Le~Hur}},\ }\bibfield
  {title} {\bibinfo {title} {{General relation between entanglement and
  fluctuations in one dimension}},\ }\href
  {https://doi.org/10.1103/PhysRevB.82.012405} {\bibfield  {journal} {\bibinfo
  {journal} {Phys. Rev. B}\ }\textbf {\bibinfo {volume} {82}},\ \bibinfo
  {pages} {012405} (\bibinfo {year} {2010})}\BibitemShut {NoStop}%
\bibitem [{\citenamefont {Song}\ \emph {et~al.}(2011)\citenamefont {Song},
  \citenamefont {Flindt}, \citenamefont {Rachel}, \citenamefont {Klich},\ and\
  \citenamefont {Le~Hur}}]{Song11}%
  \BibitemOpen
  \bibfield  {author} {\bibinfo {author} {\bibfnamefont {H.~F.}\ \bibnamefont
  {Song}}, \bibinfo {author} {\bibfnamefont {C.}~\bibnamefont {Flindt}},
  \bibinfo {author} {\bibfnamefont {S.}~\bibnamefont {Rachel}}, \bibinfo
  {author} {\bibfnamefont {I.}~\bibnamefont {Klich}},\ and\ \bibinfo {author}
  {\bibfnamefont {K.}~\bibnamefont {Le~Hur}},\ }\bibfield  {title} {\bibinfo
  {title} {{Entanglement entropy from charge statistics: Exact relations for
  noninteracting many-body systems}},\ }\href
  {https://doi.org/10.1103/PhysRevB.83.161408} {\bibfield  {journal} {\bibinfo
  {journal} {Phys. Rev. B}\ }\textbf {\bibinfo {volume} {83}},\ \bibinfo
  {pages} {161408} (\bibinfo {year} {2011})}\BibitemShut {NoStop}%
\bibitem [{\citenamefont {Song}\ \emph {et~al.}(2012)\citenamefont {Song},
  \citenamefont {Rachel}, \citenamefont {Flindt}, \citenamefont {Klich},
  \citenamefont {Laflorencie},\ and\ \citenamefont {Le~Hur}}]{Song12}%
  \BibitemOpen
  \bibfield  {author} {\bibinfo {author} {\bibfnamefont {H.~F.}\ \bibnamefont
  {Song}}, \bibinfo {author} {\bibfnamefont {S.}~\bibnamefont {Rachel}},
  \bibinfo {author} {\bibfnamefont {C.}~\bibnamefont {Flindt}}, \bibinfo
  {author} {\bibfnamefont {I.}~\bibnamefont {Klich}}, \bibinfo {author}
  {\bibfnamefont {N.}~\bibnamefont {Laflorencie}},\ and\ \bibinfo {author}
  {\bibfnamefont {K.}~\bibnamefont {Le~Hur}},\ }\bibfield  {title} {\bibinfo
  {title} {{Bipartite fluctuations as a probe of many-body entanglement}},\
  }\href {https://doi.org/10.1103/PhysRevB.85.035409} {\bibfield  {journal}
  {\bibinfo  {journal} {Phys. Rev. B}\ }\textbf {\bibinfo {volume} {85}},\
  \bibinfo {pages} {035409} (\bibinfo {year} {2012})}\BibitemShut {NoStop}%
\bibitem [{\citenamefont {Calabrese}\ \emph {et~al.}(2012)\citenamefont
  {Calabrese}, \citenamefont {Mintchev},\ and\ \citenamefont
  {Vicari}}]{Calabrese12}%
  \BibitemOpen
  \bibfield  {author} {\bibinfo {author} {\bibfnamefont {P.}~\bibnamefont
  {Calabrese}}, \bibinfo {author} {\bibfnamefont {M.}~\bibnamefont
  {Mintchev}},\ and\ \bibinfo {author} {\bibfnamefont {E.}~\bibnamefont
  {Vicari}},\ }\bibfield  {title} {\bibinfo {title} {{Exact relations between
  particle fluctuations and entanglement in Fermi gases}},\ }\href
  {https://doi.org/10.1209/0295-5075/98/20003} {\bibfield  {journal} {\bibinfo
  {journal} {Europhys. Lett.}\ }\textbf {\bibinfo {volume} {98}},\ \bibinfo
  {pages} {20003} (\bibinfo {year} {2012})}\BibitemShut {NoStop}%
\bibitem [{\citenamefont {Rachel}\ \emph {et~al.}(2012)\citenamefont {Rachel},
  \citenamefont {Laflorencie}, \citenamefont {Song},\ and\ \citenamefont
  {Le~Hur}}]{Rachel12}%
  \BibitemOpen
  \bibfield  {author} {\bibinfo {author} {\bibfnamefont {S.}~\bibnamefont
  {Rachel}}, \bibinfo {author} {\bibfnamefont {N.}~\bibnamefont {Laflorencie}},
  \bibinfo {author} {\bibfnamefont {H.~F.}\ \bibnamefont {Song}},\ and\
  \bibinfo {author} {\bibfnamefont {K.}~\bibnamefont {Le~Hur}},\ }\bibfield
  {title} {\bibinfo {title} {{Detecting Quantum Critical Points Using Bipartite
  Fluctuations}},\ }\href {https://doi.org/10.1103/PhysRevLett.108.116401}
  {\bibfield  {journal} {\bibinfo  {journal} {Phys. Rev. Lett.}\ }\textbf
  {\bibinfo {volume} {108}},\ \bibinfo {pages} {116401} (\bibinfo {year}
  {2012})}\BibitemShut {NoStop}%
\bibitem [{\citenamefont {Fr\'erot}\ and\ \citenamefont
  {Roscilde}(2015)}]{Frerot15}%
  \BibitemOpen
  \bibfield  {author} {\bibinfo {author} {\bibfnamefont {I.}~\bibnamefont
  {Fr\'erot}}\ and\ \bibinfo {author} {\bibfnamefont {T.}~\bibnamefont
  {Roscilde}},\ }\bibfield  {title} {\bibinfo {title} {{Area law and its
  violation: A microscopic inspection into the structure of entanglement and
  fluctuations}},\ }\href {https://doi.org/10.1103/PhysRevB.92.115129}
  {\bibfield  {journal} {\bibinfo  {journal} {Phys. Rev. B}\ }\textbf {\bibinfo
  {volume} {92}},\ \bibinfo {pages} {115129} (\bibinfo {year}
  {2015})}\BibitemShut {NoStop}%
\bibitem [{\citenamefont {Cr\'epel}\ \emph {et~al.}(2021)\citenamefont
  {Cr\'epel}, \citenamefont {Hackenbroich}, \citenamefont {Regnault},\ and\
  \citenamefont {Estienne}}]{Crepel21}%
  \BibitemOpen
  \bibfield  {author} {\bibinfo {author} {\bibfnamefont {V.}~\bibnamefont
  {Cr\'epel}}, \bibinfo {author} {\bibfnamefont {A.}~\bibnamefont
  {Hackenbroich}}, \bibinfo {author} {\bibfnamefont {N.}~\bibnamefont
  {Regnault}},\ and\ \bibinfo {author} {\bibfnamefont {B.}~\bibnamefont
  {Estienne}},\ }\bibfield  {title} {\bibinfo {title} {{Universal signatures of
  Dirac fermions in entanglement and charge fluctuations}},\ }\href
  {https://doi.org/10.1103/PhysRevB.103.235108} {\bibfield  {journal} {\bibinfo
   {journal} {Phys. Rev. B}\ }\textbf {\bibinfo {volume} {103}},\ \bibinfo
  {pages} {235108} (\bibinfo {year} {2021})}\BibitemShut {NoStop}%
\bibitem [{\citenamefont {Estienne}\ \emph {et~al.}(2022)\citenamefont
  {Estienne}, \citenamefont {St{\'e}phan},\ and\ \citenamefont
  {Witczak-Krempa}}]{Estienne22}%
  \BibitemOpen
  \bibfield  {author} {\bibinfo {author} {\bibfnamefont {B.}~\bibnamefont
  {Estienne}}, \bibinfo {author} {\bibfnamefont {J.-M.}\ \bibnamefont
  {St{\'e}phan}},\ and\ \bibinfo {author} {\bibfnamefont {W.}~\bibnamefont
  {Witczak-Krempa}},\ }\bibfield  {title} {\bibinfo {title} {{Cornering the
  universal shape of fluctuations}},\ }\href
  {https://doi.org/10.1038/s41467-021-27727-1} {\bibfield  {journal} {\bibinfo
  {journal} {Nat. Commun.}\ }\textbf {\bibinfo {volume} {13}},\ \bibinfo
  {pages} {287} (\bibinfo {year} {2022})}\BibitemShut {NoStop}%
\bibitem [{\citenamefont {Xavier}\ \emph {et~al.}(2018)\citenamefont {Xavier},
  \citenamefont {Alcaraz},\ and\ \citenamefont {Sierra}}]{Xavier18}%
  \BibitemOpen
  \bibfield  {author} {\bibinfo {author} {\bibfnamefont {J.~C.}\ \bibnamefont
  {Xavier}}, \bibinfo {author} {\bibfnamefont {F.~C.}\ \bibnamefont
  {Alcaraz}},\ and\ \bibinfo {author} {\bibfnamefont {G.}~\bibnamefont
  {Sierra}},\ }\bibfield  {title} {\bibinfo {title} {{Equipartition of the
  entanglement entropy}},\ }\href {https://doi.org/10.1103/PhysRevB.98.041106}
  {\bibfield  {journal} {\bibinfo  {journal} {Phys. Rev. B}\ }\textbf {\bibinfo
  {volume} {98}},\ \bibinfo {pages} {041106} (\bibinfo {year}
  {2018})}\BibitemShut {NoStop}%
\bibitem [{\citenamefont {Goldstein}\ and\ \citenamefont
  {Sela}(2018)}]{Goldstein18}%
  \BibitemOpen
  \bibfield  {author} {\bibinfo {author} {\bibfnamefont {M.}~\bibnamefont
  {Goldstein}}\ and\ \bibinfo {author} {\bibfnamefont {E.}~\bibnamefont
  {Sela}},\ }\bibfield  {title} {\bibinfo {title} {{Symmetry-Resolved
  Entanglement in Many-Body Systems}},\ }\href
  {https://doi.org/10.1103/PhysRevLett.120.200602} {\bibfield  {journal}
  {\bibinfo  {journal} {Phys. Rev. Lett.}\ }\textbf {\bibinfo {volume} {120}},\
  \bibinfo {pages} {200602} (\bibinfo {year} {2018})}\BibitemShut {NoStop}%
\bibitem [{\citenamefont {Feldman}\ and\ \citenamefont
  {Goldstein}(2019)}]{Feldman19}%
  \BibitemOpen
  \bibfield  {author} {\bibinfo {author} {\bibfnamefont {N.}~\bibnamefont
  {Feldman}}\ and\ \bibinfo {author} {\bibfnamefont {M.}~\bibnamefont
  {Goldstein}},\ }\bibfield  {title} {\bibinfo {title} {{Dynamics of
  charge-resolved entanglement after a local quench}},\ }\href
  {https://doi.org/10.1103/PhysRevB.100.235146} {\bibfield  {journal} {\bibinfo
   {journal} {Phys. Rev. B}\ }\textbf {\bibinfo {volume} {100}},\ \bibinfo
  {pages} {235146} (\bibinfo {year} {2019})}\BibitemShut {NoStop}%
\bibitem [{\citenamefont {Bonsignori}\ \emph {et~al.}(2019)\citenamefont
  {Bonsignori}, \citenamefont {Ruggiero},\ and\ \citenamefont
  {Calabrese}}]{Bonsignori19}%
  \BibitemOpen
  \bibfield  {author} {\bibinfo {author} {\bibfnamefont {R.}~\bibnamefont
  {Bonsignori}}, \bibinfo {author} {\bibfnamefont {P.}~\bibnamefont
  {Ruggiero}},\ and\ \bibinfo {author} {\bibfnamefont {P.}~\bibnamefont
  {Calabrese}},\ }\bibfield  {title} {\bibinfo {title} {{Symmetry resolved
  entanglement in free fermionic systems}},\ }\href
  {https://doi.org/10.1088/1751-8121/ab4b77} {\bibfield  {journal} {\bibinfo
  {journal} {J. Phys. A: Math. Theor.}\ }\textbf {\bibinfo {volume} {52}},\
  \bibinfo {pages} {475302} (\bibinfo {year} {2019})}\BibitemShut {NoStop}%
\bibitem [{\citenamefont {Parez}\ \emph {et~al.}(2021)\citenamefont {Parez},
  \citenamefont {Bonsignori},\ and\ \citenamefont {Calabrese}}]{Parez21a}%
  \BibitemOpen
  \bibfield  {author} {\bibinfo {author} {\bibfnamefont {G.}~\bibnamefont
  {Parez}}, \bibinfo {author} {\bibfnamefont {R.}~\bibnamefont {Bonsignori}},\
  and\ \bibinfo {author} {\bibfnamefont {P.}~\bibnamefont {Calabrese}},\
  }\bibfield  {title} {\bibinfo {title} {Quasiparticle dynamics of
  symmetry-resolved entanglement after a quench: Examples of conformal field
  theories and free fermions},\ }\href
  {https://doi.org/10.1103/PhysRevB.103.L041104} {\bibfield  {journal}
  {\bibinfo  {journal} {Phys. Rev. B}\ }\textbf {\bibinfo {volume} {103}},\
  \bibinfo {pages} {L041104} (\bibinfo {year} {2021})}\BibitemShut {NoStop}%
\bibitem [{\citenamefont {Parez}\ \emph {et~al.}()\citenamefont {Parez},
  \citenamefont {Bonsignori},\ and\ \citenamefont {Calabrese}}]{Parez21b}%
  \BibitemOpen
  \bibfield  {author} {\bibinfo {author} {\bibfnamefont {G.}~\bibnamefont
  {Parez}}, \bibinfo {author} {\bibfnamefont {R.}~\bibnamefont {Bonsignori}},\
  and\ \bibinfo {author} {\bibfnamefont {P.}~\bibnamefont {Calabrese}},\
  }\bibfield  {title} {\bibinfo {title} {Exact quench dynamics of symmetry
  resolved entanglement in a free fermion chain},\ }\href
  {https://doi.org/10.1088/1742-5468/ac21d7} {\bibfield  {journal} {\bibinfo
  {journal} {J. Stat. Mech.}\ }\textbf {\bibinfo {volume} {2021}},\ \bibinfo
  {pages} {093102}}\BibitemShut {NoStop}%
\bibitem [{\citenamefont {Fraenkel}\ and\ \citenamefont
  {Goldstein}(2021)}]{Fraenkel21}%
  \BibitemOpen
  \bibfield  {author} {\bibinfo {author} {\bibfnamefont {S.}~\bibnamefont
  {Fraenkel}}\ and\ \bibinfo {author} {\bibfnamefont {M.}~\bibnamefont
  {Goldstein}},\ }\bibfield  {title} {\bibinfo {title} {{Entanglement measures
  in a nonequilibrium steady state: Exact results in one dimension}},\ }\href
  {https://doi.org/10.21468/SciPostPhys.11.4.085} {\bibfield  {journal}
  {\bibinfo  {journal} {SciPost Phys.}\ }\textbf {\bibinfo {volume} {11}},\
  \bibinfo {pages} {085} (\bibinfo {year} {2021})}\BibitemShut {NoStop}%
\bibitem [{\citenamefont {Cornfeld}\ \emph {et~al.}(2018)\citenamefont
  {Cornfeld}, \citenamefont {Goldstein},\ and\ \citenamefont
  {Sela}}]{Cornfeld18}%
  \BibitemOpen
  \bibfield  {author} {\bibinfo {author} {\bibfnamefont {E.}~\bibnamefont
  {Cornfeld}}, \bibinfo {author} {\bibfnamefont {M.}~\bibnamefont
  {Goldstein}},\ and\ \bibinfo {author} {\bibfnamefont {E.}~\bibnamefont
  {Sela}},\ }\bibfield  {title} {\bibinfo {title} {Imbalance entanglement:
  Symmetry decomposition of negativity},\ }\href
  {https://doi.org/10.1103/PhysRevA.98.032302} {\bibfield  {journal} {\bibinfo
  {journal} {Phys. Rev. A}\ }\textbf {\bibinfo {volume} {98}},\ \bibinfo
  {pages} {032302} (\bibinfo {year} {2018})}\BibitemShut {NoStop}%
\bibitem [{\citenamefont {Parez}\ \emph {et~al.}(2022)\citenamefont {Parez},
  \citenamefont {Bonsignori},\ and\ \citenamefont {Calabrese}}]{Parez22}%
  \BibitemOpen
  \bibfield  {author} {\bibinfo {author} {\bibfnamefont {G.}~\bibnamefont
  {Parez}}, \bibinfo {author} {\bibfnamefont {R.}~\bibnamefont {Bonsignori}},\
  and\ \bibinfo {author} {\bibfnamefont {P.}~\bibnamefont {Calabrese}},\
  }\bibfield  {title} {\bibinfo {title} {Dynamics of charge-imbalance-resolved
  entanglement negativity after a quench in a free-fermion model},\ }\href
  {https://doi.org/10.1088/1742-5468/ac666c} {\bibfield  {journal} {\bibinfo
  {journal} {J. Stat. Mech.}\ }\textbf {\bibinfo {volume} {2022}},\ \bibinfo
  {pages} {053103} (\bibinfo {year} {2022})}\BibitemShut {NoStop}%
\bibitem [{\citenamefont {Noel}\ \emph {et~al.}(2022)\citenamefont {Noel},
  \citenamefont {Niroula}, \citenamefont {Zhu}, \citenamefont {Risinger},
  \citenamefont {Egan}, \citenamefont {Biswas}, \citenamefont {Cetina},
  \citenamefont {Gorshkov}, \citenamefont {Gullans}, \citenamefont {Huse},\
  and\ \citenamefont {Monroe}}]{Noel22}%
  \BibitemOpen
  \bibfield  {author} {\bibinfo {author} {\bibfnamefont {C.}~\bibnamefont
  {Noel}}, \bibinfo {author} {\bibfnamefont {P.}~\bibnamefont {Niroula}},
  \bibinfo {author} {\bibfnamefont {D.}~\bibnamefont {Zhu}}, \bibinfo {author}
  {\bibfnamefont {A.}~\bibnamefont {Risinger}}, \bibinfo {author}
  {\bibfnamefont {L.}~\bibnamefont {Egan}}, \bibinfo {author} {\bibfnamefont
  {D.}~\bibnamefont {Biswas}}, \bibinfo {author} {\bibfnamefont
  {M.}~\bibnamefont {Cetina}}, \bibinfo {author} {\bibfnamefont {A.~V.}\
  \bibnamefont {Gorshkov}}, \bibinfo {author} {\bibfnamefont {M.~J.}\
  \bibnamefont {Gullans}}, \bibinfo {author} {\bibfnamefont {D.~A.}\
  \bibnamefont {Huse}},\ and\ \bibinfo {author} {\bibfnamefont
  {C.}~\bibnamefont {Monroe}},\ }\bibfield  {title} {\bibinfo {title}
  {{Measurement-induced quantum phases realized in a trapped-ion quantum
  computer}},\ }\href {https://doi.org/10.1038/s41567-022-01619-7} {\bibfield
  {journal} {\bibinfo  {journal} {Nat. Phys.}\ }\textbf {\bibinfo {volume}
  {18}},\ \bibinfo {pages} {760} (\bibinfo {year} {2022})}\BibitemShut
  {NoStop}%
\bibitem [{\citenamefont {Koh}\ \emph {et~al.}(2022)\citenamefont {Koh},
  \citenamefont {Sun}, \citenamefont {Motta},\ and\ \citenamefont
  {Minnich}}]{Koh22}%
  \BibitemOpen
  \bibfield  {author} {\bibinfo {author} {\bibfnamefont {J.~M.}\ \bibnamefont
  {Koh}}, \bibinfo {author} {\bibfnamefont {S.-N.}\ \bibnamefont {Sun}},
  \bibinfo {author} {\bibfnamefont {M.}~\bibnamefont {Motta}},\ and\ \bibinfo
  {author} {\bibfnamefont {A.~J.}\ \bibnamefont {Minnich}},\ }\bibfield
  {title} {\bibinfo {title} {{Experimental Realization of a Measurement-Induced
  Entanglement Phase Transition on a Superconducting Quantum Processor}},\
  }\Eprint {https://arxiv.org/abs/2203.04338} {arXiv:2203.04338}  (\bibinfo
  {year} {2022})\BibitemShut {NoStop}%
\bibitem [{\citenamefont {Garratt}\ \emph {et~al.}(2022)\citenamefont
  {Garratt}, \citenamefont {Weinstein},\ and\ \citenamefont
  {Altman}}]{Garratt22}%
  \BibitemOpen
  \bibfield  {author} {\bibinfo {author} {\bibfnamefont {S.~J.}\ \bibnamefont
  {Garratt}}, \bibinfo {author} {\bibfnamefont {Z.}~\bibnamefont {Weinstein}},\
  and\ \bibinfo {author} {\bibfnamefont {E.}~\bibnamefont {Altman}},\
  }\bibfield  {title} {\bibinfo {title} {{Measurements conspire nonlocally to
  restructure critical quantum states}},\ }\Eprint
  {https://arxiv.org/abs/2207.09476} {arXiv:2207.09476}  (\bibinfo {year}
  {2022})\BibitemShut {NoStop}%
\bibitem [{\citenamefont {Richter}\ \emph {et~al.}(2022)\citenamefont
  {Richter}, \citenamefont {Lunt},\ and\ \citenamefont {Pal}}]{Richter22}%
  \BibitemOpen
  \bibfield  {author} {\bibinfo {author} {\bibfnamefont {J.}~\bibnamefont
  {Richter}}, \bibinfo {author} {\bibfnamefont {O.}~\bibnamefont {Lunt}},\ and\
  \bibinfo {author} {\bibfnamefont {A.}~\bibnamefont {Pal}},\ }\bibfield
  {title} {\bibinfo {title} {{Transport and entanglement growth in long-range
  random Clifford circuits}},\ }\Eprint {https://arxiv.org/abs/2205.06309}
  {arXiv:2205.06309}  (\bibinfo {year} {2022})\BibitemShut {NoStop}%
\bibitem [{\citenamefont {Iadecola}\ \emph {et~al.}(2022)\citenamefont
  {Iadecola}, \citenamefont {Ganeshan}, \citenamefont {Pixley},\ and\
  \citenamefont {Wilson}}]{Iadecola22}%
  \BibitemOpen
  \bibfield  {author} {\bibinfo {author} {\bibfnamefont {T.}~\bibnamefont
  {Iadecola}}, \bibinfo {author} {\bibfnamefont {S.}~\bibnamefont {Ganeshan}},
  \bibinfo {author} {\bibfnamefont {J.~H.}\ \bibnamefont {Pixley}},\ and\
  \bibinfo {author} {\bibfnamefont {J.~H.}\ \bibnamefont {Wilson}},\ }\bibfield
   {title} {\bibinfo {title} {Dynamical entanglement transition in the
  probabilistic control of chaos},\ }\Eprint {https://arxiv.org/abs/2207.12415}
  {arXiv:2207.12415}  (\bibinfo {year} {2022})\BibitemShut {NoStop}%
\bibitem [{\citenamefont {Buchhold}\ \emph {et~al.}(2022)\citenamefont
  {Buchhold}, \citenamefont {Müller},\ and\ \citenamefont
  {Diehl}}]{Buchhold22}%
  \BibitemOpen
  \bibfield  {author} {\bibinfo {author} {\bibfnamefont {M.}~\bibnamefont
  {Buchhold}}, \bibinfo {author} {\bibfnamefont {T.}~\bibnamefont {Müller}},\
  and\ \bibinfo {author} {\bibfnamefont {S.}~\bibnamefont {Diehl}},\ }\bibfield
   {title} {\bibinfo {title} {{Revealing measurement-induced phase transitions
  by pre-selection}},\ }\Eprint {https://arxiv.org/abs/2208.10506}
  {arXiv:2208.10506}  (\bibinfo {year} {2022})\BibitemShut {NoStop}%
\bibitem [{\citenamefont {Friedman}\ \emph {et~al.}(2022)\citenamefont
  {Friedman}, \citenamefont {Hart},\ and\ \citenamefont
  {Nandkishore}}]{Friedman22}%
  \BibitemOpen
  \bibfield  {author} {\bibinfo {author} {\bibfnamefont {A.~J.}\ \bibnamefont
  {Friedman}}, \bibinfo {author} {\bibfnamefont {O.}~\bibnamefont {Hart}},\
  and\ \bibinfo {author} {\bibfnamefont {R.}~\bibnamefont {Nandkishore}},\
  }\bibfield  {title} {\bibinfo {title} {{Measurement-induced phases of matter
  require adaptive dynamics}},\ }\Eprint {https://arxiv.org/abs/2210.07256}
  {arXiv:2210.07256}  (\bibinfo {year} {2022})\BibitemShut {NoStop}%
\bibitem [{\citenamefont {Iyoda}\ and\ \citenamefont {Sagawa}(2018)}]{Iyoda18}%
  \BibitemOpen
  \bibfield  {author} {\bibinfo {author} {\bibfnamefont {E.}~\bibnamefont
  {Iyoda}}\ and\ \bibinfo {author} {\bibfnamefont {T.}~\bibnamefont {Sagawa}},\
  }\bibfield  {title} {\bibinfo {title} {{Scrambling of quantum information in
  quantum many-body systems}},\ }\href
  {https://doi.org/10.1103/PhysRevA.97.042330} {\bibfield  {journal} {\bibinfo
  {journal} {Phys. Rev. A}\ }\textbf {\bibinfo {volume} {97}},\ \bibinfo
  {pages} {042330} (\bibinfo {year} {2018})}\BibitemShut {NoStop}%
\bibitem [{\citenamefont {Houdayer}\ and\ \citenamefont
  {Hartmann}(2004)}]{Houdayer04}%
  \BibitemOpen
  \bibfield  {author} {\bibinfo {author} {\bibfnamefont {J.}~\bibnamefont
  {Houdayer}}\ and\ \bibinfo {author} {\bibfnamefont {A.~K.}\ \bibnamefont
  {Hartmann}},\ }\bibfield  {title} {\bibinfo {title} {{Low-temperature
  behavior of two-dimensional Gaussian Ising spin glasses}},\ }\href
  {https://doi.org/10.1103/PhysRevB.70.014418} {\bibfield  {journal} {\bibinfo
  {journal} {Phys. Rev. B}\ }\textbf {\bibinfo {volume} {70}},\ \bibinfo
  {pages} {014418} (\bibinfo {year} {2004})}\BibitemShut {NoStop}%
\bibitem [{\citenamefont {Luitz}\ \emph {et~al.}(2015)\citenamefont {Luitz},
  \citenamefont {Laflorencie},\ and\ \citenamefont {Alet}}]{Luitz15}%
  \BibitemOpen
  \bibfield  {author} {\bibinfo {author} {\bibfnamefont {D.~J.}\ \bibnamefont
  {Luitz}}, \bibinfo {author} {\bibfnamefont {N.}~\bibnamefont {Laflorencie}},\
  and\ \bibinfo {author} {\bibfnamefont {F.}~\bibnamefont {Alet}},\ }\bibfield
  {title} {\bibinfo {title} {{Many-body localization edge in the random-field
  Heisenberg chain}},\ }\href {https://doi.org/10.1103/PhysRevB.91.081103}
  {\bibfield  {journal} {\bibinfo  {journal} {Phys. Rev. B}\ }\textbf {\bibinfo
  {volume} {91}},\ \bibinfo {pages} {081103} (\bibinfo {year}
  {2015})}\BibitemShut {NoStop}%
\end{thebibliography}%
%%%%%%%%%%%%%%%%%%%%%%%%%%%%%%%%%%%%%%%%%%%%%%%%%%%%%%%%%%%%%%%%%%%%%%%%%%%%%%%%%%%%%

\end{document}